\documentclass[11pt]{article}
\usepackage[outline]{contour}
\usepackage{graphicx}
\graphicspath{{Figures/}}
\usepackage[sort&compress, numbers]{natbib}
\usepackage{amsmath}
\usepackage{array}
\usepackage{url}
\usepackage{amssymb}
\usepackage{stmaryrd}
\usepackage{mathtools}
\usepackage{amsthm}
\usepackage{tabu}
\usepackage{xcolor,colortbl}
\usepackage{colortbl}
\usepackage{fullpage}
\usepackage[footnotesize]{caption}
\usepackage[caption=false,font=footnotesize]{subfig}
\usepackage{multirow}
\definecolor{mycolor1}{rgb}{0.83,0.83,1}
\usepackage{tikz}
\usetikzlibrary{arrows}

\usepackage{pgfplots}
\usepackage{multirow}
\usepackage{rotating}

\newtheorem{thm}{Theorem}
\newtheorem{lem}{Lemma}
\newtheorem{prop}{Proposition}

\newtheorem{remark}{Remark}
\usepackage{color}
\usepackage{hyperref}

\newcommand{\range}[1]{\llbracket #1 \rrbracket}

\newcommand{\supp}{{\rm supp}\,}

\newcommand{\subtoind}[4]{#1 \xrightleftharpoons{#4}(#2,#3)}

\DeclareMathOperator{\st}{s.t.}

\DeclareMathOperator*{\argmin}{arg\,min}

\newcommand{\bs}{\boldsymbol}
\newcommand{\bb}{\mathbb}
\newcommand{\cl}{\mathcal}

\newcommand{\ts}{\textstyle}
\newcommand{\ie}{\emph{i.e.}, }
\newcommand{\eg}{\emph{e.g.}, }
\newcommand{\etal}{\emph{et al.}}

\newcommand{\iid}{%
    \ifmmode
        \mathrm{iid}%
    \else%
        iid\xspace%
    \fi%
}

\begin{document}

\title{Close Encounters of the Binary Kind: Signal Reconstruction Guarantees for Compressive Hadamard Sampling\\ with Haar Wavelet Basis}
\author{A.~Moshtaghpour\thanks{ISPGroup, ICTEAM/ELEN, UCLouvain, Belgium 
    (\url{{amirafshar.moshtaghpour,laurent.jacques}@uclouvain.be}). AM is funded by the FRIA/FNRS. LJ funded by the F.R.S.-FNRS.} \and J.~M.~Bioucas-Dias\footnotemark[2] \and L.~Jacques\footnotemark[1]%
}
\markboth{}%
{}

\maketitle
\begin{abstract} 
We investigate the problems of 1-D and 2-D signal recovery from subsampled Hadamard measurements using Haar wavelet sparsity prior. These problems are of interest in, \eg computational imaging applications relying on optical multiplexing or single pixel imaging. However, the realization of such modalities is often hindered by the coherence between the Hadamard and Haar bases. The variable and multilevel density sampling strategies solve this issue by adjusting the subsampling process to the local and multilevel coherence, respectively, between the two bases; hence enabling successful signal recovery. In this work, we compute an explicit sample-complexity bound for Hadamard-Haar systems as well as uniform and non-uniform recovery guarantees; a seemingly missing result in the related literature. We explore the faithfulness of the numerical simulations to the theoretical results and show in a practically relevant instance, \eg single pixel camera, that the target signal can be obtained from a few Hadamard measurements.
\end{abstract}
	{\noindent {\em Keywords: Hadamard transform, Haar wavelet, variable density sampling, compressive sensing.}
\section{Introduction} \label{sec:intro} 
The theory of Compressed Sensing (CS), introduced by Donoho \cite{donoho2006compressed} and Cand\`es and Tao \cite{candes2006near}, is now a versatile sampling paradigm 
in many real-world applications, \eg Magnetic Resonance Imaging (MRI) \cite{lustig2008compressed}, fluorescence microscopy~\cite{studer2012compressive,roman2014asymptotic}, and imaging~\cite{duarte2008single}. Mathematically, CS considers the problem of recovering a signal $\bs x \in \bb C^N$ from $M$ noisy measurements 
\begin{equation}
\bs y = \bs A \bs x + \bs n \in \bb C^M. \label{eq: cs acquisition model}
\end{equation}
In \eqref{eq: cs acquisition model}, the matrix $\bs A \in \bb C^{M\times N}$ approximates the physical sensing process of $\bs x$, and $\bs n$ denotes an additive noise vector. A typical goal in CS is minimizing the number of measurements $M$ while guaranteeing the quality of the signal recovery. This is indeed a critical aspect of the applications of CS. For example, the number of measurements in MRI application can be translated into the X-ray dose, which has to be minimized. 

In this paper, we tackle an important problem in the applications of CS theory: recovering a signal from subsampled Hadamard measurements using the Haar wavelet sparsity basis. In particular, in a wide range of imaging modalities, \eg optical multiplexing or single pixel camera~\cite{duarte2008single}, the sensing process can be modeled as taking measurements from the Hadamard transform. Moreover, considering the Haar wavelet basis paves the way to study other wavelet bases in combination with the Hadamard matrix. In this context, the main question becomes: how to design an optimum sampling strategy for subsampling the Hadamard measurements. Note that in this paper, the term ``optimum'' refers to a sampling strategy that minimizes the required number of measurements without degrading the quality of the signal reconstruction. 

Traditional CS relying on orthonormal sensing systems \cite{candes2007sparsity} suggests selecting the rows of the sensing matrix (in our case, the Hadamard matrix) uniformly at random, \ie according to a Uniform Density Sampling (UDS). Unfortunately, this approach fails when the target signal is sparse or compressible in a basis, called sparsity basis, that is too coherent with the sensing basis (see Sec.~\ref{sec:CS}). One example of this failure is the Hadamard-Haar system, where the sensing basis (Hadamard) is maximally coherent with the sparsity basis (Haar wavelet). This drawback is often called the ``coherence barrier'' in the literature \cite{adcock2015quest}.

Nevertheless, this barrier can be broken. Several empirical~\cite{baldassarre2015structured,moshtaghpour2018compressive,moshtaghpour2018compressive2} and theoretical evidences~\cite{adcock2017breaking,krahmer2014stable,adcock2015quest} suggest using a non-uniform density sampling strategy~\cite{krahmer2014stable,adcock2017breaking,puy2011variable}, which densifies the subsampling of the lower Hadamard frequencies, to obtain superior signal reconstruction quality. In a general context, Krahmer and Ward~\cite{krahmer2014stable} and Adcock~\etal~\cite{adcock2017breaking} arguably began to replace the notion of global coherence with its local versions, \ie \textit{local coherence}  and \textit{multilevel coherence} parameters, respectively. The idea in these works is to discriminate the elements of the sensing basis (\eg Hadamard) in favor of those that are highly coherent with all the elements of the sparsity basis (\eg Haar). 

Although there are other versions of non-uniform sampling strategies, \eg \cite{puy2011variable,bigot2016analysis,boyer2015compressed,li2017compressed}, we here focus on the framework of Krahmer and Ward~\cite{krahmer2014stable}, called here Variable Density Sampling (VDS), and the one of Adcock~\etal~\cite{adcock2017breaking}, called here Multilevel Density Sampling (MDS). These allow us to derive suitable sampling strategies for Hadamard-Haar systems. Note that the term ``VDS'' is often used in the literature for other non-uniform density sampling strategies, while we here use VDS and MDS terms to distinguish the frameworks of \cite{krahmer2014stable} and \cite{adcock2017breaking,adcock2015quest}. 

An important aspect of CS theory is the difference between \textit{uniform} and \textit{non-uniform}\footnote{A word of caution. Uniform and non-uniform density sampling strategies are related to the way the rows of a matrix are subsampled and should not be confused with uniform and non-uniform recovery guarantees.} (or fixed signal) recovery guarantees \cite[Chapter 9]{foucart2013mathematical}. The former is tightly connected to the Restricted Isometry Property (RIP) \cite{candes2005decoding}. Essentially, a uniform recovery guarantee claims that a single draw of the sampling matrix is, with high probability, sufficient for the recovery of \textit{all} sparse signals.  A non-uniform recovery guarantee asserts that a single draw of the sampling matrix is, with high probability, sufficient for recovery of a \textit{fixed} sparse signal. 

This paper derives both uniform and non-uniform recovery guarantees for Hadamard-Haar systems and associated optimum sampling strategies. For uniform (and non-uniform) guarantee we resort to the VDS framework of Krahmer and Ward \cite{krahmer2014stable} (resp. MDS framework of Adcock~\etal~\cite{adcock2017breaking}). 

\subsection{Related Works}\label{sec:related works}

\paragraph{Non-uniform density sampling (theory and application):}
The idea of non-uniform density sampling dates back to the emergence of CS. Donoho in~\cite{donoho2006compressed} proposed a two-level sampling approach for recovering wavelet coefficients, where the coarse wavelet scale coefficients are fully sampled while the remaining coefficients are subsampled with UDS strategy. This idea was later extended by Tsaig and Donoho in~\cite{tsaig2006extensions} to a multiscale setup. Puy \etal~\cite{puy2011variable} advocated a convex optimization procedure for minimizing the coherence between the sensing and sparsity bases. A non-uniform density sampling approach is proposed by Wang and Arce in~\cite{wang2010variable} based on the statistical models of natural images. Bigot \etal~\cite{bigot2016analysis} introduced the notion of block sampling for CS, based on acquiring the blocks of measurements instead of isolated measurements; see also a similar study of Polak \etal~\cite{polak2015performance}. Boyer \etal~incorporated the idea of block sampling with structured sparsity in \cite{boyer2015compressed}, whose stable and robust recovery guarantee was later proved by Adcock \etal~in~\cite{adcock2018oracle}. A RIP-based recovery guarantee is presented by Krahmer and Ward~\cite{krahmer2014stable} based on the notion of \textit{random bounded orthonormal systems} introduced in~\cite{rauhut2010compressive,foucart2013mathematical}. The sampling strategy in~\cite{krahmer2014stable} is controlled by the local coherence between the sensing and sparsity bases. Adcock \etal~\cite{adcock2017breaking} provided a novel MDS scheme based on the local sparsity and multilevel coherence between the sensing and sparsity bases. A generalization of the RIP for MDS strategy of \cite{adcock2017breaking} in finite dimensions (and infinite dimensions) has been analyzed by Li and Adcock in~\cite{li2017compressed} (resp. Adcock~\etal,~\cite{adcock2019uniform}). 

Most of the works above also tackled the problem of signal recovery from subsampled Fourier measurements using wavelet sparsity basis (Fourier-Wavelet system), \eg \cite{puy2011variable,krahmer2014stable,boyer2015compressed,adcock2017breaking,adcock2016note}. In this context, the applications of non-uniform density sampling have shown promising results in MRI~\cite{lustig2007sparse,lustig2008compressed,roman2014asymptotic} and interferometric hyperspectral imaging \cite{moshtaghpourcoded,moshtaghpour2018multilevel,moshtaghpour2018,moshtaghpour2017a,moshtaghpour2016itwist}.

\paragraph{Imaging applications of the Hadamard transform:}
The Hadamard matrix has become an emerging element in many computational imaging applications relying on optical multiplexing or single pixel imaging, such as Hadamard spectroscopy~\cite{studer2012compressive,roman2014asymptotic,davis2004hyperspectral}, lensless camera~\cite{huang2013lensless}, 3-D video imaging~\cite{zhang20163d}, laser-based failure-analysis \cite{sun2009obic}, compressive holography~\cite{clemente2013compressive}, single pixel Fourier transform interferometry \cite{jin2017hyperspectral,moshtaghpour2018compressive,moshtaghpour2018compressive2}, digital holography~\cite{martinez2017single}, intracranial electroencephalogram acquisition~\cite{baldassarre2015structured}, single pixel camera~\cite{zhang2017hadamard}, and micro-optoelectromechanical systems~\cite{deverse2004application}.

Most of the works above have already been designed based on different non-uniform sampling schemes, that can be categorized, based on the sampling designs, in four groups, \textit{i.e.,} the rows of the Hadamard matrix are selected with respect to \textit{(i)} UDS~\cite{davis2004hyperspectral}, \textit{(ii)} MDS~\cite{baldassarre2015structured,roman2014asymptotic,moshtaghpour2018compressive}, \textit{(iii)} low-pass sampling where the first $M$ rows are selected~\cite{zhang20163d,martinez2017single}, and \textit{(iv)} half-half sampling where the first $M/2$ rows are always selected and the other $M/2$ rows are selected uniformly at random among the rest of the rows \cite{studer2012compressive}. Although all these works have obtained high quality signal recovery, they do not provide an explicit recovery guarantee. 

Moreover, other contributions, \eg on video compressive sensing~\cite{wakin2006compressive}, 3-D imaging~\cite{sun20133d}, remote sensing~\cite{ma2009single}, terahertz imaging \cite{chan2008single}, and single pixel camera \cite{duarte2008single,takhar2006new}, which utilize random binary patterns (\eg Bernoulli matrices) can potentially adopt Hadamard sensing with no hardware burden.

\paragraph{Hadamard-wavelet systems (theory):}
The recovery of 1-D signals that are sparse in an orthonormal wavelet basis (from subsampled Hadamard measurements) has been studied in \cite{antun2016coherence} in the context of MDS. The problem of signal reconstruction from the Hadamard (or binary) measurements has recently received attention in other contexts than CS, \eg in generalized sampling methods where the goal is to recover an infinite-dimensional signal from a full set of measurements (without subsampling) via a linear reconstruction. In this context, there exist several works where the sampling space is assumed to be the domain of Hadamard transform and the reconstruction takes place in the span of some wavelet basis (see, \eg \cite{hansen2017sampling,adcock2019uniform} and \cite{calderbank2019} for a survey). In this work, however, we address the recovery of both 1-D and 2-D finite-dimensional signals using the Haar wavelet sparsity basis. To the best of our knowledge, four other papers have addressed the relationship between the 1-D Hadamard and 1-D Haar wavelet bases \cite{fino1972relations,falkowski1996walsh,thompson2016compressive,rafiq2016haar}. In the next section, as well as Table~\ref{tab:related works}, we compare our contribution with the state-of-the-art works.
\begin{table}[t]\footnotesize
	\begin{center}\noindent
		\scalebox{0.78}{
			\begin{tabular}{
				|l                                   
				!{\vrule width 1.5pt} m{4.5mm}       
				|m{4.5mm}                            
				|m{4.5mm}                            
				!{\vrule width 1.5pt} m{3.8mm}       
				|m{3.8mm}                            
				|m{3.8mm}                            
				|m{3.8mm}                            
				!{\vrule width 1.5pt}m{3.2mm}        
				|m{3.2mm}                            
				|m{3.2mm}                            
				!{\vrule width 1.5pt}m{8.7mm}        
				|m{8.7mm}                            
				!{\vrule width 1.5pt}m{7.5mm}        
				|m{7.5mm}                            
				!{\vrule width 1.5pt}m{5.8mm}        
				|m{5.8mm}                            
				!{\vrule width 1.5pt}m{4.5mm}        
				|m{4.5mm}|}                          
				\hline
				                                                                                   &  
				\multicolumn{3}{c!{\vrule width 1.5pt}}{\multirow{2}{*}{\textbf{Sensing basis}}}   &  
			    \multicolumn{4}{c!{\vrule width 1.5pt}}{\multirow{2}{*}{\textbf{Sparsity basis}}} &  
			    \multicolumn{3}{c!{\vrule width 1.5pt}}{\textbf{Signal}}                           &  
			    \multicolumn{2}{c!{\vrule width 1.5pt}}{\textbf{Sampling}}                         &  
			    \multicolumn{2}{c!{\vrule width 1.5pt}}{\multirow{2}{*}{\textbf{Signal type}}}                     &  
			    \multicolumn{2}{c!{\vrule width 1.5pt}}{\textbf{Recovery}}                         &  
			    \multicolumn{2}{c|}{\multirow{2}{*}{\textbf{Context}}}                \tabularnewline 

				                                                                    &  
				\multicolumn{3}{c!{\vrule width 1.5pt}}{}                           &  
			    \multicolumn{4}{c!{\vrule width 1.5pt}}{}                           &  
			    \multicolumn{3}{c!{\vrule width 1.5pt}}{\textbf{dimension}}         &  
			    \multicolumn{2}{c!{\vrule width 1.5pt}}{\textbf{strategy}}          &  
			    \multicolumn{2}{c!{\vrule width 1.5pt}}{}                           &  
			    \multicolumn{2}{c!{\vrule width 1.5pt}}{\textbf{guarantee}}         &  
			    \multicolumn{2}{c|}{\multirow{2}{*}{}}                 \tabularnewline 
				\cline{2-19}
				                                                                    &  
				\centering \rotatebox[origin=c]{90}{any orthonormal}                &  
				\centering \rotatebox[origin=c]{90}{Fourier}                        &  
				\centering \rotatebox[origin=c]{90}{Hadamard}                       &  
				\centering \rotatebox[origin=c]{90}{any orthonormal}                &  
				\centering \rotatebox[origin=c]{90}{Haar wavelet}                   &  
				\centering \rotatebox[origin=c]{90}{Daubechies wavelet}           &  
				\centering \rotatebox[origin=c]{90}{\, orthogonal wavelet \,}      &  
				\centering \rotatebox[origin=c]{90}{1-D (vector)}                   &  
				\centering \rotatebox[origin=c]{90}{2-D (matrix)}                   &  
				\centering \rotatebox[origin=c]{90}{d-D (tensor)}                   &  
				\centering \rotatebox[origin=c]{90}{VDS}                            &  
				\centering \rotatebox[origin=c]{90}{MDS}                            &  
				\centering \rotatebox[origin=c]{90}{finite-dimensional}                    &  
				\centering \rotatebox[origin=c]{90}{infinite-dimensional}                  &  
				\centering \rotatebox[origin=c]{90}{uniform}                        &  
				\centering \rotatebox[origin=c]{90}{non-uniform}                    &  
				\centering \rotatebox[origin=c]{90}{\,compressive sensing\,}        &  
				\centering \rotatebox[origin=c]{90}{\,generalized sampling\,}\tabularnewline 
				\hline
				\multirow{1}{*}{Krahmer and Ward \cite{krahmer2014stable}} 
					& \centering\checkmark & \centering\checkmark &                      & \centering\checkmark & \centering\checkmark &                      
					&                      & \centering\checkmark & \centering\checkmark &                      & \centering\checkmark &                      
					& \centering\checkmark &                      & \centering\checkmark &                      & \centering\checkmark &    				   
					\tabularnewline
				\hline
				\multirow{1}{*}{Adcock~\etal~\cite{adcock2017breaking}} 
					& \centering\checkmark & \centering\checkmark &                      & \centering\checkmark &                      &                      
					& \centering\checkmark & \centering\checkmark &                      &                      &                      & \centering\checkmark 
					& \centering\checkmark &                      &                      & \centering\checkmark & \centering\checkmark & 					
					\tabularnewline
				\hline
				Adcock~\etal~\cite{adcock2016note}
					&                      & \centering\checkmark &                      &                      & \centering\checkmark &                      
					&					   & \centering\checkmark &                      &                      &                      & \centering\checkmark 
					& \centering\checkmark &                      &                      & \centering\checkmark & \centering\checkmark & 
					\tabularnewline
				\hline
				\multirow{1}{*}{Li and Adcock~\cite{li2017compressed}}
					& \centering\checkmark & \centering\checkmark &                      & \centering\checkmark & \centering\checkmark &                      
					&					   & \centering\checkmark &                      &                      &                      & \centering\checkmark 
					& \centering\checkmark &                      & \centering\checkmark &                      & \centering\checkmark &                      
					\tabularnewline
				\hline
				\multirow{1}{*}{Antun \cite{antun2016coherence}}
					&                      &                      & \centering\checkmark &                      & \centering\checkmark & \centering\checkmark  
					&					   & \centering\checkmark &                      &                      &                      & \centering\checkmark 
					& \centering\checkmark &                      &                      & \centering\checkmark & \centering\checkmark &                      
					\tabularnewline
				\hline
				\multirow{1}{*}{Adcock~\etal~\cite{adcock2019uniform}}
					& \centering\checkmark &                      & \centering\checkmark & \centering\checkmark & \centering\checkmark &                      
					& \centering\checkmark & \centering\checkmark &                      &                      &                      & \centering\checkmark 
					&					   & \centering\checkmark & \centering\checkmark &                      & \centering\checkmark &                      
					\tabularnewline
				\hline
				Hansen and Thesing~\cite{hansen2018}
					&                      &                      & \centering\checkmark &                      &                      &                      
					& \centering\checkmark &                      &                      & \centering\checkmark & \centering ---       & \centering ---       &                      &\centering\checkmark  & \centering ---       & \centering ---       &                      &\centering\checkmark 
					\tabularnewline
				\hline
				Thesing and Hansen~\cite{thesing2018}
					&                      &                      & \centering\checkmark &                      & \centering\checkmark &                      
					&					   &                      &                      & \centering\checkmark & \centering ---       & \centering ---       &                      &\centering\checkmark  & \centering ---       & \centering ---       &                      &\centering\checkmark 
					\tabularnewline
				\hline
				Hansen and Terhaar~\cite{hansen2017sampling}
					&                      &                      &\centering\checkmark  &                      &                      &                      &\centering\checkmark  &                      &                      &\centering\checkmark  & \centering ---       & \centering ---       &                      &\centering\checkmark  & \centering ---       & \centering ---       &                      &\centering\checkmark
					\tabularnewline
				\hline
				Thesing and Hansen~\cite{thesing2019}
					&                      &                      & \centering\checkmark &                      &                      &\centering\checkmark  &
					& \centering\checkmark &                      &                      &                      & \centering\checkmark 
					&     				   & \centering\checkmark &                      & \centering\checkmark & \centering\checkmark &   
					\tabularnewline
				\hline
				\rowcolor{blue!10}
					\rowcolor{blue!10}\multirow{-1}{*}{This work: Thm.~\ref{thm:uniform guarantee for Hadamard-Haar system}}
					&                      &                      & \centering\checkmark &                      & \centering\checkmark &                     
					& 					   & \centering\checkmark & \centering\checkmark &                      & \centering\checkmark &                      
					& \centering\checkmark &                      & \centering\checkmark &                      & \centering\checkmark &
					\tabularnewline
				\hline
				\rowcolor{blue!10}
					\rowcolor{blue!10}\multirow{-1}{*}{This work: Thm.~\ref{thm:Non-uniform guarantee for Hadamard-Haar system}}
					&                      &                      & \centering\checkmark &                      & \centering\checkmark &                      
					&					   & \centering\checkmark & \centering\checkmark &                      &                      & \centering\checkmark 
					& \centering\checkmark &                      &                      & \centering\checkmark & \centering\checkmark &
					\tabularnewline
				\hline
			\end{tabular}
		}
	\end{center}
	\caption{Comparison between the state-of-the-art works and our contribution. In this table, we consider those contributions in the field of CS that are based on the local coherence (developed by Krahmer and Ward in \cite{krahmer2014stable}) and multilevel coherence (developed by Adcock~\etal~in~\cite{adcock2017breaking}) parameters.}
	\label{tab:related works}
\end{table}              
\subsection{Our Contributions}
The main contributions of this paper are the followings:
\begin{itemize}
\item We provide uniform and non-uniform recovery guarantees for compressive Hadamard-Haar systems that are stable with respect to non-sparse signals and robust to the measurement noise. We build our analysis upon \cite{krahmer2014stable} (for the uniform guarantee) and upon \cite{adcock2017breaking} (for the non-uniform guarantee).
\item The results cover the recovery of 1-D and 2-D signals. In the latter case, we treat two constructions of the 2-D Haar basis, \ie using the tensor product and multi-resolution analysis. We will prove that either construction results in a different optimum sampling strategy. 
\item By computing the exact values of the local and multilevel coherence parameters for Hadamard-Haar systems, we provide tight sample complexity bounds relatively to the VDS and MDS frameworks. 
\end{itemize}
This work thus provides explicit CS strategies for Hadamard-Haar systems; an association that was seemingly not covered by the related literature (see, \eg Table~\ref{tab:related works}). Our main results are presented in Thm.~\ref{thm:uniform guarantee for Hadamard-Haar system} (for the uniform recovery guarantee) and in Thm.~\ref{thm:Non-uniform guarantee for Hadamard-Haar system} (for the non-uniform guarantee). Note that this Hadamard-Haar sensing system has recently been applied to the CS of hyperspectral data with single pixel imaging (when the light illumination is spatially coded with Hadamard system) \cite{moshtaghpour2018compressive2,moshtaghpour2018compressive}, with improved recovery performances compared to UDS strategy. During the finalization of this paper, we became aware of this recent survey of Calderbank~\etal~\cite{calderbank2019} that pursue similar objectives. Compared to our result in Thm.~\ref{thm:Non-uniform guarantee for Hadamard-Haar system}, Thm.~5.8 in \cite{calderbank2019}, which is stated from \cite{thesing2019}, covers the problem of infinite-dimensional signal recovery using Daubechies wavelets. Moreover, Fig.~3 in~\cite{calderbank2019} advocates the same structure as in Fig.~\ref{fig:structure HadHaar after reordering}-right for infinite-dimensional 2-D Hadamard-Haar system. However, the mathematical expression in Prop.~\ref{prop: structure of 2d-had-haar}-(ii) for modeling those structures is original.

\subsection{Paper Organization}
This paper is organized as follows. We first provide a summary of the CS theory in Sec.~\ref{sec:CS}, emphasizing on sensing strategies exploiting orthonormal bases (\eg Fourier or Hadamard), as well as the uniform and non-uniform signal recovery guarantees. In Sec.~\ref{sec:main results}, after delivering a short introduction to the Hadamard and Haar wavelet bases, we present our main results in Thm.~\ref{thm:uniform guarantee for Hadamard-Haar system} and Thm.~\ref{thm:Non-uniform guarantee for Hadamard-Haar system}. Note that the technical proofs are postponed to Sec.~\ref{sec:proofs}. Finally, we conduct a series of numerical tests in Sec.~\ref{sec:simulations}, which confirms the efficiency of our analysis. 
\subsection{Notations}
Domain dimensions are represented by capital letters, \eg $K,M,N$. Vectors and matrices are denoted by bold symbols. For a matrix $\bs{U} = [\bs{u}_1, \dots, \bs{u}_{N_2}] \in \bb{C}^{N_1 \times N_2}$, $\bs{u} = {\rm vec}(\bs{U}) \coloneqq [\bs{u}_1^\top, \hdots, \bs{u}_{N_2}^\top]^\top \in \bb{C}^{N_1N_2}$ corresponds to the folded vector representation of $\bs{U}$. When this is clear from the context, we assimilate a matrix in $\bb{C}^{N_1\times N_2 }$ with its vectorized version, \ie identifying $\bs{U}$ with $\bs{u}$. For any matrix (or vector) $\bs V \in \bb C^{M\times N}$, $\bs V^\top$ and $\bs V^*$ represent the transposed and the conjugate transpose of $\bs V$, respectively, and $\bs{V} \otimes \bs{W}$ denotes the Kronecker product of two matrices $\bs V$ and $\bs W$. The $\ell_p$-norm of $\bs{u}$ reads $\|\bs{u}\|_p \coloneqq  (\sum_i |u_i|^p)^{1/p}$, for $p\geq 1$, with $\|\bs u\| \coloneqq \|\bs u\|_2$. For a matrix $\bs{U}$, $\|\bs{U}\|_{p,q} \coloneqq  \max_{\bs x}\{\|\bs U \bs x\|_q~\st~\|\bs x\|_p =1\}$, for $p,q\geq 1$.  The identity matrix of dimension $N$ is represented as $\bs{I}_N$. Similarly, $\bs 1_N $ (or $\bs 0_N$) describes a vector of length $N$ with all components equal 1 (resp. 0); when the value of $N$ is clear from the text we simply write $\bs 1$ (resp. $\bs 0$) for simplicity.  By an abuse of convention, unless expressed differently, we consider that a set (or a subset) of indices is actually a \textit{multiset}, \ie the repetition and ordering of the elements are allowed; the set cardinality thus considers the total number of (non-unique) multiset elements. For a subset $\Omega =\{\omega_j\}_{j=1}^{M} \subset \range{N}\coloneqq \{1, \dots, N\}$ of cardinality $|\Omega|$, the restriction operator is denoted by $\bs P_\Omega \in \{0,1\}^{M\times N}$ with $(\bs P_{\Omega} \bs x)_j = x_{\omega_j}$. Moreover, the mask operator is represented by $\bar{\bs P}_\Omega \coloneqq \bs P^\top_\Omega \bs P_\Omega\in \{0,1\}^{N\times N}$ with $(\bar{\bs P}_{\Omega} \bs x)_j = x_j$ if $j \in \Omega$, and zero otherwise. The concatenation of two sets $\cl S= \{s_i\}_{i=1}^M$ and $\cl T = \{t_i\}_{i=1}^N$ is denoted by $\cl S\,\underline{\cup}\,\cl T\coloneqq \{s_1,\cdots,s_M,t_1,\cdots,t_N\}$. We thus have $\bs P_{\cl S\underline{\cup}\cl T} \bs u = [\bs P^\top_{\cl S}, \bs P^\top_{\cl T}]^\top\bs u$. We define $\range{N}_0 \coloneqq \{0\}~\underline{\cup}~\range{N} $. The floor function $\lfloor u \rfloor$ outputs the greatest integer less than or equal to $u$. We use the asymptotic relations $f \lesssim g$ (or $f \gtrsim g$), if $f \le c\,g$ (resp. $g \le c\, f$) for two functions $f$ and $g$ and some value $c>0$ independent of their parameters. In order to go back and forth between the 1-D and 2-D index representations $l \in \range{N_1N_2}$ and $(l_1,l_2) \in \range{N_1} \times \range{N_2}$, respectively, we use the relation $\subtoind{l}{l_1}{l_2}{N_1,N_2}$, meaning that $k = l_1 + N_1(l_2-1)$, $l_1 = (l-1\!\!\!\mod N_1)+1$, and $l_2 = \lfloor (l-1)/N_1\rfloor +1$; when $N_1 = N_2=N$ we simply write $\subtoind{l}{l_1}{l_2}{N}$. 
The Cartesian product with \textit{lexicographical} ordering of two sets $\cl S_1 \subset \range{N_1}$ and $\cl S_2 \subset \range{N_2}$ is denoted by $\cl S_1 \times \cl S_2$. Moreover, $\overline{\cl S_1 \times \cl S_2} \coloneqq \{i \in \range{N_1N_2}:  i \xrightleftharpoons{N_1,N_2}(j,k), j\in \cl S_1,~ k\in \cl S_2\}$ converts the 2-tuple elements of $\cl S_1 \times \cl S_2$ to 1-tuple elements . 
\section{Compressed Sensing for Orthonormal Bases}\label{sec:CS}
We summarize here uniform and non-uniform recovery guarantees for CS of signals acquired from partial measurements in bounded orthonormal systems (\eg Fourier and Hadamard). We begin by recalling some limitations of the coherence-based analysis between two orthonormal systems, one used for the signal sensing and the other one for the sparse signal representation.

\paragraph{The need for non-uniform density sampling:} Let $\bs x \in \bb C^N$ be a signal to be recovered from noisy compressive measurements
\begin{equation}
\bs y = \bs P_{\Omega} \bs \Phi^* \bs x +\bs n \in \bb C^M,
\end{equation}
where the sensing basis $\bs \Phi \in \bb C^{N \times N}$ is an orthonormal basis (\eg Hadamard), $\Omega \subset \range{N}$ is a set (or multiset) of indices chosen at random with $|\Omega| = M \ll N$, and $\bs n$ is an additive observation noise. 

In order to estimate $\bs x$ from $\bs y$, CS theory requires three ingredients: \textit{(i)} low-complexity prior information of the signal, \textit{(ii)} efficient subsampling strategy of $\range{N}$ for defining $\Omega$, and \textit{(iii)} a non-linear recovery algorithm which takes into account the low-complexity prior model. A typical low-complexity prior in real-world applications, \eg image processing and hyperspectral imaging, is that the target signal $\bs x$ has a $K$-sparse or compressible (\ie well-approximated by a sparse signal) representation in a general orthonormal basis $\bs \Psi \in \bb C^{N\times N}$ (\eg Haar wavelet), \ie $|{\rm supp}(\bs \Psi^* \bs x)| \le K$. In this context, by-now-traditional results in CS theory (\eg \cite[Cor.~12.38]{foucart2013mathematical} or \cite{candes2007sparsity}) state that in order to reconstruct $K$-sparse signals, one must draw 
\begin{equation}\label{eq:sample complexity uds}
M \gtrsim N \mu^2(\bs \Phi^*\bs \Psi) K \log(N)
\end{equation}
elements of $\Omega$ uniformly at random in $\range{N}$, where
\begin{equation}\label{eq:mutual coherenec}
\mu(\bs U)\coloneqq \max_{1\le i,j\le N}|u_{i,j}| \in [1/\sqrt{N},1],
\end{equation}
is the coherence of $\bs U \in \bb C^{N \times N}$. For coherent sensing and sparsity bases, \eg Hadamard and Haar wavelet bases, respectively, $\mu(\bs \Phi^*\bs \Psi) =1$ and thus, $M\gtrsim N$ measurements are required for signal recovery, which prevents (at least in theory) any compression. This limitation is indeed related to the construction of $\Omega$, \ie according to a Uniform Density Sampling (UDS).

As explained in Sec.~\ref{sec:related works}, there exist fundamentally-different theoretical results that suggest a non-uniform density sampling in the case of coherent bases, with the same idea of breaking the coherence barrier (\eg \cite{adcock2017breaking,puy2011variable,krahmer2014stable,boyer2015compressed}). In this work we restrict our analysis to the VDS scheme of Krahmer and Ward~\cite{krahmer2014stable} (for uniform guarantee) and the MDS scheme of Adcock~\etal~\cite{adcock2017breaking} (for non-uniform guarantee).

\paragraph{VDS and uniform recovery guarantee:} Krahmer and Ward showed in~\cite{krahmer2014stable} that the sample-complexity bound in \eqref{eq:sample complexity uds} can be modified by assigning higher sampling probability to the columns of the sensing basis $\bs \Phi$ (or equivalently, to the rows of $\bs \Phi^*$) that are highly coherent with the columns of the sparsity basis $\bs \Psi$. They have thus defined the \textit{local coherence}
 \begin{equation}\label{eq:local coherence}
 \mu_l^{\rm loc} = \mu_l^{\rm loc}(\bs \Phi^* \bs \Psi) \coloneqq \max_{1\le j\le N} |(\bs \Phi^*\bs \Psi)_{l,j}|\in [1/\sqrt{N},1],~~~\forall l\in \range{N},
 \end{equation}
with $\max_l \mu^{\rm loc}_l(\bs \Phi^* \bs \Psi) = \mu(\bs \Phi^* \bs \Psi)$. In the sequel, we will denote the vector $\bs \mu^{\rm loc} \coloneqq [\mu^{\rm loc}_1,\cdots,\mu^{\rm loc}_N]^\top$ formed by all local coherences. Krahmer and Ward~\cite{krahmer2014stable} proved that this quantity determines a sufficient condition on the construction of the subsampling set $\Omega$ for most classes of recovery algorithms, \eg convex optimization~\cite{candes2005decoding}, thresholding~\cite{blumensath2009iterative}, and greedy~\cite{Zhangopm} strategies. Therefore, we leverage Thm.~5.2 in \cite{krahmer2014stable} and combine it with Thm.~2.1 in \cite{cai2014sparse} in order to provide the following self-contained recovery guarantee.
\begin{thm}[VDS and uniform recovery guarantee for CS, adapted from \cite{krahmer2014stable} and \cite{cai2014sparse}]\label{thm:vds}
	Let $\bs{\Phi} \in \bb{C}^{N \times N}$ and $\bs{\Psi} \in \bb{C}^{N \times N}$ be orthonormal sensing and sparsity bases, respectively, with $\mu^{\rm loc}_l(\bs \Phi^* \bs \Psi) \le \kappa_l$ for some values $\kappa_l \in \bb R_+$. Let us define $\bs \kappa \coloneqq [\kappa_1,\cdots,\kappa_N]^\top$. Fix $\epsilon \in (0,1]$ and $ \delta <1/\sqrt{2}$ and suppose $K \gtrsim \log(N)$,
  \begin{equation}
    M \gtrsim \delta^{-2}\|\bs \kappa\|^2 K \log (\epsilon^{-1}),\label{eq:sample complexity vds}
  \end{equation}
  and choose $M$ (possibly not distinct) indices $l \in \Omega \subset \range{N}$ i.i.d. with respect to the probability distribution $\eta$ on $\range{N} $ given by 
  \begin{equation}
 \ts \eta(l) \coloneqq \frac{\kappa_l^2}{\|\bs \kappa\|^2}.\label{eq:pmf vds}
  \end{equation}
  Consider the diagonal matrix $\bs{D}={\rm diag}(\bs{d}) \in \bb{R}^{M \times M}$ with $d_j = {1}/{\sqrt{\eta(\Omega_j)}}$, $j \in \range{M}$. With probability exceeding $1-\epsilon$, for all $\bs{x} \in \bb{C}^N$ observed through the noisy CS model $\bs{y} = \bs P_{\Omega}\bs \Phi^* \bs{x}+\bs{n}$ with  $\|\bs D\bs n\| \le\varepsilon\sqrt{M}$, the solution $\hat{\bs x}$ of the program
  \begin{equation}
    \hat{\bs x}\ =\ \argmin_{\bs u \in \bb{C}^{N}} \|\bs \Psi^*\bs u\|_1\ \st\ \frac{1}{\sqrt{M}}\| \bs D(\bs y-\bs P_{\Omega}\bs \Phi^* \bs u)\| \le \varepsilon, \label{eq:BPDN uniform}
  \end{equation}
satisfies 
  	\begin{equation*}
  	\|\bs x-\hat{\bs x}\| \le \textstyle c_1\frac{\sigma_K(\bs \Psi^* \bs x)_1}{\sqrt{K}} + c_2\varepsilon,~~~c_1 = 2\frac{\delta+\sqrt{\delta(1/\sqrt{2}-\delta)}}{1-\sqrt{2}\delta}, c_2 = \ts \frac{2\sqrt{2(1+\delta)}}{1-\sqrt{2}\delta},
  	\end{equation*}
  	where $\sigma_K(\bs{u})_1 \coloneqq\|\bs{u}-\cl H_K(\bs{u})\|_1$ is the best $K$-term approximation error (in the $\ell_1$ sense), and $\cl H_K$ is the hard thresholding operator that maps all but the $K$ largest-magnitude entries of the argument to zero. In particular, the reconstruction is exact, \ie $\hat{\bs x}=\bs x$, if $\bs x$ is $K$-sparse and $\varepsilon$ = 0. Note that for $\delta = 1/3$, $c_1 < 2.58$, and $c_2 < 6.18$.
\end{thm}
This recovery guarantee is uniform in the sense that a single construction of the measurement matrix $\bs P_\Omega \bs \Phi^*$ with respect to the sample-complexity bound in \eqref{eq:sample complexity vds} and sampling pmf in \eqref{eq:pmf vds} is sufficient to ensure (with high probability) the recovery of all sparse vectors. This result is of interest in those applications of CS where sparsity (or compressibility) of the target signal is the only possible prior knowledge. In the following we describe a method, which takes into account local sparsity of the target signal. 	
\paragraph{MDS and non-uniform recovery guarantee}
Adcock and co-authors \cite{adcock2017breaking} have also advocated that the global notion of coherence in \eqref{eq:mutual coherenec} and sparsity must be replaced by a proper local versions in order to obtain a better subsampling strategy. 

To fix the ideas, we first introduce the CS setup proposed in \cite{adcock2017breaking}. For a fixed $r \in \bb N$ we decompose the signal (or sparsity) domain $\range{N}$ into $r$ disjoint \textit{sparsity levels} $\cl S \coloneqq\{\cl S_1,\dots,\cl S_r\}$ such that $\bigcup_{l=1}^r \cl S_l = \range{N}$. Given a vector of sparsity parameters $\bs k = [k_1,\dots,k_r]^\top \in \bb N^r$, a vector $\bs s\in \bb C^N$ is called $(\cl S,\bs k)$-sparse-in-level, and we write $\bs s \in \Sigma_{\cl S,\bs k}$, if $|\supp \bs P_{\cl S_l}\bs s| \le k_l$ for all $l \in \range{r}$. For an arbitrary vector $\bs s$, its $(\cl S,\bs k)$-approximation error is denoted by $\sigma_{\cl S,\bs k}(\bs s) \coloneqq \min\{\|\bs s-\bs z\|_1 : \bs z \in \Sigma_{\cl S,\bs k}\} = \sum_l \sigma_{k_l}(\bs P_{\cl S_l} \bs s)_1$. We quickly observe that the sparsity-in-level model reduces to the global sparsity model by setting $r = 1$ and $\cl S = \range{N}$.

Similar to the sparsity domain, we decompose the sampling domain $\range{N}$ into $r$ disjoint \textit{sampling levels} defined as $\cl W \coloneqq \{\cl W_1,\dots,\cl W_r\}$ with $\bigcup_{l=1}^r \cl W_l = \range{N}$. Given $\bs m = [m_1,\dots,m_r]^\top \in \bb N^r$, the set $\Omega_{\cl W,\bs m} \coloneqq \bigcup_{t=1}^{r} \Omega_t$ provides an MDS scheme, or $(\cl W, \bs m)$-MDS, if, for each $1\leq t\leq r$, $\Omega_t  \subseteq \cl W_t$, $|\Omega_t| = m_t \le |\cl W_t|$, and if the entries of $\Omega_t$ are chosen uniformly at random (without replacement) in $\cl W_t$. 

We further need to define two quantities controlling the sample-complexity bound in MDS scheme (see below). Given an orthonormal matrix $\bs U \in\bb C^{N\times N}$ and local sparsity values $\bs k$, the $t^{\rm th}$ \textit{relative sparsity} is defined as
\begin{equation}
\label{eq:relative sparsity}
\ts  K_t^{\cl W, \cl S}(\bs U, \bs k) = \max_{\bs z \in \Sigma_{\cl S,\bs k} :\, \|\bs z\|_\infty\le 1} \|\bs P_{\cl W_t} \bs U\bs z\|^2.
\end{equation}

In the cases where the computation of the exact relative sparsity values given in \eqref{eq:relative sparsity} is not feasible, one can instead upper bound it as stated in the next lemma, which is adapted from \cite[Eq. 13]{adcock2016note}.
\begin{lem}\label{lem:relative sparsity} 
$\ts  \sqrt{K_t^{\cl W, \cl S}(\bs U, \bs k)} \le \sum_{l=1}^{|\cl S|} \|\bs P_{\cl W_t} \bs U \bs P^\top_{\cl S_l}\|_{2,2} \sqrt{k_l}$.
\end{lem}
Moreover, the $(t,l)^{\rm th}$ \textit{multilevel coherence} of $\bs U$ with respect to the sampling and sparsity levels $\cl W$ and $\cl S$, respectively, is defined as 
\begin{equation}
  \label{eq:multilevel coherence}
  \ts \mu_{t,l}^{\cl W,\cl S}(\bs U) \coloneqq \mu(\bs P_{\cl W_t}\bs U)\,\mu(\bs P_{\cl W_t}\bs U\bs P^\top_{\cl S_l}).
\end{equation}
Within this context, the following guarantee can be reformulated from \cite[Thm.~4.4]{adcock2017breaking}.

\begin{thm}[MDS and non-uniform recovery guarantee for CS, adapted from \cite{adcock2017breaking}]\label{thm:mds}
	Let $\bs{\Phi} \in \bb{C}^{N \times N}$ and $\bs{\Psi} \in \bb{C}^{N \times N}$ be orthonormal sensing and sparsity bases, respectively. Fix sampling and sparsity levels $\cl W$ and $\cl S$, respectively. Let $\Omega = \Omega_{\cl W,\bs m}$ be a $(\cl W,\bs m)$-MDS and $(\cl S,\bs k)$ be any pair such that the following holds: for $0<\epsilon\le {\rm exp}(-1)$, $K = \|\bs k\|_1$, $\hat{m}_t$ is such that for all $l \in \range{|\cl S|}$,
	\begin{equation}\label{eq:sample complexity mds 2}
	   \ts 1\ \gtrsim\ \sum_{t=1}^{r}\Big( \left(\frac{|\cl W_t|}{\hat{m}_t}-1\right)\ \mu_{t,l}^{\cl W,\cl S}(\bs \Phi^*\bs \Psi)\, \ts  K_t^{\cl W, \cl S}(\bs \Phi^*\bs \Psi, \bs k)\Big),	
	\end{equation} and $m_t$ such that for all $t \in \range{|\cl W|}$,
	\begin{align}
		m_t &\gtrsim \hat{m}_t\,\log(K\epsilon^{-1})\,\log(N), \label{eq:sample complexity mds 3}\\
	   \ts	m_t &\gtrsim |\cl W_t| \, \big(\sum_{l=1}^{r}\mu_{t,l}^{\cl W,\cl S}(\bs \Phi^*\bs \Psi) \, k_l\big)\,\log(K\epsilon^{-1})\,\log(N).\label{eq:sample complexity mds 1}
	\end{align}
	Given the noisy CS measurements $\bs{y} = \bs P_{\Omega}\bs \Phi^* \bs{x}+\bs{n}$ with  $\|\bs n\| \le\varepsilon$, suppose that $\hat{\bs x}\in \bb C^N$ is a minimizer of 
	\begin{equation}\label{eq:bpdn non-uniform}
		\ts \hat{\bs x} = \argmin_{\bs u \in \bb C^{N}} \|\bs \Psi^*\bs u\|_1 \ \st \ \|\bs y -\bs P_{\Omega} \bs \Phi^*\bs u\| \le \varepsilon.
	\end{equation}	
	Then, with probability exceeding $1-\epsilon$, we have
	\begin{equation}
	   \ts	\|\bs x-\hat{\bs x}\| \le c_1\,\sigma_{\cl S,\bs k}(\bs \Psi^*\bs x) +c_2 (1+C\sqrt{K}) \, \varepsilon\sqrt{q},\label{eq:MLS error bound}
	\end{equation}
	where $q\coloneqq\max_t \frac{|\cl W_t|}{m_t}$ for some constant $0< c_1 \le 22, 0< c_2\le 11 $, and where $0\le C \le 3\sqrt{6} + \frac{4\sqrt{6}\sqrt{\log(6N\epsilon^{-1})}}{\log(N)}$.
\end{thm}
We remark the main differences between this theorem and Thm.~\ref{thm:vds}. First, Thm.~\ref{thm:mds} provides a non-uniform recovery guarantee: the measurement matrix $\bs P_\Omega \bs \Phi^*$ satisfying the conditions of the proposition needs (in theory) to be redrawn when a new vector is to be recovered. Second, the MDS scheme requires a prior information about the local sparsity of the target signal; while the VDS scheme in Prop.~\ref{thm:vds} does not need such information. Third, the parameter $\varepsilon$ in optimization program \eqref{eq:bpdn non-uniform} is a bound on the observation noise power, while in the VDS scheme \eqref{eq:BPDN uniform} it is a bound on the weighted noise power. However, we showed in \cite[Thm.~2.5]{moshtaghpour2018} that one can determine a bound on $\|\bs D \bs n\|$, which holds with controllable probability, and that depends on the $\|\bs n\|$, $\|\bs n\|_\infty$ (or on estimations bounding these quantities with high probability) and a parameter fixed by the pmf defining the VDS scheme.

\section{The compressive sensing Hadamard-Haar problem}\label{sec:cs Hadamard-haar poblem}
We now focus on a special case of the CS settings concerned by Thm.~\ref{thm:vds} and Thm.~\ref{thm:mds}, where the sensing and sparsity bases are set as Hadamard and Haar wavelet bases, respectively.  Before presenting the main results, we first recall the definitions of the 1-D, 2-D anisotropic, and 2-D isotropic Haar wavelet bases and Paley-ordered Hadamard matrix. These definitions are useful to develop the machinery of our contributions.
\subsection{Haar and Hadamard Bases}\label{sec:definitions}
\paragraph{1-D Discrete Haar Wavelet (DHW) basis:}
	Fix $N = 2^r$ for some $r \in \bb{N}$. The DHW basis of $\bb R^N$ consists of $N$ vectors
	$$
	\{\psi_j^{\rm 1d}\}_{j=1}^{N}\coloneqq \{\bar{h}\} \cup \{h^{(1)}_{s,p} : 0\le s \le r-1, 0 \le p  \le 2^s-1\} \subset \bb R^{N},
	$$
	where, for $\tau \in \range{N-1}_0, \bar{h}(\tau):= 2^{-r/2}$ is the constant (scaling) function and $h^{(1)}_{s,p}(\tau)\coloneqq 2^\frac{s-r}{2} h(2^{s-r} \tau - p)$ is the wavelet function at scale (or resolution) $s$ and position $p$, with $h(\tau)$ equals 1, -1, and 0 over $[0,1/2)$, $[1/2,1)$, and $\bb R\backslash[0,1)$, respectively (see~\cite[Page 2]{mallat2008wavelet}, \cite[Page 6]{stollnitz1995wavelets}, or \cite{adcock2016note}), \ie 
	\begin{equation}\label{eq:def:dhw}
		h^{(1)}_{s,p}(\tau) = \left\{\begin{matrix}
		2^{\frac{s-r}{2}}, & {\rm for~} p 2^{r-s} \le \tau < (p +\frac{1}{2})2^{r-s},\\ 
		-2^{\frac{s-r}{2}}, & {\rm for~}(p +\frac{1}{2}) 2^{r-s} \le \tau < (p +1)2^{r-s},\\ 
		0,& {\rm \normalfont otherwise}.
		\end{matrix}\right.
	\end{equation}
	In a matrix form, DHW basis in $\bb R^{N \times N}$ can be constructed~\cite{fino1972relations,falkowski1996walsh} from the recursive relation
	\begin{equation}\label{eq:def:dhw matrix}
		\bs \Psi_{\rm dhw} \coloneqq \bs W^{(1)}_{r} \coloneqq \frac{1}{\sqrt{2}}\left[\bs W^{(1)}_{r-1} \otimes \begin{bmatrix} 1  \\ 1 \end{bmatrix},\bs I_{2^{r-1}} \otimes \begin{bmatrix} 1  \\ -1 \end{bmatrix}       \right]
		,~~~ {\rm with~}\bs W^{(1)}_{0} \coloneqq  [1],
	\end{equation}
	which collects in its columns all the vectors of $\{\psi_j^{\rm 1d}\}_{j=1}^{N}$ (see Lemma~\ref{lem:dhw matrix}).
	
	In order to extend the DHW basis to the 2-D Haar wavelet basis, we need to define $N-1$ window functions
		\begin{equation}\label{eq:def:dhw0}
			h^{(0)}_{s,p}(\tau) = \left\{
				\begin{matrix}
					2^{\frac{s-r}{2}}, & {\rm for~} p 2^{r-s} \le \tau < (p +1)2^{r-s},\\ 
					0,& {\rm \normalfont otherwise},
				\end{matrix}\right.
		\end{equation}
	for the resolution $0\le s \le r-1$ and position $0 \le p\le  2^s-1$ parameters. Similar to the construction of the DHW basis in \eqref{eq:def:dhw matrix}, we define the matrix
	\begin{equation}\label{eq:def:dhw0 matrix}
		\bs W^{(0)}_{r} \coloneqq \frac{1}{\sqrt{2}}\left[\bs W^{(0)}_{r-1} \otimes \begin{bmatrix} 1  \\ 1 \end{bmatrix},\bs I_{2^{r-1}} \otimes \begin{bmatrix} 1  \\ 1 \end{bmatrix}       \right],~~~ {\rm with~} \bs W^{(0)}_{0} \coloneqq  [1],
	\end{equation}
	Which collects in its first column the vector $\bar{h}$ and in the other columns all the vectors $\{h^{(0)}_{s,p}\}$ (see Lemma~\ref{lem:dhw matrix}).
	
	Associated with the DHW basis, the \textit{1-D dyadic levels} $\cl T^{\rm 1d}\coloneqq \{\cl T^{\rm 1d}_l\}_{l=0}^r$ gather coefficient indices with identical wavelet levels; they are defined as 
	\begin{equation}\label{eq:1d_dyadic_levels}
	\mathcal{T}_l^{\rm 1d}\coloneqq   \range{2^l}\backslash \range{2^{l-1}},~~~{\rm for~} l\in\range{r},~~~{\rm and~}\cl T_0^{\rm 1d} \coloneqq \{1\},
	\end{equation}
	 with cardinality $|\cl T_l^{\rm 1d}| = 2^{l-1}$, for $l \in \range{r}$. We also define left-complement of dyadic levels as $\cl T_{{<l}}^{\rm 1d} \coloneqq  \bigcup_{j=0}^{l-1}\cl T_{j}^{\rm 1d} = \range{2^{l-1}}$ for $l \in \range{r}$ and $\cl T_{<0}^{\rm 1d} \coloneqq  \emptyset$. These levels are important to isolate the indices of the columns (components) of $\bs \Psi_{\rm dhw}=\bs W^{(1)}_r$ (resp. $(\bs \Psi_{\rm dhw})^\top \bs x$) associated with a given scale, as well as those of $\bs W^{(0)}_r$
	
	\begin{lem}\label{lem:dhw matrix}
	For $l\in \range{r}_0$ and $a \in \{0,1\}$, the matrix $\bs W^{(a)}_r \bs P^\top_{\cl T^{\rm 1d}_l} \in \bb R^{2^r \times |\cl T^{\rm 1d}_l|}$ collects in its columns all the vectors $\{h^{(a)}_{l-1,p}\}_{p=0}^{2^{l-1}-1}$, if $l\in \range{r}$, and if $l=0$, it collects $\bar{h}$ in its single column.
	\end{lem}
	\begin{proof}
	See Sec.~\ref{proof:lem:dhw matrix}.
	\end{proof}

There exist two natural ways to construct a 2-D wavelet basis from a 1-D basis, \ie by tensor product of two 1-D bases, and by following a multi-resolution analysis (see \cite[Sec.~7.7]{mallat2008wavelet}, \cite{beylkin1991fast}, or \cite{nowak1999wavelet}), which amounts to multiplying all possible pairs of wavelet and scaling functions sharing the same resolution. We describe below those two approaches for 2-D Haar wavelet construction.
\paragraph{2-D Anisotropic Discrete Haar Wavelet (ADHW) basis:} 
	For the first approach, the tensor product of two DHW bases leads to an anisotropic 2-D DHW basis. For $N = 2^r$ and some $r \in \bb{N}$, we consider the scaling and wavelet functions $\bar{h}$ and $h^{(1)}_{s,p}$ defined above, and we build the ADHW basis  of $\bb R^{N^2}$ as 
\begin{align*}
\{\psi_j^{\rm aniso}\}_{j=1}^{N^2}\coloneqq \{\psi^{\rm 1d}_{j_1}\psi^{\rm 1d}_{j_2}: \subtoind{j}{j_1}{j_2}{N}\},
\end{align*}
which provides $N^2$ possible functions. This basis is of interest for image compression \cite{devore1992image}, sparsity basis for MRI images \cite{boyer2015compressed}, and sparsity basis for monochromatic images in fluorescence spectroscopy \cite{moshtaghpour2018}. In particular, Neumann and von Sachs \cite{neumann1997wavelet} showed that if a multi-dimensional signal has different degrees of smoothness in different directions, the tensor wavelet construction is a better choice for signal estimation. 
	
	In a matrix form, the ADHW basis in $\bb R^{N^2 \times N^2}$ can be constructed \cite{mallat2008wavelet,nowak1999wavelet} as
		\begin{equation*}
	 \bs \Psi_{\rm adhw} \coloneqq  \bs \Psi_{\rm dhw} \otimes \bs \Psi_{\rm dhw},
		\end{equation*}
where $\bs \Psi_{\rm adhw}$ collects in its columns all the vectors of $\{\psi_j^{\rm aniso}\}_{j=1}^{N^2}$. Associated with the ADHW basis, we define the \textit{2-D anisotropic wavelet levels} $\cl T^{\rm aniso} \coloneqq \{\cl T_{l}^{\rm aniso}\}_{l=1}^{r^2}$ where $\cl T_{l}^{\rm aniso}\coloneqq  \overline{\cl T^{\rm 1d}_{l_1} \times \cl T^{\rm 1d}_{l_2}}$, for $l \in \range{(r+1)^2}$ and $l_1,l_2 \in \range{r}_0$, with the relation $\subtoind{l}{l_1+1}{l_2+1}{r+1}$ and hence $|\cl T^{\rm aniso}_l| = |\cl T^{\rm 1d}_{l_1}|\cdot |\cl T^{\rm 1d}_{l_2}|$. These levels thus gather the indices of wavelet coefficients associated with the constant resolution (see the illustration on Fig.~\ref{fig:levels}-right for $N=8$).
\begin{remark}\label{rem:aniso_level_adhw}
According to the construction of $\bs \Psi_{\rm adhw}$ and $\cl T^{\rm aniso}$, one can use Lemma~\ref{lem:dhw matrix} and Lemma~\ref{lem:kronecker_cartesian_relation} to show that 
$$
\bs \Psi_{\rm adhw}\bs P^\top_{\cl T_{l}^{\rm aniso}} = \left(\bs \Psi_{\rm dhw}\bs P^\top_{\cl T_{l_2}^{\rm 1d}}\right) \otimes \left(\bs \Psi_{\rm dhw}\bs P^\top_{\cl T_{l_1}^{\rm 1d}}\right),~{\rm for~} \subtoind{l}{l_1+1}{l_2+1}{r+1}.
$$
\end{remark}
\paragraph{2-D Isotropic Discrete Haar Wavelet (IDHW) basis:}
The second type of the 2-D DHW basis is built from a multi-resolution analysis \cite{mallat2008wavelet}. Fix $N = 2^r$ for some $r \in \bb{N}$. Let $\bar{h}$, $h^{(1)}$, and $h^{(0)}$ be the scaling, wavelet, and window functions defined above.  Following~\cite{mallat2008wavelet} the IDHW basis $\{\psi_j^{\rm iso}\}_{j=1}^{N^2}$ of $\bb R^{N^2}$ consists of the functions
	$$
	\{\phi^{(00)}\}\cup \{\phi^{(ab)}_{s,(p_1,p_2)} : 0\le s \le r-1, 0\le p_1,p_2 \le 2^{s}-1,~(a,b) \in \{0,1\}^2\backslash \{0,0\}\},
	$$
	such that
	\begin{align}
		&\phi^{(00)}(\tau_1,\tau_2) = \bar{h}(\tau_1) \bar{h}(\tau_2),\\
		&\phi^{(ab)}_{s,(p_1,p_2)}(\tau_1,\tau_2) = h^{(a)}_{s,p_1}(\tau_1) h^{(b)}_{s,p_2}(\tau_2),
	\end{align}
	where $0 \le s \le r-1$ and $0\le p_1,p_2\le 2^s-1$ are the resolution and position indices, respectively, \ie there are $N^2$ possible functions.
	\begin{figure}
		\centering
		\begin{minipage}{0.49\linewidth}
			\centering
			\scalebox{1}{
%
%
\begin{tikzpicture}

\begin{axis}[%
width=1.3in,
height=1.3in,
at={(0in,0in)},
scale only axis,
point meta min=0,
point meta max=1,
axis on top,
xmin=0.5,
xmax=640.5,
xtick = {40,120,200,280,360,440,520,600},
xticklabels = {1,2,3,4,5,6,7,8},
xlabel = {second dimension index $(j)$},
ticklabel style={font=\fontsize{8}{1}\selectfont},
xlabel style={at = {(0.5,0.08)},font=\fontsize{10}{1}\selectfont},
tick align=inside,
y dir=reverse,
ymin=0.5,
ymax=640.5,
ytick =  {40,120,200,280,360,440,520,600},
yticklabels = {1,2,3,4,5,6,7,8},
ylabel = {first dimension index $(i)$},
ylabel style={at = {(0.2,0.5)},font=\fontsize{10}{1}\selectfont},
axis background/.style={fill=white}
]
\addplot [forget plot] graphics [xmin=0.5,xmax=640.5,ymin=0.5,ymax=640.5] {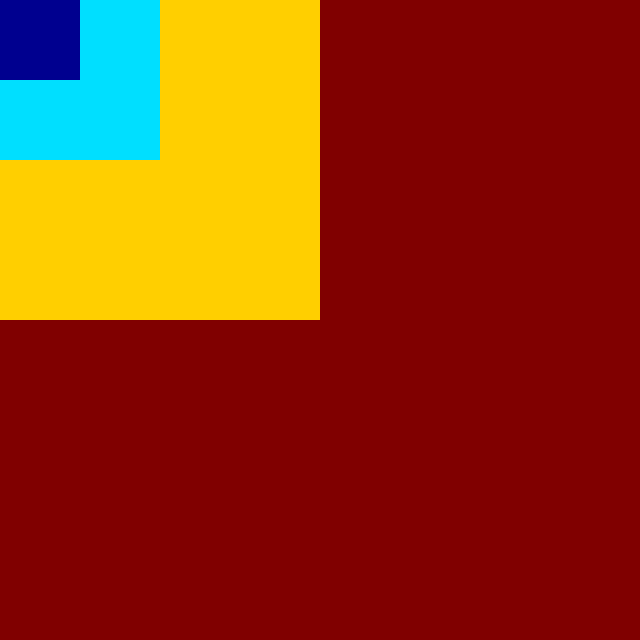};
%
%
%

\draw[color = white, fill = none] (axis cs:-5.5,320.5) -- (axis cs:320.5,320.5);
\draw[color = white, fill = none] (axis cs:320.5,-5.5) -- (axis cs:320.5,320.5);

\draw[color = white, fill = none] (axis cs:-5.5,160.5) -- (axis cs:160.5,160.5);
\draw[color = white, fill = none] (axis cs:160.5,-5.5) -- (axis cs:160.5,160.5);

\draw[color = white, fill = none] (axis cs:-5.5,80.5) -- (axis cs:80.5,80.5);
\draw[color = white, fill = none] (axis cs:80.5,-5.5) -- (axis cs:80.5,80.5);
\end{axis}
\end{tikzpicture}%
}
		\end{minipage}
		\begin{minipage}{0.49\linewidth}
			\centering
			\scalebox{1}{
%
%
\begin{tikzpicture}

\begin{axis}[%
width=1.3in,
height=1.3in,
at={(0in,0in)},
scale only axis,
point meta min=0,
point meta max=1,
axis on top,
xmin=0.5,
xmax=640.5,
xtick = {40,120,200,280,360,440,520,600},
xticklabels = {1,2,3,4,5,6,7,8},
xlabel = {second dimension index $(j)$},
ticklabel style={font=\fontsize{8}{1}\selectfont},
xlabel style={at = {(0.5,0.08)},font=\fontsize{9}{1}\selectfont},
tick align=inside,
y dir=reverse,
ymin=0.5,
ymax=640.5,
ytick =  {40,120,200,280,360,440,520,600},
yticklabels = {1,2,3,4,5,6,7,8},
ylabel = {first dimension index $(i)$},
ylabel style={at = {(0.2,0.5)},font=\fontsize{9}{1}\selectfont},
axis background/.style={fill=white}
]
\addplot [forget plot] graphics [xmin=0.5,xmax=640.5,ymin=0.5,ymax=640.5] {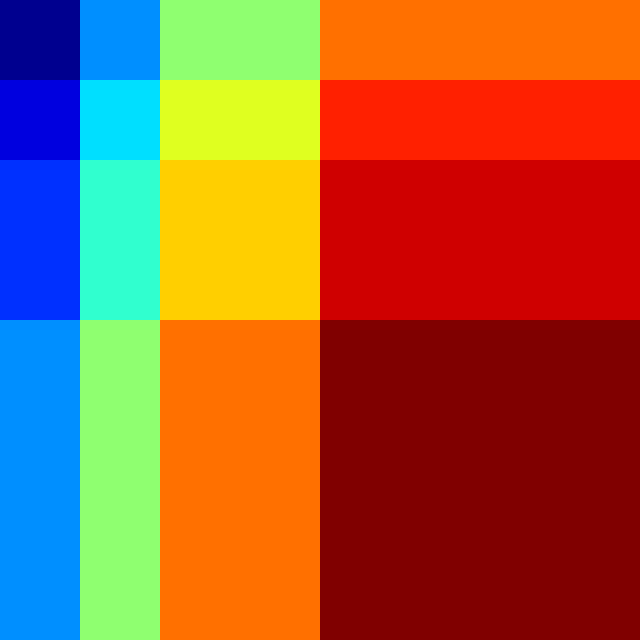};

\draw[color = white, fill = none] (axis cs:-5.5,320.5) -- (axis cs:641.5,320.5);
\draw[color = white, fill = none] (axis cs:320.5,-5.5) -- (axis cs:320.5,641.5);

\draw[color = white, fill = none] (axis cs:-5.5,160.5) -- (axis cs:641.5,160.5);
\draw[color = white, fill = none] (axis cs:160.5,-5.5) -- (axis cs:160.5,641.5);

\draw[color = white, fill = none] (axis cs:-5.5,80.5) -- (axis cs:641.5,80.5);
\draw[color = white, fill = none] (axis cs:80.5,-5.5) -- (axis cs:80.5,641.5);
\end{axis}
\end{tikzpicture}%
}
		\end{minipage}
		\caption{An example of the 2-D isotropic wavelet levels $\cl T^{\rm iso}_l$ for $l\in \range{r}_0$ (left) versus 2-D anisotropic wavelet levels $\cl T^{\rm aniso}_l$ for $l \in \range{(r+1)^2}$ (right) with $N = 8$ (or $r =3$). Each area represents the subset of pairs of indices $(i,j)$ that belong to a level $l$.}
		\label{fig:levels}
	\end{figure}
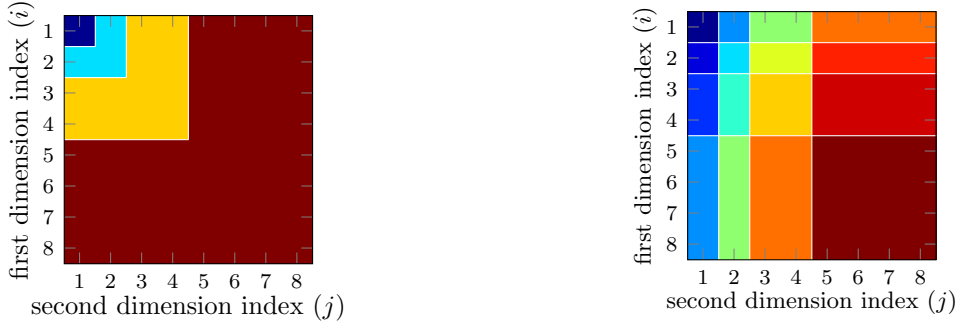	

To construct the orthonormal matrix $\bs \Psi_{\rm idhw} \in \bb R^{N^2\times N^2}$ associated with the 2-D IDHW basis, we leverage the 1-D partitions $\cl T^{\rm 1d}_l$ defined above so that the column ordering of $\bs \Psi_{\rm idhw}$ will ease any further column selection (\eg in Sec.~\ref{sec:main results}).

We first define the submatrices
$$
\ts \bs \Psi^{(ab)}_{l} \coloneqq \left(\bs W^{(a)}\bs P^\top_{\cl T^{\rm 1d}_l}\right) \otimes \left(\bs W^{(b)}\bs P^\top_{\cl T^{\rm 1d}_l}\right)  \in \bb R^{N^2 \times |\cl T^{\rm 1d}_l|^2},~l \in \range{r},~(a,b)\in \{0,1\}^2\backslash\{(0,0)\}
$$
and $\bs \Psi^{(00)}_0 \coloneqq \bs 1_{N^2}$. For each level $l$, these submatrices clearly contain all the functions $\{\phi^{(ab)}_{l-1,(p_1,p_2)}: p_1,p_2 \in \range{2^{l}-1}_0$. Moreover, since $\cl T^{\rm 1d}_{<l} = \range{2^{l-1}}$ and $\cl T^{\rm 1d}_{l} = \range{2^l}/\range{2^{l-1}}$, the following disjoint sets
$$
\cl T^{(00)}_0\coloneqq \overline{\cl T^{\rm 1d}_{0} \times\cl T^{\rm 1d}_{{0}}}
,~ \cl T^{(11)}_l \coloneqq\overline{\cl T^{\rm 1d}_{l} \times\cl T^{\rm 1d}_{{l}}}
,~ \cl T^{(10)}_l \coloneqq\overline{\cl T^{\rm 1d}_{<l} \times\cl T^{\rm 1d}_{l}}
,~ \cl T^{(01)}_l \coloneqq\overline{\cl T^{\rm 1d}_{l} \times\cl T^{\rm 1d}_{{<l}}},
$$
are such that $|\cl T^{(00)}_0| =1$, $|\cl T^{(11)}_l| = |\cl T^{(01)}_l| = |\cl T^{(10)}_l| = |\cl T^{\rm 1d}_l|^2 = 2^{2(l-1)}$, and 
$$
\cl T^{(00)}_0  \cup \bigcup\limits_{l \in \range{r}}\left(\cl T^{(01)}_l \cup \cl T^{(11)}_l \cup \cl T^{(10)}_l\right) = \range{N^2}.
$$
 Therefore, as illustrated in Fig.~\ref{fig:levels}-left, we can order the columns of $\bs \Psi_{\rm idhw}$ such that, for the 2-D isotropic wavelet levels $\cl T^{\rm iso}\coloneqq\{\cl T^{\rm iso}_l\}_{l=0}^r$ defined by 
\begin{equation}
	\cl T^{\rm iso}_0 \coloneqq \cl T^{(00)}_0,~{\rm and~} \cl T_{l}^{\rm iso}\coloneqq \cl T^{(01)}_l\,\underline{\cup}\,  \cl T^{(11)}_l \,\underline{\cup}\, \cl T^{(10)}_l, ~\l \in \range{r}_0,\label{eq:2-D isotropic wavelet levels}
\end{equation}
we have $\bs \Psi_{\rm idhw}\bs P^\top_{\cl T^{\rm iso}_0} = \bs \Psi^{(00)}_0$, $\bs \Psi_{\rm idhw}\bs P^\top_{\cl T^{(ab)}_l} = \bs \Psi^{(ab)}_l$, and $\bs \Psi_{\rm idhw}\bs P^\top_{\cl T^{\rm iso}_l} = [\bs \Psi^{(01)}_l,\bs \Psi^{(11)}_l,\bs \Psi^{(10)}_l]$ for $l \in \range{r}$ and $(a,b)\in \{0,1\}^2\backslash\{(0,0)\}$.
\paragraph{(Paley-ordered) Hadamard matrix:}
	We now present an important family of orthogonal matrices introduced by J. Hadamard \cite{hadamard}, \ie the Hadamard matrix, that has appeared in various fields, \eg coding theory \cite{popa1990classification}, harmonic analysis \cite{kolounzakis2006complex}, and optics \cite{jin2017hyperspectral}. There exist mainly three constructions of the Hadamard matrix, each with specific row ordering, called ordinary (or Sylvester)-, sequency-, and Paley-ordered Hadamard matrix \cite{sylvester1867lx,zhihua1983ordering}. In this paper, we focus only on the Paley-ordered Hadamard matrix. But all our results are clearly extendable to the other two constructions after proper reordering (see \cite[Chapter 4]{antun2016coherence} for the row ordering). 

	Given $r \in \bb{N}$, the $2^r \times 2^r$ Hadamard matrix \cite{horadam2012hadamard,falkowski1996walsh} $\bs \Phi_{\rm had} \coloneqq \bs H_r \in \{\pm 2^{-r/2}\}^{2^r \times 2^r}$ is defined by
		\begin{equation}\label{eq:def:hadamard_matrix}
		 \bs H_{r} \coloneqq \frac{1}{\sqrt{2}}\left[\bs H_{r-1} \otimes \begin{bmatrix} 1  \\ 1 \end{bmatrix},\bs H_{r-1} \otimes \begin{bmatrix} 1  \\ -1 \end{bmatrix}\right]
		,\bs H_{0} \coloneqq [1].
		\end{equation}
	Note that this recurrence relation bears some resemblance with the one of the Haar wavelet basis in \eqref{eq:def:dhw matrix}. Moreover, from \eqref{eq:def:hadamard_matrix}, we can easily show that $\bs \Phi_{\rm had}$ is symmetric, \ie $\bs \Phi_{\rm had}^\top = \bs \Phi_{\rm had}$. The Hadamard transformation of a signal $\bs x \in \bb C^{N}$ with $N=2^r$ reads $\bs z = \bs \Phi_{\rm had}^\top\bs x$. For 2-D signals, the Hadamard basis is defined by $\bs \Phi_{\rm 2had} \coloneqq \bs \Phi_{\rm had} \otimes \bs \Phi_{\rm had} \in \bb R^{N^2\times N^2}$ so that the Hadamard transformation of a matrix $\bs X \in \bb C^{N\times N}$ is $\bs Z = \bs \Phi_{\rm had}^\top \bs X \bs \Phi_{\rm had}$, or equivalently ${\rm vec}(\bs Z) = \bs \Phi_{\rm 2had}^\top {\rm vec}(\bs X)$. 
	\begin{remark}\label{rem:iso_aniso_level_2had}
	Following the definition of $\bs \Phi_{\rm 2had}$, $\cl T^{\rm aniso}$, and $\cl T^{\rm iso}$ and using Lemma.~\ref{lem:kronecker_cartesian_relation} we directly deduce that, for $l_1,l_2\in\range{r}_0$ and $l\in\range{(r+1)^2}$,
	\begin{align*}
	\bs P_{\cl T^{\rm aniso}_l} \bs \Phi_{\rm 2had}^\top &= \left(\bs P_{\cl T^{\rm 1d}_{l_2}} \bs \Phi_{\rm had}^\top\right) \otimes \left(\bs P_{\cl T^{\rm 1d}_{l_1}} \bs \Phi_{\rm had}^\top\right), \subtoind{l}{l_1+1}{l_2+1}{r+1},
	\end{align*}
	and for $l\in \range{r}_0$,
	\begin{align*}
	\bs P_{\cl T^{\rm iso}_l} \bs \Phi_{\rm 2had}^\top &=
	\begin{bmatrix} \bs P_{\overline{\cl T^{\rm 1d}_{{l}}  \times \cl T^{\rm 1d}_{<l}}}  \left(\bs \Phi_{\rm had}\otimes \bs \Phi_{\rm had}\right)
	\\
	\bs P_{\overline{\cl T^{\rm 1d}_{{l}}  \times \cl T^{\rm 1d}_{l}}}  \left(\bs \Phi_{\rm had}\otimes \bs \Phi_{\rm had}\right)
	\\
	\bs P_{\overline{\cl T^{\rm 1d}_{{<l}}  \times \cl T^{\rm 1d}_{l}}}  \left(\bs \Phi_{\rm had}\otimes \bs \Phi_{\rm had}\right)
	\end{bmatrix}
	=	
	\begin{bmatrix} 	(\bs P_{\cl T^{\rm 1d}_{<l}} \bs \Phi_{\rm had}) \otimes (\bs P_{\cl T^{\rm 1d}_{t}} \bs \Phi_{\rm had})
	\\
	(\bs P_{\cl T^{\rm 1d}_l} \bs \Phi_{\rm had}) \otimes (\bs P_{\cl T^{\rm 1d}_{l}} \bs \Phi_{\rm had})
	\\
	(\bs P_{\cl T^{\rm 1d}_l} \bs \Phi_{\rm had}) \otimes (\bs P_{\cl T^{\rm 1d}_{<l}} \bs \Phi_{\rm had})
	\end{bmatrix}.
	\end{align*}
	\end{remark}
\subsection{Main Results}\label{sec:main results}
Equipped with the definitions above, we are now ready to develop our main results. To do so, we need to calculate the local coherence~\eqref{eq:local coherence}, multilevel coherence~\eqref{eq:multilevel coherence}, and relative sparsity~\eqref{eq:relative sparsity} for the Hadamard-Haar systems in one and two dimensions. Note that the proofs of this section are all postponed to Sec.~\ref{sec:proofs}. 

We start with the following crucial proposition; it captures a particular recursive block structure of the Hadamard-Haar matrix obtained by multiplying the 1-D Hadamard and Haar matrices.
\begin{prop}\label{prop: structure of had-haar}
	Given the integer $r\ge 0$ and defining the Hadamard-Haar matrix $\bs U^{(a)}_r \coloneqq \bs H^\top_r\bs W^{(a)}_r$ for $a \in \{0,1\}$, we observe that $\bs U^{(1)}_{0} = \bs U^{(0)}_{0} = [1]$, and for $r\ge 1$,
	\begin{subequations}
	\begin{align*}
	\begin{split}
		  \ts \bs U^{(1)}_{r}= 
		\begin{bmatrix} 
			\bs U^{(1)}_{r-1} & \bs 0 \\ \bs 0 &  \bs H_{r-1}
		\end{bmatrix},
	\end{split}
	\begin{split}
		  \bs U^{(0)}_{r} = 
		\begin{bmatrix} 
			\bs U^{(0)}_{r-1} &  \bs H_{r-1} \\ \bs 0 & \bs 0
		\end{bmatrix}.
	\end{split}
	\end{align*}
	\end{subequations}
	In particular, the matrix $\bs U^{(1)}_r$ is clearly symmetric, and $\bs U^{(1)}_{r}$ and $\bs U^{(0)}_{r}$ contain the structure illustrated in Fig.~\ref{fig:structure in hadamard haar}.
	\end{prop}
\begin{proof}
See Sec.~\ref{proof:prop:structure of had-haar}.
\end{proof}
\begin{figure}
	\centering
	\begin{minipage}{0.4\linewidth}\centering
		\includegraphics[width=0.8\columnwidth]{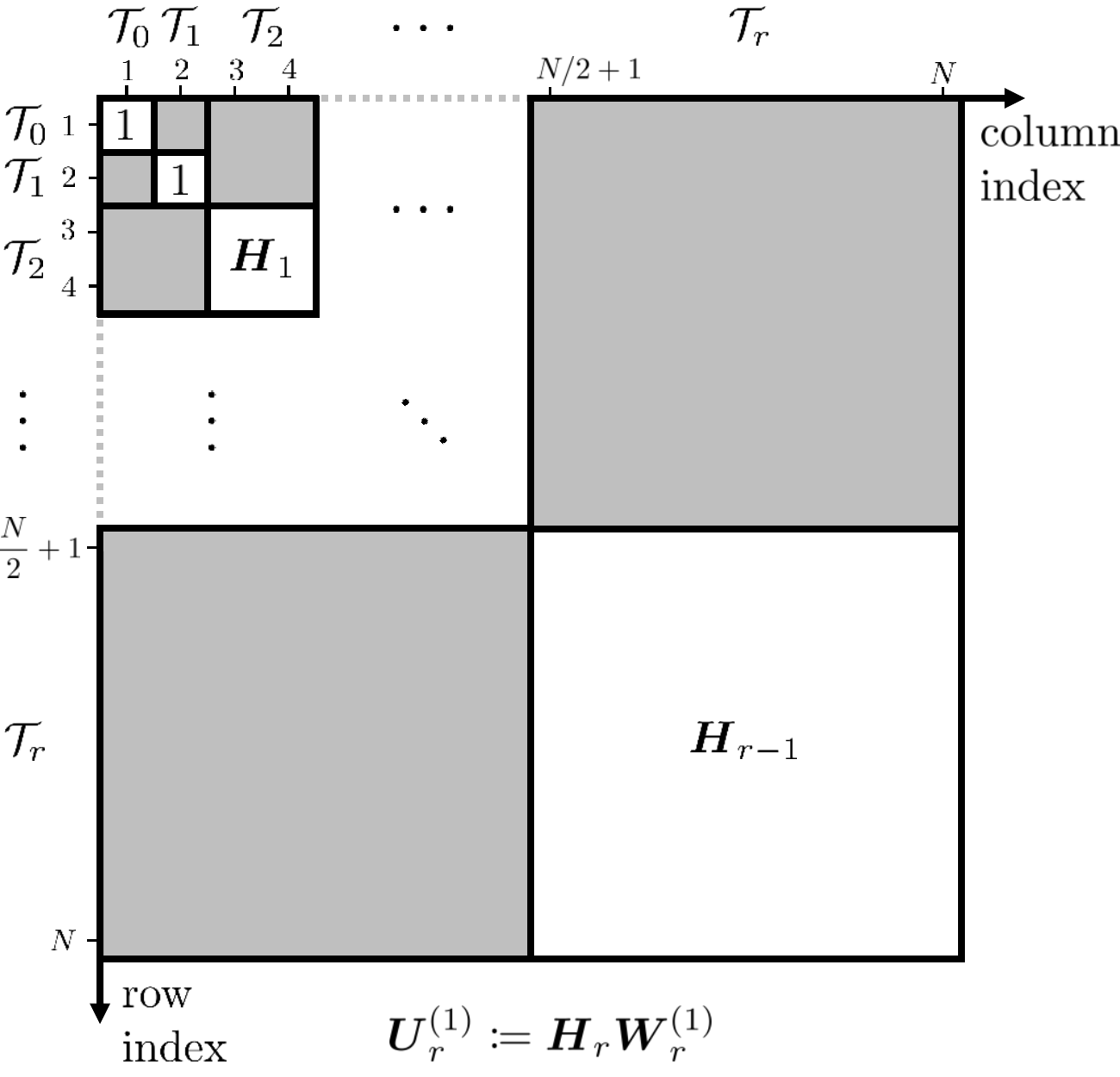}
	\end{minipage}
	\centering
	\begin{minipage}{0.4\linewidth}\centering
		\includegraphics[width=0.8\columnwidth]{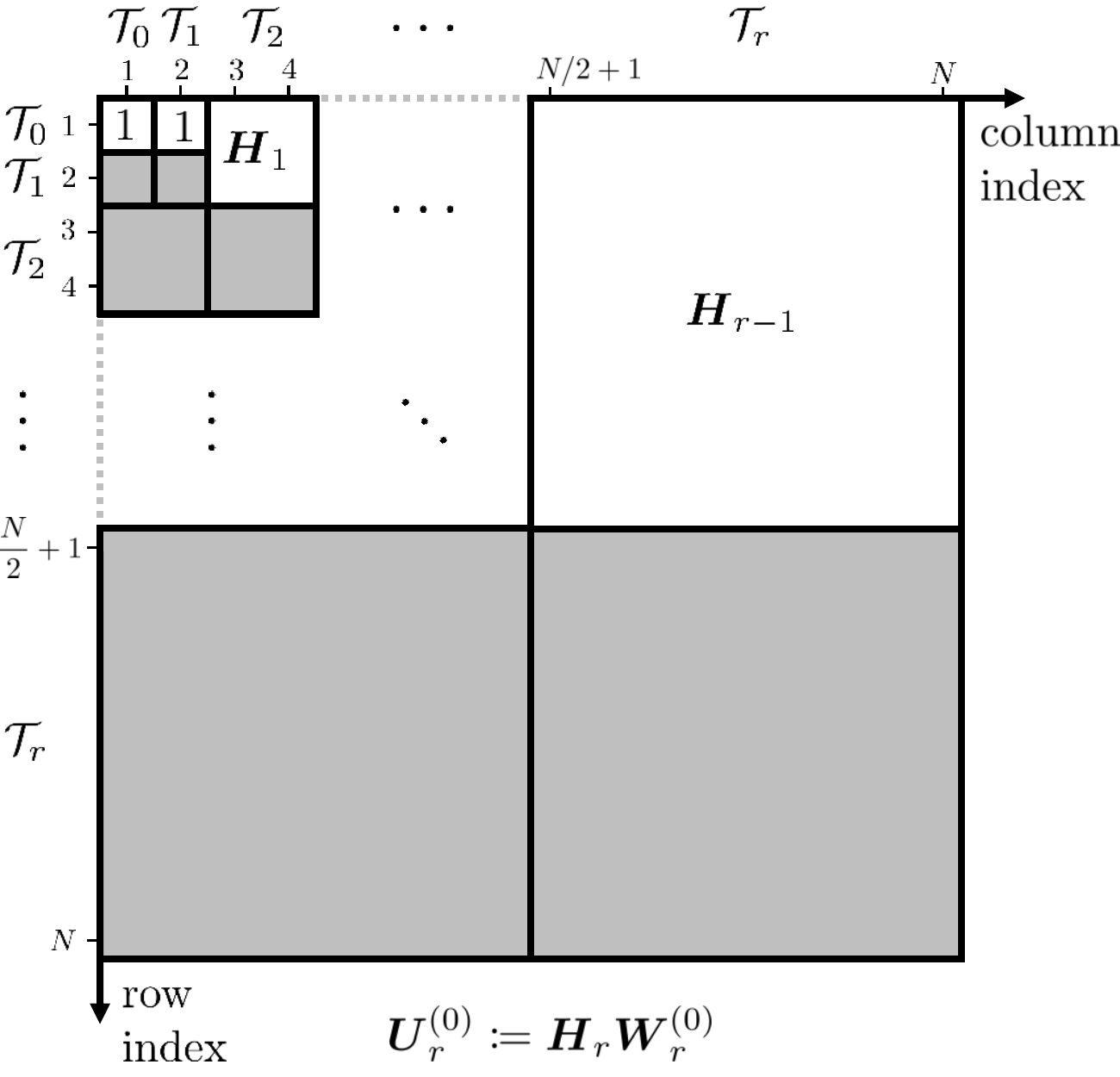}
	\end{minipage}
	\caption{Block structure of the matrices $\bs H_{r}\bs W^{(1)}_{r}$ (left) and  $\bs H_{r}\bs W^{(0)}_{r}$ (right) where $\cl T_l = \cl T^{\rm 1d}_l$ for $l\in \range{r}_0$. Gray color represents zero value.}
	\label{fig:structure in hadamard haar}
\end{figure}
\begin{remark}\label{rem:structure of had-haar}
In the context of Prop.~\ref{prop: structure of had-haar}, from the definition of $\cl T_l = \cl T^{\rm 1d}_l$ in \eqref{eq:1d_dyadic_levels} and block structure of $\bs U^{(0)}_r$ and $\bs U^{(1)}_r$ unfolded in Fig.~\ref{fig:structure in hadamard haar}, we easily deduce the following relations:
\\
for $t,l \in \range{r}_0$, we have
\begin{subequations}
	\begin{align}
		\begin{split}\label{eq:rem:structure of had-haar_1}
			&\bs P_{\cl T_t}\bs U^{(1)}_{r} \bs P^\top_{\cl T_l} = 
			\begin{cases}
				\bs H_{(t-1)_+}, & {\rm if}~ t=l,\\
				\bs 0, & {\rm otherwise},
			\end{cases}
		\end{split}\\
		\begin{split}\label{eq:rem:structure of had-haar_2}
			&\bs P_{\cl T_{{<t}}}\bs U^{(1)}_{r} \bs P^\top_{\cl T_l} = 
				\begin{cases}
					\bs U^{(1)}_r\bs P^\top_{\cl T_l}, & {\rm if}~ t>l,\\
					\bs 0, & {\rm otherwise},
				\end{cases}
		\end{split}\\
		\begin{split}\label{eq:rem:structure of had-haar_3}
			&\bs P_{\cl T_t}\bs U^{(0)}_{r} \bs P^\top_{\cl T_t} = 
				\begin{cases}
					1, & {\rm if}~ t = 0,\\
					\bs 0, & {\rm otherwise},
				\end{cases}
		\end{split}\\
		\begin{split}\label{eq:rem:structure of had-haar_4}
			&\bs P_{\cl T_{<t}}\bs U^{(0)}_{r} \bs P^\top_{\cl T_t} = \bs H_{t-1}, {~\rm for~} t >0,
		\end{split}
	\end{align}
	 where $(u)_+ \coloneqq \max(u,0)$.
\end{subequations}
\end{remark}
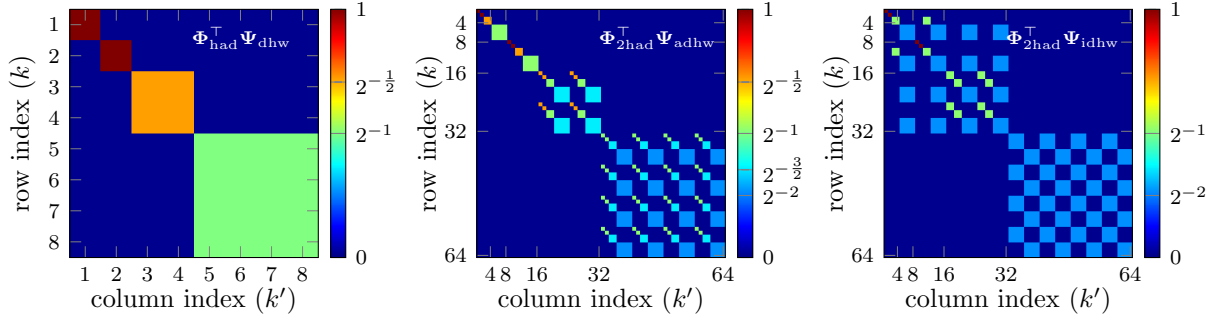
\begin{figure}[t]
	\centering
	\begin{minipage}{0.32\linewidth}
		\centering
%
%
\begin{tikzpicture}

\begin{axis}[%
width=1.3in,
height=1.3in,
at={(0in,0in)},
scale only axis,
point meta min=0,
point meta max=1,
axis on top,
xmin=0.5,
xmax=640.5,
xtick = {40,120,200,280,360,440,520,600},
xticklabels = {1,2,3,4,5,6,7,8},
xlabel = {column index $(k')$},
ticklabel style={font=\fontsize{8}{1}\selectfont},
xlabel style={at = {(0.5,0.08)},font=\fontsize{10}{1}\selectfont},
tick align=inside,
y dir=reverse,
ymin=0.5,
ymax=640.5,
ytick = {40,120,200,280,360,440,520,600},
yticklabels = {1,2,3,4,5,6,7,8},
ylabel = {row index $(k)$},
ylabel style={at = {(0.18,0.5)},font=\fontsize{10}{1}\selectfont},
colormap/jet,
colorbar,
colorbar style={at = {(1.05,1)},ytick={0,0.5,0.7071,1},yticklabels={0,$2^{-1}$,$2^{-\frac{1}{2}}$,1},font=\fontsize{5}{1}\selectfont,height=1.3in,width=0.2cm}, 
]
\addplot [forget plot] graphics [xmin=0.5,xmax=640.5,ymin=0.5,ymax=640.5] {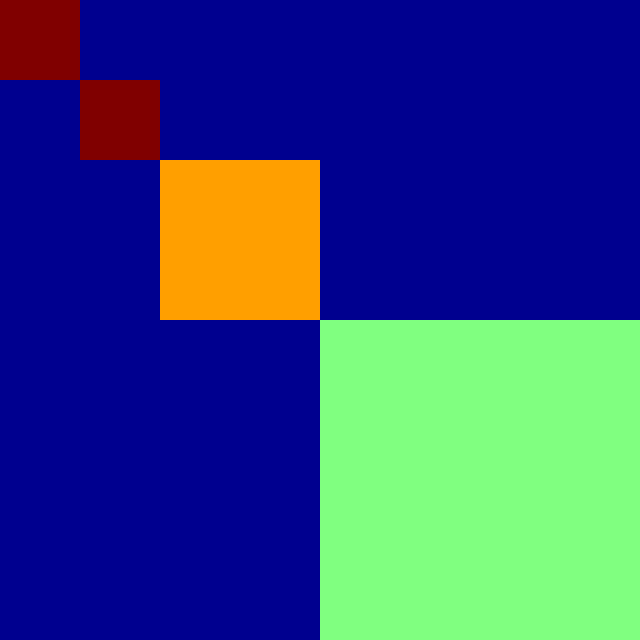};
\node[fill = none,text=white, draw=none, anchor=north west] at (rel axis cs:0.45,0.97) {\fontsize{7}{1}\selectfont$\bs \Phi^\top_{\rm had} \bs \Psi_{\rm dhw}$};
\end{axis}
\end{tikzpicture}%
	\end{minipage}
	\begin{minipage}{0.32\linewidth}
		\centering
%
%
\begin{tikzpicture}

\begin{axis}[%
width=1.3in,
height=1.3in,
at={(0in,0in)},
scale only axis,
point meta min=0,
point meta max=1,
axis on top,
xmin=0.5,
xmax=640.5,
xtick = {35,75,155,315,635},
xticklabels = {4,8,16,32,64},
xlabel = {column index $(k')$},
ticklabel style={font=\fontsize{8}{1}\selectfont},
xlabel style={at = {(0.5,0.08)},font=\fontsize{10}{1}\selectfont},
tick align=inside,
y dir=reverse,
ymin=0.5,
ymax=640.5,
ytick = {35,85,165,315,635},
yticklabels = {4,8,16,32,64},
ylabel = {row index $(k)$},
ylabel style={at = {(0.18,0.5)},font=\fontsize{10}{1}\selectfont},
colormap/jet,
colorbar,
colorbar style={at = {(1.05,1)},ytick={0,0.25,0.3536,0.5,0.7071,1},yticklabels={0,$2^{-2}$,$2^{-\frac{3}{2}}$,$2^{-1}$,$2^{-\frac{1}{2}}$,1},font=\fontsize{5}{1}\selectfont,height=1.3in,width=0.2cm}, 
]
\addplot [forget plot] graphics [xmin=0.5,xmax=640.5,ymin=0.5,ymax=640.5] {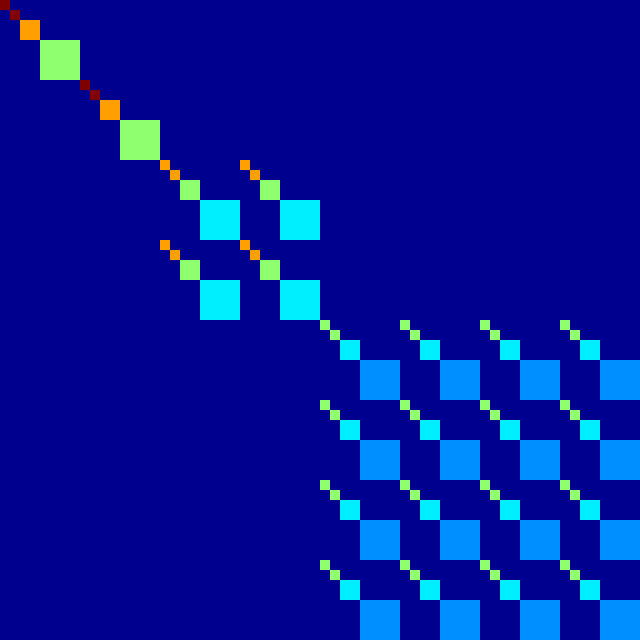};
\node[fill = none,text=white, draw=none, anchor=north west] at (rel axis cs:0.45,0.97) {\fontsize{7}{1}\selectfont$\bs \Phi^\top_{\rm 2had} \bs \Psi_{\rm adhw}$};
\end{axis}
\end{tikzpicture}%
	\end{minipage}
	\begin{minipage}{0.32\linewidth}
		\centering
%
%
\begin{tikzpicture}

\begin{axis}[%
width=1.3in,
height=1.3in,
at={(0in,0in)},
scale only axis,
point meta min=0,
point meta max=1,
axis on top,
xmin=0.5,
xmax=640.5,
xtick = {35,75,155,315,635},
xticklabels = {4,8,16,32,64},
xlabel = {column index $(k')$},
ticklabel style={font=\fontsize{8}{1}\selectfont},
xlabel style={at = {(0.5,0.08)},font=\fontsize{10}{1}\selectfont},
tick align=inside,
y dir=reverse,
ymin=0.5,
ymax=640.5,
ytick = {35,85,165,315,635},
yticklabels = {4,8,16,32,64},
ylabel = {row index $(k)$},
ylabel style={at = {(0.18,0.5)},font=\fontsize{10}{1}\selectfont},
colormap/jet,
colorbar,
colorbar style={at = {(1.05,1)},ytick={0,0.25,0.5,1},yticklabels={0,$2^{-2}$,$2^{-1}$,1},font=\fontsize{5}{1}\selectfont,height=1.3in,width=0.2cm}, 
]
\addplot [forget plot] graphics [xmin=0.5,xmax=640.5,ymin=0.5,ymax=640.5] {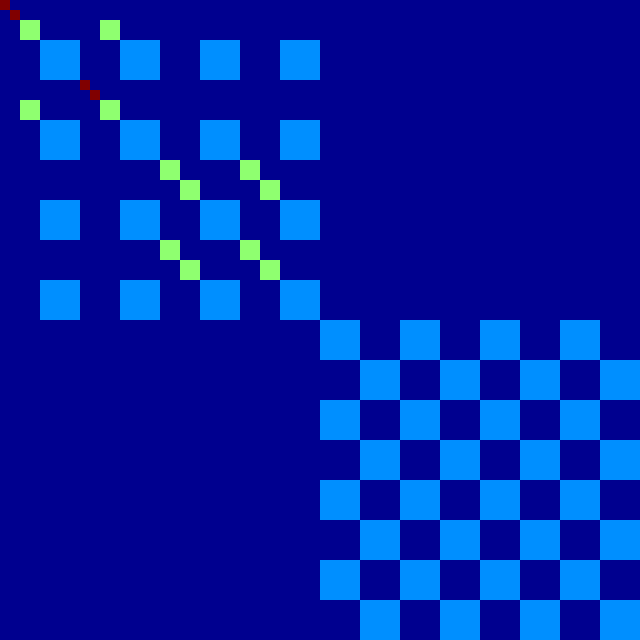};
\node[fill = none,text=white, draw=none, anchor=north west] at (rel axis cs:0.45,0.97) {\fontsize{7}{1}\selectfont$\bs \Phi^\top_{\rm 2had} \bs \Psi_{\rm idhw}$};
\end{axis}
\end{tikzpicture}%
	\end{minipage}
	\caption{Structure of the matrix $|(\bs \Phi^\top_{\rm had} \bs \Psi_{\rm dhw})_{k,k'}|$ (left), $|(\bs \Phi^\top_{\rm 2had} \bs \Psi_{\rm adhw})_{k,k'}|$ (middle), and $|(\bs \Phi^\top_{\rm 2had} \bs \Psi_{\rm idhw})_{k,k'}|$ (right) for $N=8$. We observe that the figure in the middle is the Kronecker product of the matrix on the left with itself. This is actually the consequence of the construction of the 2-D Hadamard matrix and ADHW basis using the Kronecker product.}
	\label{fig:structure HadHaar}
\end{figure}
\noindent Noting that $|(\bs H_r)_{k,k'}| = 2^{-r/2}$ for $r\ge 0$ and $k,k' \in \range{2^r}$, Fig.~\ref{fig:structure HadHaar}-left confirms the result in Prop.~\ref{prop: structure of had-haar} for $N=8$.\medskip

We now focus on the 2-D Hadamard-Haar systems to extract a similar structure.
\begin{prop}\label{prop: structure of 2d-had-haar}
	Given an integer $r\ge 0$, we observe that \\
	\begin{subequations}
	(i) for $\subtoind{t}{t_1+1}{t_2+1}{r+1},~\subtoind{l}{l_1+1}{l_2+1}{r+1}$, $t_1,t_2,l_1,l_2 \in \range{r}_0$,
	\begin{align}\label{eq:prop: structure of 2d-had-haar_aniso}
	\ts \bs P_{\cl T^{\rm aniso}_t}\bs \Phi^\top_{\rm 2had} \bs \Psi_{\rm adhw}\bs P^\top_{\cl T^{\rm aniso}_l}
		= \begin{cases}
		\bs H_{(t_2-1)_+}  \otimes \bs H_{(t_1-1)_+}, & {\rm if~}t_1=l_1,~~t_2 = l_2,\\
		\bs 0, & {\rm otherwise,}
		\end{cases} 
	\end{align}
	(ii) $\ts \bs P_{\cl T^{\rm iso}_0}\bs \Phi^\top_{\rm 2had} \bs \Psi_{\rm idhw}\bs P^\top_{\cl T^{\rm iso}_0} = 1$, and for $t,l \in \range{r}_0$ with $(t,l) \ne (0,0)$,
	\begin{align}\label{eq:prop: structure of 2d-had-haar_iso}
	\ts \bs P_{\cl T^{\rm iso}_t}\bs \Phi^\top_{\rm 2had} \bs \Psi_{\rm idhw}\bs P^\top_{\cl T^{\rm iso}_l} = 
		\begin{cases}
			\bs I_3 \otimes \big(\bs H_{(t-1)} \otimes \bs H_{(t-1)}\big), & {\rm if}~t=l,\\
			\bs 0, & {\rm otherwise}.
		\end{cases}
	\end{align}
	\end{subequations}
	\end{prop}
\begin{proof}
See Sec.~\ref{proof:prop:structure of 2d-had-haar}.
\end{proof}
Fig.~\ref{fig:structure HadHaar}-middle and -right depict the structure of the 2-D Hadamard-Haar matrices obtained by multiplying the 2-D Hadamard and Haar matrices. Prop.~\ref{prop: structure of 2d-had-haar} provides a meaningful expression for those structures. We emphasize that the key aspects in the proof of this proposition is the design of the 2-D isotropic and anisotropic levels, as well as the specific column ordering of the IDHW matrix explained in Sec.~\ref{sec:definitions}.

The scaling relations in Prop.~\ref{prop: structure of had-haar} and Prop.~\ref{prop: structure of 2d-had-haar} allow us to determine the local and multilevel coherence of the Hadamard-Haar systems; a result that is at the heart of the proofs of Thm.~\ref{thm:uniform guarantee for Hadamard-Haar system} and Thm.~\ref{thm:Non-uniform guarantee for Hadamard-Haar system}.
\begin{prop}[Local coherence of Hadamard-Haar systems]\label{prop:Local coherence of Had-haar}
	Given integers $r\ge 1$ and $N=2^r$, the following equalities hold:\\
	\begin{subequations}
	(i) for the 1-D Hadamard-Haar system: for $l\in\range{N}$,
	\begin{align}\label{eq:local_coherence_hadhaar_1d}
	&\begin{cases}
		\mu_l^{\rm loc}(\bs \Phi_{\rm had}^\top \bs \Psi_{\rm dhw}) = \min\left(1,2^{-\frac{\left \lfloor \log_2 (l-1) \right \rfloor}{2}}\right),\\
		\|\bs \mu^{\rm loc}(\bs \Phi_{\rm had}^\top \bs \Psi_{\rm dhw})\|^2 = \log_2(N)+1,
	\end{cases}
	\end{align}
	(ii) for the 2-D isotropic Hadamard-Haar system: for $l \overset{N}{\rightleftharpoons}(l_1,l_2)$,
	\begin{align}\label{eq:local_coherence_hadhaar_2d_iso}
	&\begin{cases}
		\mu_l^{\rm loc}(\bs \Phi_{\rm 2had}^\top \bs \Psi_{\rm idhw}) = \min\left(1,2^{-\left \lfloor \log_2 (\max (l_1,l_2)-1) \right \rfloor}\right),\\
		\|\bs \mu^{\rm loc}(\bs \Phi_{\rm 2had}^\top \bs \Psi_{\rm idhw})\|^2 = 3\log_2(N)+1,
	\end{cases}
	\end{align}
	(iii) for the 2-D anisotropic Hadamard-Haar system: for $l \overset{N}{\rightleftharpoons}(l_1,l_2)$,
	\begin{align}\label{eq:local_coherence_hadhaar_2d_aniso}
	\begin{cases}
	\mu_l^{\rm loc}(\bs \Phi_{\rm 2had}^\top \bs \Psi_{\rm adhw}) = \min\left(1,2^{-\frac{\left \lfloor \log_2 (l_1-1) \right \rfloor}{2}}\right) \cdot \min\left(1,2^{-\frac{\left \lfloor \log_2 (l_2-1) \right \rfloor}{2}}\right),\\
	\|\bs \mu^{\rm loc}(\bs \Phi_{\rm 2had}^\top \bs \Psi_{\rm adhw})\|^2 = \big(\log_2(N)+1\big)^2.
	\end{cases}
	\end{align}
	\end{subequations}
\end{prop}
\begin{proof}
	See Sec.~\ref{proof:prop:Local coherence of Had-haar}.
\end{proof}
The exact values of the local coherence are illustrated in Fig.\ref{fig:LocalCoherenceHadamardHaar} for $N=8$. We thus observe that those values are well-controlled in Prop.~\ref{prop:Local coherence of Had-haar}, while the global coherence of the Hadamard-Haar systems is equal to one. Since the value of the local coherence is closed-form in all the three cases considered in Prop.~\ref{prop:Local coherence of Had-haar}, following the argument of Thm.~\ref{thm:vds}, we can set the upper bounds $\kappa_l$ to $\mu^{\rm loc}_l$ to characterize the associated systems in the following theorem.
\begin{figure}[t]
	\centering
	\begin{minipage}{0.32\linewidth}
		\centering
%
%
\begin{tikzpicture}

\begin{axis}[%
width=1.3in,
height=1.3in,
at={(6.944in,3.488in)},
scale only axis,
point meta min=0,
point meta max=1,
axis on top,
xmin=0.5,
xmax=640.5,
xtick ={0},
xticklabels = {0},
xlabel = {$l_2$},
xlabel style={at = {(0.5,0.05)},font=\fontsize{10}{1}\selectfont,color = white},
y dir=reverse,
ymin=0.5,
ymax=640.5,
ytick = {40,120,200,280,360,440,520,600},
yticklabels = {1,2,3,4,5,6,7,8},
ylabel = {$l$},
ylabel style={at = {(0.18,0.5)},font=\fontsize{10}{1}\selectfont},
tick label style={font=\fontsize{8}{1}\selectfont},
colormap/jet,
colorbar,
colorbar style={at = {(1.05,1)},ytick={0,0.5,0.7071,1},yticklabels={0,$2^{-1}$,$2^{-\frac{1}{2}}$,1},font=\fontsize{5}{1}\selectfont,height=1.3in,width=0.2cm}, 
]
\addplot [forget plot] graphics [xmin=0.5,xmax=640.5,ymin=0.5,ymax=640.5] {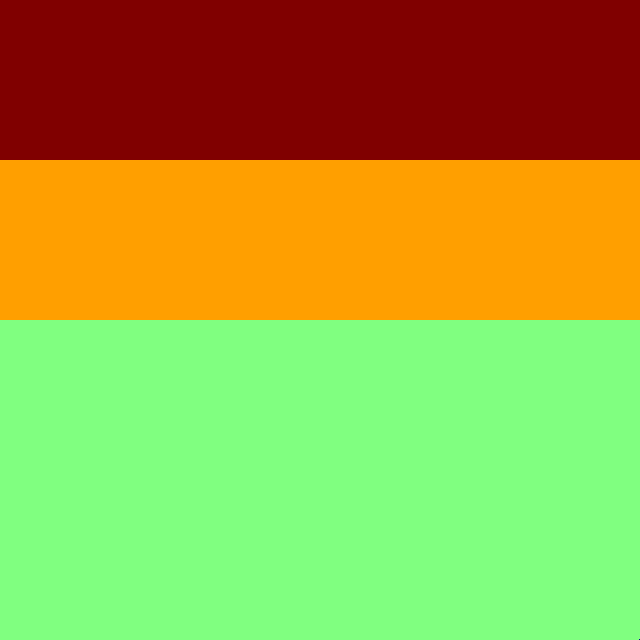};
\node[fill = none,text=black, draw=none, anchor=north west] at (rel axis cs:0.02,0.2) {\fontsize{8}{1}\selectfont $\|\bs \mu^{\rm loc}\|^2 = 4$};
\end{axis}
\end{tikzpicture}%
	\end{minipage}
	\begin{minipage}{0.32\linewidth}
		\centering
%
%
\begin{tikzpicture}

\begin{axis}[%
width=1.3in,
height=1.3in,
at={(6.944in,3.488in)},
scale only axis,
point meta min=0,
point meta max=1,
axis on top,
xmin=0.5,
xmax=640.5,
xtick = {40,120,200,280,360,440,520,600},
xticklabels = {1,2,3,4,5,6,7,8},
xlabel = {$l_2$},
ticklabel style={font=\fontsize{8}{1}\selectfont},
xlabel style={at = {(0.5,0.05)},font=\fontsize{10}{1}\selectfont},
tick align=inside,
y dir=reverse,
ymin=0.5,
ymax=640.5,
ytick = {40,120,200,280,360,440,520,600},
yticklabels = {1,2,3,4,5,6,7,8},
ylabel = {$l_1$},
ylabel style={at = {(0.18,0.5)},font=\fontsize{10}{1}\selectfont},
tick label style={font=\fontsize{8}{1}\selectfont},
colormap/jet,
colorbar,
colorbar style={at = {(1.05,1)},ytick={0,0.25,0.3536,0.5,0.7071,1},yticklabels={0,$2^{-2}$,$2^{-\frac{3}{2}}$,$2^{-1}$,$2^{-\frac{1}{2}}$,1},font=\fontsize{5}{1}\selectfont,height=1.3in,width=0.2cm}, 
]
\addplot [forget plot] graphics [xmin=0.5,xmax=640.5,ymin=0.5,ymax=640.5] {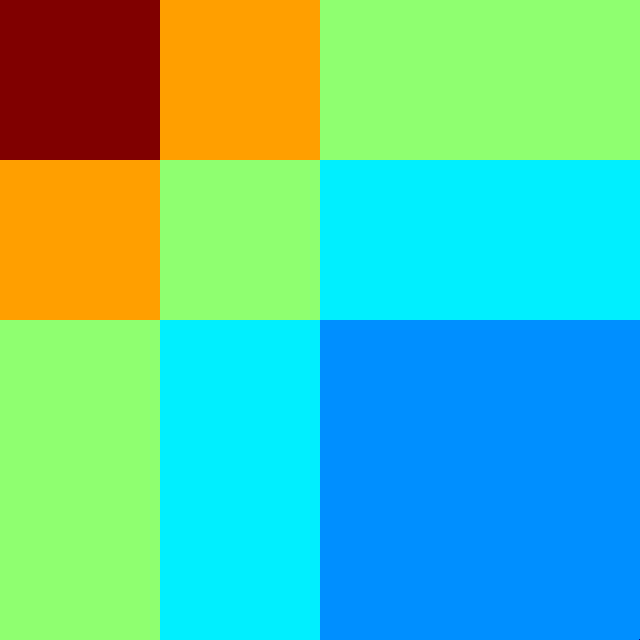};
\node[fill = none,text=black, draw=none, anchor=north west] at (rel axis cs:0.02,0.2) {\fontsize{8}{1}\selectfont $\|\bs \mu^{\rm loc}\|^2 = 16$};
\end{axis}
\end{tikzpicture}%
	\end{minipage}
	\begin{minipage}{0.32\linewidth}
		\centering
%
%
\begin{tikzpicture}

\begin{axis}[%
width=1.3in,
height=1.3in,
at={(6.944in,3.488in)},
scale only axis,
point meta min=0,
point meta max=1,
axis on top,
xmin=0.5,
xmax=640.5,
xtick = {40,120,200,280,360,440,520,600},
xticklabels = {1,2,3,4,5,6,7,8},
xlabel = {$l_2$},
ticklabel style={font=\fontsize{8}{1}\selectfont},
xlabel style={at = {(0.5,0.05)},font=\fontsize{10}{1}\selectfont},
tick align=inside,
y dir=reverse,
ymin=0.5,
ymax=640.5,
ytick = {40,120,200,280,360,440,520,600},
yticklabels = {1,2,3,4,5,6,7,8},
ylabel = {$l_1$},
ylabel style={at = {(0.18,0.5)},font=\fontsize{10}{1}\selectfont},
tick label style={font=\fontsize{8}{1}\selectfont},
colormap/jet,
colorbar,
colorbar style={at = {(1.05,1)},ytick={0,0.25,0.5,1},yticklabels={0,$2^{-2}$,$2^{-1}$,1},font=\fontsize{5}{1}\selectfont,height=1.3in,width=0.2cm}, 
]
\addplot [forget plot] graphics [xmin=0.5,xmax=640.5,ymin=0.5,ymax=640.5] {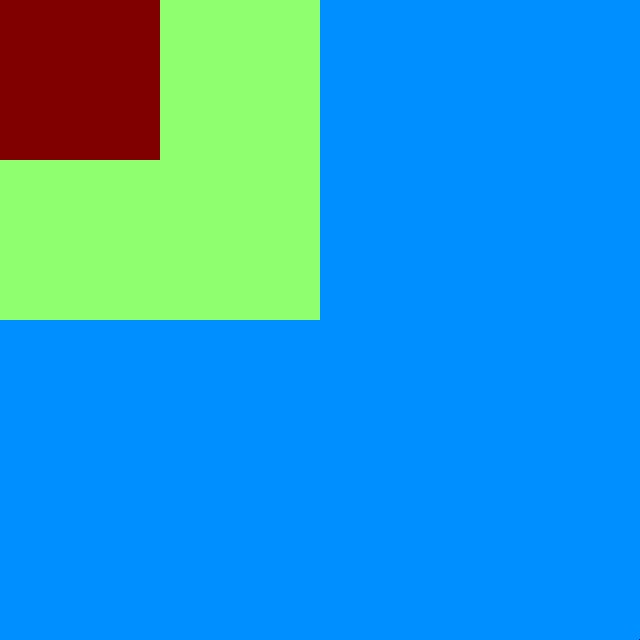};
\node[fill = none,text=black, draw=none, anchor=north west] at (rel axis cs:0.02,0.2) {\fontsize{8}{1}\selectfont $\|\bs \mu^{\rm loc}\|^2 = 10$};
\end{axis}
\end{tikzpicture}%
	\end{minipage}
	\caption{The exact local coherence values for $\mu^{\rm loc}_{l}(\bs \Phi^\top_{\rm had} \bs \Psi_{\rm dhw})$  (left), $\mu^{\rm loc}_{l}(\bs \Phi^\top_{\rm 2had} \bs \Psi_{\rm adhw})$ (middle), and $\mu^{\rm loc}_{l}(\bs \Phi^\top_{\rm 2had} \bs \Psi_{\rm idhw})$ (right) for $N=8$, with $l_1$ and $l_2$ defined in Prop.~\ref{prop:Local coherence of Had-haar}. The values shown here are equal to the estimated values in Prop.~\ref{prop:Local coherence of Had-haar}. The block structure of these figures, as represented by the constant color areas fits the definition of the wavelet levels, \ie with 1-D dyadic (left), 2-D anisotropic (middle), and 2-D isotropic (right) levels.}
	\label{fig:LocalCoherenceHadamardHaar}
\end{figure}
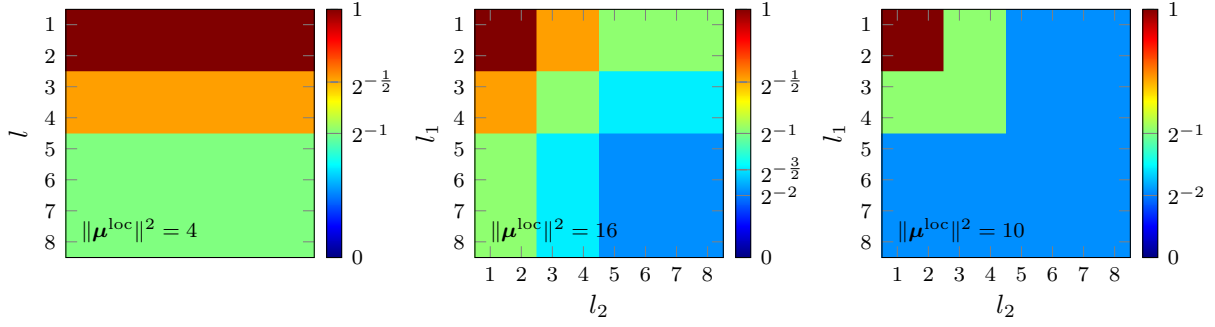
\begin{thm}[Uniform guarantee for Hadamard-Haar systems]\label{thm:uniform guarantee for Hadamard-Haar system}
	Fix $N= 2^r$ for some integer $r\in \bb N$. We provide below, for three Hadamard-Haar systems $(\bs \Phi,\bs \Psi)$ in one and two dimensions, the sample-complexity bound and sampling pmf ensuring~\eqref{eq:sample complexity vds} and \eqref{eq:pmf vds} in Thm.~\ref{thm:vds}:\\
	\begin{subequations}
	(i) for the 1-D Hadamard-Haar system: $\bs \Phi = \bs \Phi_{\rm had} \in \bb R^{N \times N},~\bs \Psi = \bs \Psi_{\rm dhw} \in \bb R^{N\times N}$,
	\begin{align}\label{eq:vds_hadhaar_1d}
	    \ts M \gtrsim K \log(N)\log(\epsilon^{-1}), \text{~and~}
	    ~\eta(l) = \frac{\min\left(1,2^{-\left \lfloor \log_2 (l-1) \right \rfloor}\right)}{\log_2(N)+1},~ l\in\range{N},
	\end{align}
	(ii)~for~the 2-D~isotropic~Hadamard-Haar~system: $\bs \Phi = \bs \Phi_{\rm 2had} \in \bb R^{N^2 \times N^2},~\bs \Psi = \bs \Psi_{\rm idhw} \in \bb R^{N^2\times N^2}\!\!$,
	\begin{align}\label{eq:vds_hadhaar_2d_iso}
		 \ts M \gtrsim K\log(N) \log(\epsilon^{-1}),\text{~and~}
		~\eta(l) = \frac{\min\left(1,2^{-2\left \lfloor \log_2 (\max (l_1,l_2)-1) \right \rfloor}\right)}{3\log_2(N)+1},~ l \overset{N}{\rightleftharpoons}(l_1,l_2),
	\end{align}
	(iii)~for~the 2-D~anisotropic~Hadamard-Haar~system: $\bs \Phi = \bs \Phi_{\rm 2had} \in \bb R^{N^2 \times N^2}\!\!,~\bs \Psi = \bs \Psi_{\rm adhw} \in \bb R^{N^2\times N^2}\!\!\!$,
	\begin{align}\label{eq:vds_hadhaar_2d_aniso}
		 \ts M \gtrsim K\log^2(N) \log(\epsilon^{-1}),\text{~and~}
		~\eta(l) = \frac{\min\left(1,2^{-\left \lfloor \log_2 (l_1-1) \right \rfloor}\right)\cdot \min \left(1,2^{-\left \lfloor \log_2 (l_2-1) \right \rfloor}\right)}{(\log_2(N)+1)^2},~l \overset{N}{\rightleftharpoons}(l_1,l_2).
	\end{align}
	\end{subequations}
\end{thm}
According to this theorem, the optimal sampling pmf $\eta(l)$ is a non-increasing function of $l$. Since $\eta(l) \propto (\mu_l^{\rm loc})^2$, (up to a normalization factor $\ts \|\bs \mu^{\rm loc}\|^2$) the values in Fig.~\ref{fig:LocalCoherenceHadamardHaar} indicate the decay behavior of the sampling pmf. In all the Hadamard-Haar systems, the total number of measurements $M$ is on the order of global sparsity $K$. However, following the computation of the local coherence values in Prop.~\ref{prop:Local coherence of Had-haar}, it could be noticed that the use of the UDS strategy gives $M \gtrsim NK\log(\epsilon^{-1})\log^\alpha(N)$ for some $\alpha \in \{1,2\}$. Moreover, the required number of measurements in \eqref{eq:vds_hadhaar_2d_aniso} is larger than the one in \eqref{eq:vds_hadhaar_2d_iso} by a $\log(N)$ factor: for those signals that have the same sparsity in IDHW and ADHW bases, \ie $ \sigma_K(\bs \Psi^\top_{\rm idhw}\bs x)_1 \approx \sigma_K(\bs \Psi^\top_{\rm adhw}\bs x)_1$, by considering IDHW basis as the sparsity basis we would require smaller number of measurements for signal recovery. \medskip

We now turn our attention to the non-uniform guarantee. Following the sample-complexity bounds \eqref{eq:sample complexity mds 1} and \eqref{eq:sample complexity mds 2}, for a fixed signal and fixed sensing and sparsity bases, the efficiency of the MDS scheme relies on \textit{(i)} a suitable partitioning of the sampling and sparsity domains and \textit{(ii)} the ability to estimate the accurate multilevel coherence and relative sparsity values. \medskip

One way to design the sampling and sparsity levels is to leverage the structure of the Hadamard-Haar systems observed in Prop.~\ref{prop: structure of had-haar} and Prop.~\ref{prop: structure of 2d-had-haar}. To visualize those structure, one can properly permute the columns and the rows of the matrices in Fig.~\ref{fig:structure HadHaar} according to specific wavelet levels, \eg the 1-D dyadic, 2-D isotropic, or 2-D anisotropic, and obtain the matrices in Fig.~\ref{fig:structure HadHaar after reordering}. Each white rectangle in Fig.~\ref{fig:structure HadHaar after reordering} centered at the index $(t,l)$ corresponds to a single partition. Note that the horizontal and vertical axis in Fig.~\ref{fig:structure HadHaar after reordering} denotes the sparsity and sampling level index, respectively; while the axis in Fig.~\ref{fig:structure HadHaar} represent the column and row indices. The observed structures in Fig.~\ref{fig:structure HadHaar after reordering}, specially the ones related to the ADHW and IDHW bases, confirms the statements of Prop.~\ref{prop: structure of had-haar} and Prop.~\ref{prop: structure of 2d-had-haar}. With these structures in mind, we can now compute the following values for multilevel coherence and relative sparsity in different Hadamard-Haar systems.
\begin{figure}
	\centering
	\begin{minipage}{0.32\linewidth}
		\centering
%
%
\begin{tikzpicture}

\begin{axis}[%
width=1.3in,
height=1.3in,
at={(0in,0in)},
scale only axis,
point meta min=0,
point meta max=1,
axis on top,
xmin=0.5,
xmax=640.5,
xtick = {40,120,240,480},
xticklabels = {0,1,2,3},
xlabel = {sparsity level $(l)$},
ticklabel style={font=\fontsize{8}{1}\selectfont},
xlabel style={at = {(0.5,0.08)},font=\fontsize{10}{1}\selectfont},
tick align=inside,
y dir=reverse,
ymin=0.5,
ymax=640.5,
ytick = {40,120,240,480},
yticklabels = {0,1,2,3},
ylabel = {sampling level $(t)$},
ylabel style={at = {(0.22,0.5)},font=\fontsize{10}{1}\selectfont},
colormap/jet,
colorbar,
colorbar style={at = {(1.05,1)},ytick={0,0.5,0.7071,1},yticklabels={0,$2^{-1}$,$2^{-\frac{1}{2}}$,1},font=\fontsize{5}{1}\selectfont,height=1.3in,width=0.2cm},  
]
\addplot [forget plot] graphics [xmin=0.5,xmax=640.5,ymin=0.5,ymax=640.5] {Fig_HadamardHaar_1d-1.png};
\draw[color = white, fill = none] (axis cs:-5.5,320.5) -- (axis cs:641.5,320.5);
\draw[color = white, fill = none] (axis cs:320.5,-5.5) -- (axis cs:320.5,641.5);

\draw[color = white, fill = none] (axis cs:-5.5,160.5) -- (axis cs:641.5,160.5);
\draw[color = white, fill = none] (axis cs:160.5,-5.5) -- (axis cs:160.5,641.5);

\draw[color = white, fill = none] (axis cs:-5.5,80.5) -- (axis cs:641.5,80.5);
\draw[color = white, fill = none] (axis cs:80.5,-5.5) -- (axis cs:80.5,641.5);
\end{axis}
\end{tikzpicture}%
	\end{minipage}
	\begin{minipage}{0.32\linewidth}
		\centering
		\vspace{1mm}
		\scalebox{1}{
%
%
\begin{tikzpicture}

\begin{axis}[%
width=1.3in,
height=1.3in,
at={(0in,0in)},
scale only axis,
point meta min=0,
point meta max=1,
axis on top,
xmin=0.5,
xmax=640.5,
xtick = {5,60,140,280,560},
xticklabels = {1,4,8,12,16},
xlabel = {sparsity level $(l)$},
ticklabel style={font=\fontsize{8}{1}\selectfont},
xlabel style={at = {(0.5,0.08)},font=\fontsize{10}{1}\selectfont},
tick align=inside,
y dir=reverse,
ymin=0.5,
ymax=640.5,
ytick = {5,60,140,280,560},
yticklabels = {1,4,8,12,16},
ylabel = {sampling level $(t)$},
ylabel style={at = {(0.20,0.5)},font=\fontsize{10}{1}\selectfont},
colormap/jet,
colorbar,
colorbar style={at = {(1.05,1)},ytick={0,0.25,0.3536,0.5,0.7071,1},yticklabels={0,$2^{-2}$,$2^{-\frac{3}{2}}$,$2^{-1}$,$2^{-\frac{1}{2}}$,1},font=\fontsize{5}{1}\selectfont,height=1.3in,width=0.2cm}, 
]
\addplot [forget plot] graphics [xmin=0.5,xmax=640.5,ymin=0.5,ymax=640.5] {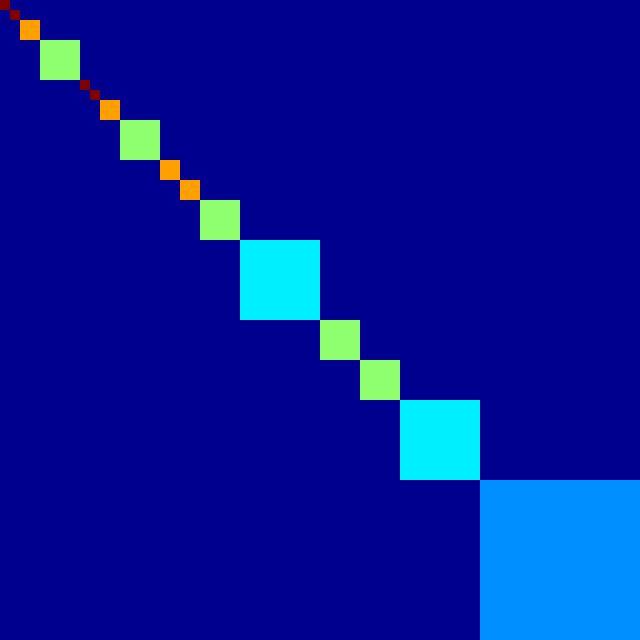};
\draw[color = white, fill = none] (axis cs:-5.5,480.5) -- (axis cs:641.5,480.5);
\draw[color = white, fill = none] (axis cs:480.5,-5.5) -- (axis cs:480.5,641.5);

\draw[color = white, fill = none] (axis cs:-5.5,400.5) -- (axis cs:641.5,400.5);
\draw[color = white, fill = none] (axis cs:400.5,-5.5) -- (axis cs:400.5,641.5);

\draw[color = white, fill = none] (axis cs:-5.5,360.5) -- (axis cs:641.5,360.5);
\draw[color = white, fill = none] (axis cs:360.5,-5.5) -- (axis cs:360.5,641.5);

\draw[color = white, fill = none] (axis cs:-5.5,320.5) -- (axis cs:641.5,320.5);
\draw[color = white, fill = none] (axis cs:320.5,-5.5) -- (axis cs:320.5,641.5);

\draw[color = white, fill = none] (axis cs:-5.5,240.5) -- (axis cs:641.5,240.5);
\draw[color = white, fill = none] (axis cs:240.5,-5.5) -- (axis cs:240.5,641.5);

\draw[color = white, fill = none] (axis cs:-5.5,200.5) -- (axis cs:641.5,200.5);
\draw[color = white, fill = none] (axis cs:200.5,-5.5) -- (axis cs:200.5,641.5);

\draw[color = white, fill = none] (axis cs:-5.5,180.5) -- (axis cs:641.5,180.5);
\draw[color = white, fill = none] (axis cs:180.5,-5.5) -- (axis cs:180.5,641.5);

\draw[color = white, fill = none] (axis cs:-5.5,160.5) -- (axis cs:641.5,160.5);
\draw[color = white, fill = none] (axis cs:160.5,-5.5) -- (axis cs:160.5,641.5);

\draw[color = white, fill = none] (axis cs:-5.5,120.5) -- (axis cs:641.5,120.5);
\draw[color = white, fill = none] (axis cs:120.5,-5.5) -- (axis cs:120.5,641.5);

\draw[color = white, fill = none] (axis cs:-5.5,100.5) -- (axis cs:641.5,100.5);
\draw[color = white, fill = none] (axis cs:100.5,-5.5) -- (axis cs:100.5,641.5);

\draw[color = white, fill = none] (axis cs:-5.5,90.5) -- (axis cs:641.5,90.5);
\draw[color = white, fill = none] (axis cs:90.5,-5.5) -- (axis cs:90.5,641.5);

\draw[color = white, fill = none] (axis cs:-5.5,80.5) -- (axis cs:641.5,80.5);
\draw[color = white, fill = none] (axis cs:80.5,-5.5) -- (axis cs:80.5,641.5);

\draw[color = white, fill = none] (axis cs:-5.5,40.5) -- (axis cs:641.5,40.5);
\draw[color = white, fill = none] (axis cs:40.5,-5.5) -- (axis cs:40.5,641.5);

\draw[color = white, fill = none] (axis cs:-5.5,20.5) -- (axis cs:641.5,20.5);
\draw[color = white, fill = none] (axis cs:20.5,-5.5) -- (axis cs:20.5,641.5);

\draw[color = white, fill = none] (axis cs:-5.5,10.5) -- (axis cs:641.5,10.5);
\draw[color = white, fill = none] (axis cs:10.5,-5.5) -- (axis cs:10.5,641.5);
\end{axis}
\end{tikzpicture}%
}
	\end{minipage}
	\begin{minipage}{0.32\linewidth}
		\centering
		\scalebox{1}{
%
%
\begin{tikzpicture}

\begin{axis}[%
width=1.3in,
height=1.3in,
at={(0in,0in)},
scale only axis,
point meta min=0,
point meta max=1,
axis on top,
xmin=0.5,
xmax=640.5,
xtick = {5,25,100,400},
xticklabels = {,1,2,3},
xlabel = {sparsity level $(l)$},
ticklabel style={font=\fontsize{8}{1}\selectfont},
xlabel style={at = {(0.5,0.08)},font=\fontsize{10}{1}\selectfont},
tick align=inside,
y dir=reverse,
ymin=0.5,
ymax=640.5,
ytick = {5,25,100,400},
yticklabels = {,1,2,3},
ylabel = {sampling level $(t)$},
ylabel style={at = {(0.22,0.5)},font=\fontsize{10}{1}\selectfont},
colormap/jet,
colorbar,
colorbar style={at = {(1.05,1)},ytick={0,0.25,0.5,1},yticklabels={0,$2^{-2}$,$2^{-1}$,1},font=\fontsize{6}{1}\selectfont,height=1.3in,width=0.2cm}, 
]
\addplot [forget plot] graphics [xmin=0.5,xmax=640.5,ymin=0.5,ymax=640.5] {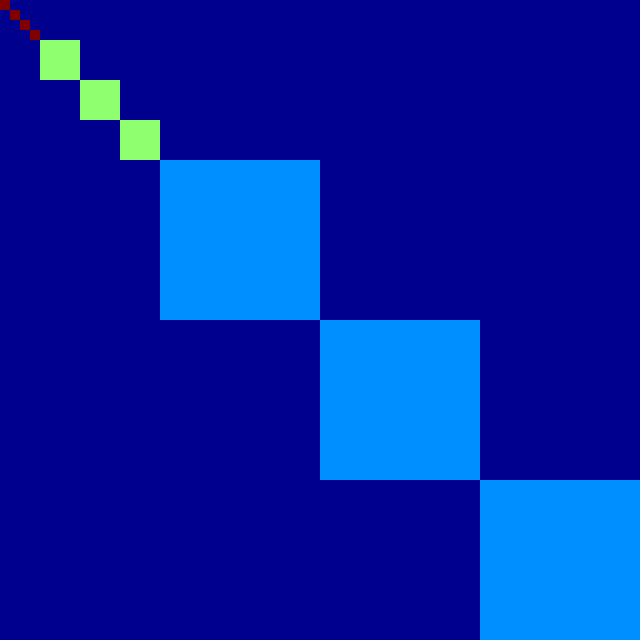};
\draw[color = white, fill = none] (axis cs:160.5,160.5)rectangle (axis cs:641.5,641.5);
\draw[color = white, fill = none] (axis cs:40.5,40.5)rectangle (axis cs:160.5,160.5);
\draw[color = white, fill = none] (axis cs:-5.5,-5.5)rectangle (axis cs:40.5,40.5);
\draw[color = white, fill = none] (axis cs:-5.5,160.5)rectangle (axis cs:40.5,641.5);
\draw[color = white, fill = none] (axis cs:160.5,-5.5)rectangle (axis cs:641.5,40.5);

\draw[color = white, fill = none] (axis cs:-5.5,-5.5)rectangle (axis cs:10.5,641.5);
\draw[color = white, fill = none] (axis cs:-5.5,-5.5)rectangle (axis cs:641.5,10.5);
\end{axis}
\end{tikzpicture}%
}
	\end{minipage}
	\caption{Rearrangement of the rows and columns of the matrices shown in Fig.~\ref{fig:structure HadHaar} with respect to the sampling and sparsity levels. Each white rectangle centered at $(t,l)$ corresponds to the $(t,l)^{\rm th}$ block, \ie $\bs P_{\cl T^{\rm 1d}_t}\bs \Phi^\top_{\rm had} \bs \Psi_{\rm dhw} \bs P^\top_{\cl  T^{\rm 1d}_l}$ (left), $\bs P_{\cl T^{\rm aniso}_t}\bs \Phi^\top_{\rm 2had} \bs \Psi_{\rm adhw} \bs P^\top_{\cl  T^{\rm aniso}_l}$ (middle), and $\bs P_{\cl T^{\rm iso}_t}\bs \Phi^\top_{\rm 2had} \bs \Psi_{\rm idhw} \bs P^\top_{\cl  T^{\rm iso}_l}$ (right).}
	\label{fig:structure HadHaar after reordering}
\end{figure}
\begin{prop}[Multilevel coherence and relative sparsity of Hadamard-Haar systems]\label{prop:Multilevel coherence of Had-haar}
	Fix integers $r$ and $N=2^r$. We consider the levels $\cl T^{\rm 1d}$, $\cl T^{\rm iso}$, and $\cl T^{\rm aniso}$ defined above and, for each of them, a vector $\bs k$ whose size equals the number of levels. Then, the following holds:\\
	\begin{subequations}
	(i) for the 1-D Hadamard-Haar system: for $t, l \in \range{r}_0$,
	\begin{align}\label{eq:multilevel_coherence_hadhaar_1d}
			\begin{cases}
			 \mu_{t,l}^{\cl T^{\rm 1d},\cl T^{\rm 1d}}(\bs \Phi^\top_{\rm had}\bs \Psi_{\rm dhw}) = 2^{-(t-1)_+}\cdot \delta_{t,l},\\
        	 K_{t}^{\cl T^{\rm 1d},\cl T^{\rm 1d}}(\bs \Phi^\top_{\rm had}\bs \Psi_{\rm dhw},\bs k) \le k_t,
			\end{cases}
     \end{align}
     (ii) for the 2-D isotropic Hadamard-Haar system: for $t, l \in \range{r}_0$,
     \begin{align}\label{eq:multilevel_coherence_hadhaar_2d_iso}
		\begin{cases}
				 \mu_{t,l}^{\cl T^{\rm iso},\cl T^{\rm iso}}(\bs \Phi^\top_{\rm 2had}\bs \Psi_{\rm idhw}) = 2^{-2(t-1)_+}\cdot \delta_{t,l},\\
		         K_{t}^{\cl T^{\rm iso},\cl T^{\rm iso}}(\bs \Phi^\top_{\rm 2had}\bs \Psi_{\rm idhw},\bs k) \le k_t,
		\end{cases}
	\end{align}
	(iii) for the 2-D anisotropic Hadamard-Haar system: for $\subtoind{t}{t_1+1}{t_2+1}{r+1},~\subtoind{l}{l_1+1}{l_2+1}{r+1}$,
	\begin{align}\label{eq:multilevel_coherence_hadhaar_2d_aniso}
		\begin{cases}
				 \ts \mu_{t,l}^{\cl T^{\rm aniso},\cl T^{\rm aniso}}(\bs \Phi^\top_{\rm 2had}\bs \Psi_{\rm adhw}) = 2^{-(t_1-1)_+}\cdot 2^{-(t_2-1)_+} \cdot\delta_{t_1,l_1}\cdot \delta_{t_2,l_2},\\
		         \ts K_{t}^{\cl T^{\rm aniso},\cl T^{\rm aniso}}(\bs \Phi^\top_{\rm 2had}\bs \Psi_{\rm adhw},\bs k) \le  k_{t},
		\end{cases}
	\end{align}
	\end{subequations}
	where the relative sparsity $K_t^{\cl W,\cl S}$ and the multilevel coherence $\mu_{t,l}^{\cl W, \cl S}$ are defined in \eqref{eq:relative sparsity} and \eqref{eq:multilevel coherence}, respectively, and where $\delta_{t,l}$ is a Kronecker function, \ie $\delta_{k,l}=1$ if $t=l$ (and zero, otherwise).
\end{prop}
\begin{proof}
See Sec.~\ref{proof:prop:Multilevel coherence of Had-haar}.
\end{proof}
\begin{figure}[t]
	\centering
	\begin{minipage}{0.32\linewidth}
	\centering
	\scalebox{0.95}{
%
%
\begin{tikzpicture}

\begin{axis}[%
width=1.3in,
height=1.3in,
at={(0in,0in)},
scale only axis,
point meta min=0,
point meta max=1,
axis on top,
xmin=0.5,
xmax=640.5,
xtick = {80,240,400,560},
xticklabels = {0,1,2,3},
xlabel = {sparsity level $(l)$},
ticklabel style={font=\fontsize{8}{1}\selectfont},
xlabel style={at = {(0.5,0.08)},font=\fontsize{10}{1}\selectfont},
tick align=inside,
y dir=reverse,
ymin=0.5,
ymax=640.5,
ytick = {80,240,400,560},
yticklabels = {0,1,2,3},
ylabel = {sampling level $(t)$},
ylabel style={at = {(0.18,0.5)},font=\fontsize{10}{1}\selectfont},
tick label style={font=\fontsize{8}{1}\selectfont},
colormap/jet,
colorbar,
colorbar style={at = {(1.05,1)},ytick={0,0.25,0.5,1},yticklabels={0,$2^{-2}$,$2^{-1}$,1},font=\fontsize{5}{1}\selectfont,height=1.3in,width=0.2cm}, 
]
\addplot [forget plot] graphics [xmin=0.5,xmax=640.5,ymin=0.5,ymax=640.5] {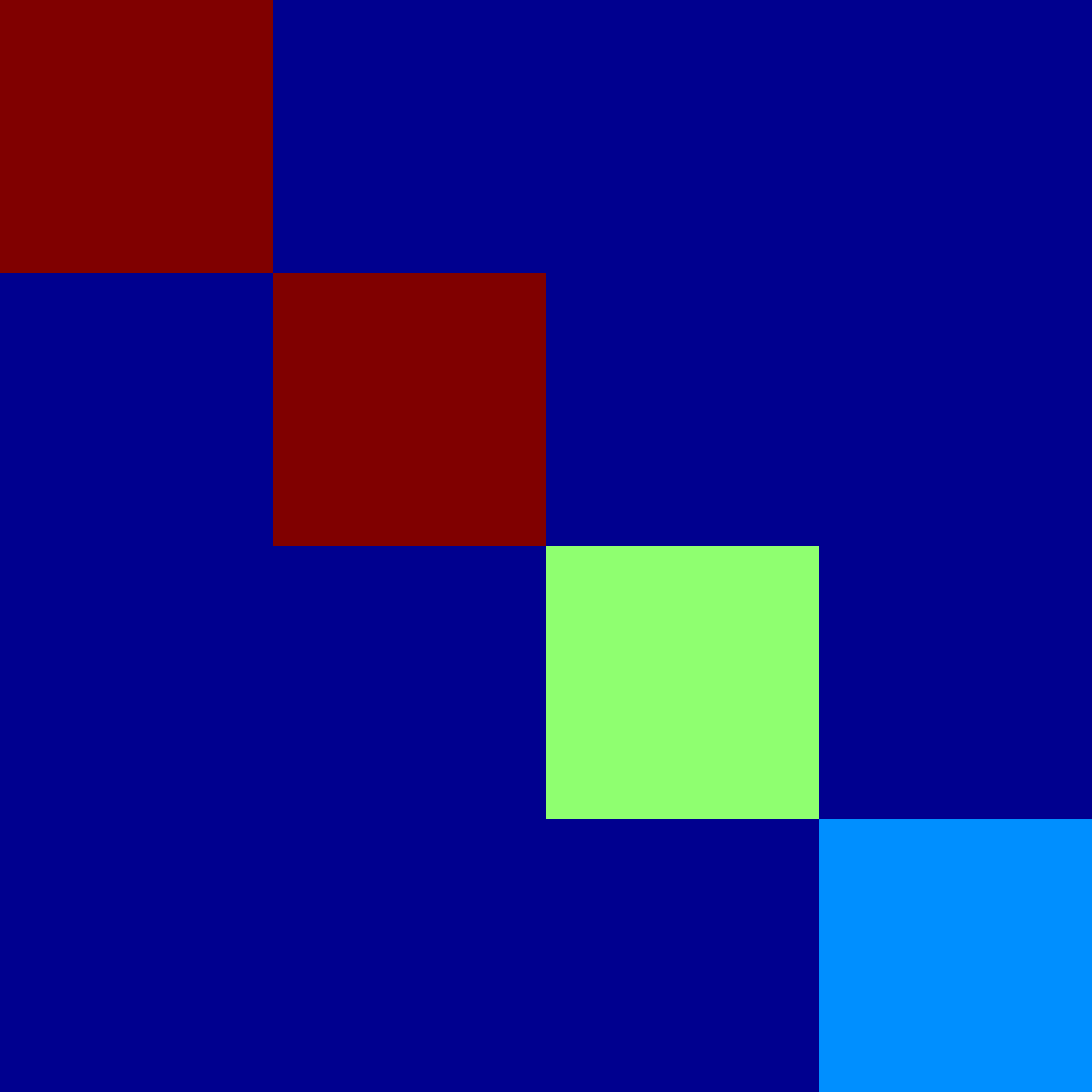};
\end{axis}
\end{tikzpicture}%
}
	\end{minipage}
	\begin{minipage}{0.32\linewidth}
	\centering
	\scalebox{0.95}{
%
%
\begin{tikzpicture}

\begin{axis}[%
width=1.3in,
height=1.3in,
at={(0in,0in)},
scale only axis,
point meta min=0,
point meta max=1,
axis on top,
xmin=0.5,
xmax=1600.5,
xtick = {50,250,450,650,850,1050,1250,1450},
xticklabels = {1,3,5,7,9,11,13,15},
xlabel = {sparsity level $(l)$},
ticklabel style={font=\fontsize{8}{1}\selectfont},
xlabel style={at = {(0.5,0.08)},font=\fontsize{10}{1}\selectfont},
tick align=inside,
y dir=reverse,
ymin=0.5,
ymax=1600.5,
ytick = {50,250,450,650,850,1050,1250,1450},
yticklabels = {1,3,5,7,9,11,13,15},
ylabel = {sampling level $(t)$},
ylabel style={at = {(0.19,0.5)},font=\fontsize{10}{1}\selectfont},
tick label style={font=\fontsize{8}{1}\selectfont},
colormap/jet,
colorbar,
colorbar style={at = {(1.05,1)},ytick={0.0625,0.125,0.25,0.5,1},yticklabels={$2^{-4}$,$2^{-3}$,$2^{-2}$,$2^{-1}$,1},font=\fontsize{5}{1}\selectfont,height=1.3in,width=0.2cm}, 
]
\addplot [forget plot] graphics [xmin=0.5,xmax=1600.5,ymin=0.5,ymax=1600.5] {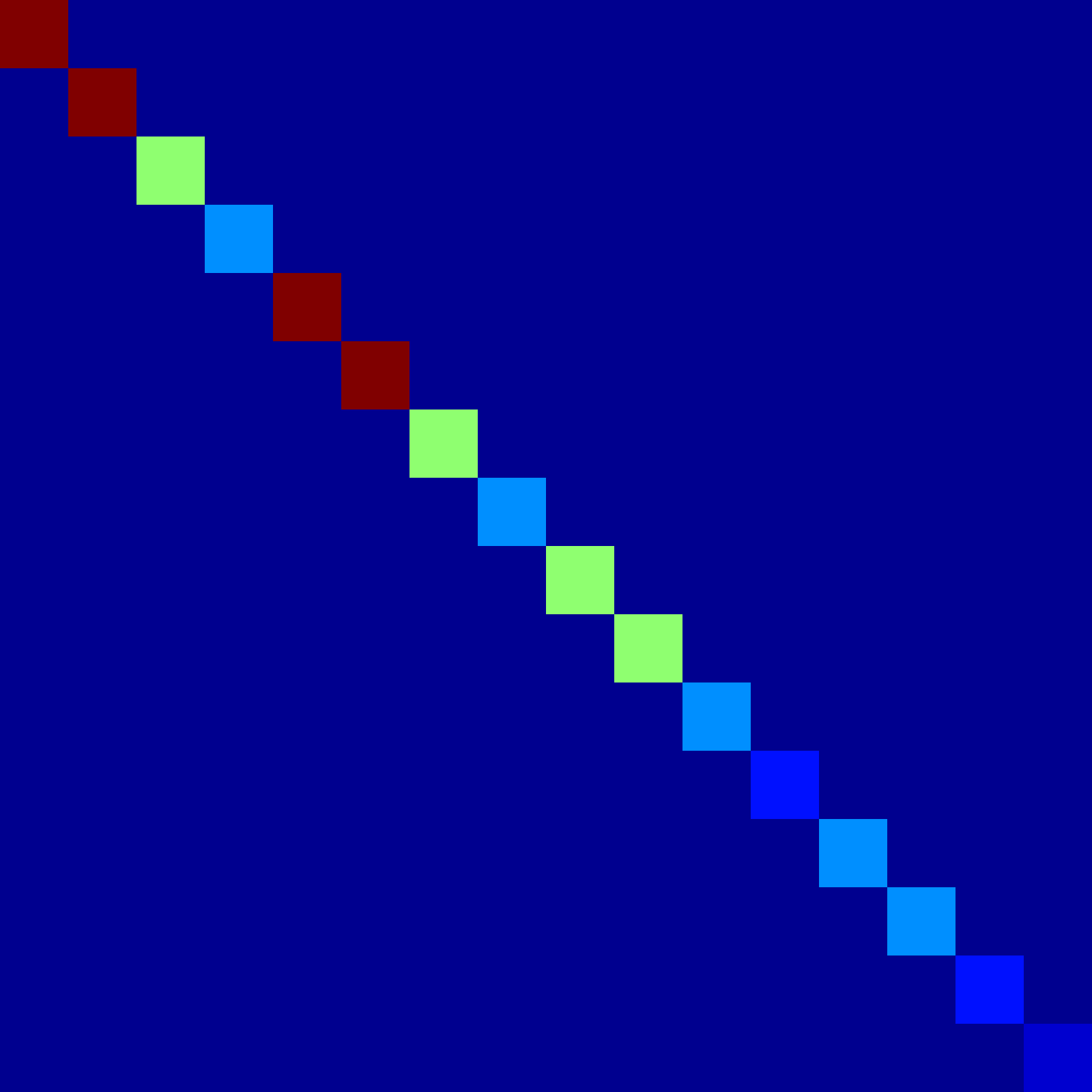};
\end{axis}
\end{tikzpicture}%
}
	\end{minipage}
	\begin{minipage}{0.32\linewidth}
	\centering
	\scalebox{0.95}{
%
%
\begin{tikzpicture}

\begin{axis}[%
width=1.3in,
height=1.3in,
at={(0in,0in)},
scale only axis,
point meta min=0,
point meta max=1,
axis on top,
xmin=0.5,
xmax=400.5,
xtick = {50,150,250,350},
xticklabels = {0,1,2,3},
xlabel = {sparsity level $(l)$},
ticklabel style={font=\fontsize{8}{1}\selectfont},
xlabel style={at = {(0.5,0.08)},font=\fontsize{10}{1}\selectfont},
tick align=inside,
y dir=reverse,
ymin=0.5,
ymax=400.5,
ytick = {50,150,250,350},
yticklabels = {0,1,2,3},
ylabel = {sampling level $(t)$},
ylabel style={at = {(0.18,0.5)},font=\fontsize{10}{1}\selectfont},
tick label style={font=\fontsize{8}{1}\selectfont},
colormap/jet,
colorbar,
colorbar style={at = {(1.05,1)},ytick={0.0625,0.25,1},yticklabels={$2^{-4}$,$2^{-2}$,1},font=\fontsize{5}{1}\selectfont,height=1.3in,width=0.2cm}, 
]
\addplot [forget plot] graphics [xmin=0.5,xmax=400.5,ymin=0.5,ymax=400.5] {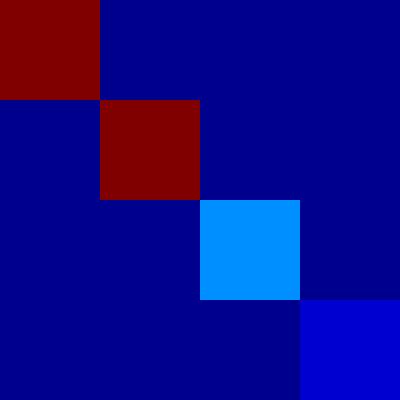};
\end{axis}
\end{tikzpicture}%
}
	\end{minipage}
	\caption{The exact multilevel coherence values for Hadamard-Haar systems with $N=8$: (left) $\mu_{t,l}^{\cl T^{\rm 1d},\cl T^{\rm 1d}}(\bs \Phi_{\rm had}^\top\bs \Psi_{\rm dhw})$, (middle) $\mu_{t,l}^{\cl T^{\rm iso},\cl T^{\rm iso}}(\bs \Phi_{\rm 2had}^\top\bs \Psi_{\rm adhw})$, and (right) $\mu_{t,l}^{\cl T^{\rm iso},\cl T^{\rm iso}}(\bs \Phi_{\rm 2had}^\top\bs \Psi_{\rm idhw})$. The multilevel coherence values in this figure confirm our estimations in Prop.~\ref{prop:Multilevel coherence of Had-haar}.}
	\label{fig:MultilevelCoherenceHadamardHaar}
\end{figure}
According to Prop.~\ref{prop:Multilevel coherence of Had-haar}, the multilevel coherence is an exponentially-decreasing function of the level index (see also Fig.~\ref{fig:MultilevelCoherenceHadamardHaar} for an illustration of the multilevel coherence for $N=8$). Moreover, as an advantage of our sampling and sparsity levels design, the multilevel coherence of the Hadamard-Haar systems at level $(t,l)$ vanishes when $t\ne l$ and thus, the sample-complexity bounds \eqref{eq:sample complexity mds 1} and \eqref{eq:sample complexity mds 2} become 
\begin{align*}
\ts	m_t &\gtrsim |\cl W_t| \,\mu_{t,t}^{\cl W,\cl S}(\bs \Phi^\top\bs \Psi) \, k_t\,\log(K\epsilon^{-1})\,\log(N),\\
\ts 1\ & \gtrsim\ \left(\frac{|\cl W_l|}{\hat{m}_l}-1\right)\ \mu_{l,l}^{\cl W,\cl S}(\bs \Phi^\top\bs \Psi)\, \ts  K_l^{\cl W, \cl S}(\bs \Phi^\top\bs \Psi, \bs k).
\end{align*}
If we ignore the second sample-complexity bound, the first bound relates the number of measurements $m_t$ at level $t$ to the sparsity value $k_t$ at the same level $t$ (and not to the sparsity values at the other levels). This is exactly as one expects when the matrix $\bs \Phi^\top\bs \Psi$ is block-diagonal (see \cite[Sec.~4.2.1]{adcock2017breaking} for more insights) and an application of the sample-complexity bound \eqref{eq:sample complexity uds} on every block gives the sufficient conditions on the number of measurements. 

We are now ready to combine the proposition above with Thm.~\ref{thm:mds} and present the following non-uniform recovery guarantees of Hadamard-Haar systems.
\begin{thm}[Non-uniform guarantee for Hadamard-Haar systems]\label{thm:Non-uniform guarantee for Hadamard-Haar system}
	Given $N= 2^r$ for some integer $r\in \bb N$, if we fix 
	\begin{equation}\label{eq:mds_hadhaar}
	 m_t \gtrsim k_t \log(K\epsilon^{-1})\log(N)
	\end{equation}
	with either:\\
	(i) for the 1-D Hadamard-Haar system:
	 $$
	 t \in \range{r}_0, \bs \Phi = \bs \Phi_{\rm had} \in \bb R^{N \times N},~\bs \Psi = \bs \Psi_{\rm dhw} \in \bb R^{N\times N}, \cl W = \cl T = \cl T^{\rm 1d};
	 $$
	(ii) for the 2-D isotropic Hadamard-Haar system: \newline
	$$
	t \in \range{r}_0, \bs \Phi = \bs \Phi_{\rm 2had} \in \bb R^{N^2 \times N^2},~\bs \Psi = \bs \Psi_{\rm idhw} \in \bb R^{N^2\times N^2}, \cl W = \cl T = \cl T^{\rm iso}; 
	$$
	(iii) for the 2-D anisotropic Hadamard-Haar system: \newline
	 $$
	 t\in\range{(r+1)^2}, \bs \Phi = \bs \Phi_{\rm 2had} \in \bb R^{N^2 \times N^2},~\bs \Psi = \bs \Psi_{\rm adhw} \in \bb R^{N^2\times N^2}, \cl W = \cl T = \cl T^{\rm aniso}; 
	 $$
	then~\eqref{eq:sample complexity mds 1} and~\eqref{eq:sample complexity mds 2} in Thm.~\ref{thm:mds} are satisfied.
\end{thm}
\begin{proof}
See Sec.~\ref{proof:thm:Non-uniform guarantee for Hadamard-Haar system}.
\end{proof}
It is worth mentioning that Thm.~\ref{thm:Non-uniform guarantee for Hadamard-Haar system} provides the tightest sample-complexity bounds, since the multilevel coherence values that lead to these estimates are accurately computed in Prop.~\ref{proof:prop:Multilevel coherence of Had-haar}. We observe in Thm.~\ref{thm:Non-uniform guarantee for Hadamard-Haar system} that the local number of measurements $m_t$ for the covered Hadamard-Haar systems is on the order of the corresponding local sparsity $k_t$. A similar observation has recently been made for the infinite-dimensional Hadamard-Haar system in \cite[Thm.~4.13]{adcock2019uniform}. Unlike the observation in \eqref{eq:mds_hadhaar}, for an arbitrary orthonormal wavelet basis the local number of measurements $m_t$ scales as a linear combination of the local sparsities (see in \cite{thesing2019} or \cite[Thm. 5.8]{calderbank2019}), which is due to the fact that the Hadamard-wavelet system is not exactly block-diagonal.

\begin{remark}\label{rem:mds}
One can question how to set the local number of measurements $m_t$  given the local sparsity values $k_l$ and the total number of measurements $M$. We provide an approach for the 1-D signal recovery problem that is easily extendable to the 2-D cases. The idea here is based on the fact that the local number of measurements in \eqref{eq:mds_hadhaar} can be written as $m_t = C k_t$ for $t \in \range{r}_0$ with $C>0$ independent of $t$ and $k_t$. Therefore, the total number of measurements is $M = \sum_t m_t = C K$ where $K$ is the total sparsity value. Therefore, up to a rounding error, the local number of measurements reads 
$$
m_t = \frac{M}{K} k_t,~t\in\range{r}_0.
$$
\end{remark}

\section{Numerical results}\label{sec:simulations}

In this section we carry out several simulations to verify the obtained theoretical results in Thm.~\ref{thm:uniform guarantee for Hadamard-Haar system} and Thm.~\ref{thm:Non-uniform guarantee for Hadamard-Haar system}. In the first set of simulations we address the problem of 1-D signal recovery from subsampled Hadamard measurements and later we focus on the 2-D signal recovery problem, which is associated with single pixel imaging application of CS. 

The general setup of the simulations is as follows. Given a ground truth signal $\bs x \in \bb C^{N}$ we follow the sensing model \eqref{eq: cs acquisition model} for some dimensions and sensing bases to be specified later, where we suppose the noise components $n_l \sim_{\rm i.i.d.} \cl N(0,\sigma)$ and $\sigma$ is fixed with respect to the desired Signal-to-Noise Ratio (SNR) $\coloneqq
20 \log_{10}(\|\bs x\|/(\sigma\sqrt{N}))$ in dB.
For all the experiments we report the Signal-to-Reconstruction Error (SRE) in dB, \ie
\begin{equation*}
{\rm SRE} \coloneqq 20 \log_{10} \bb E_{\rm e} \|\bs x\|/\|\bs x - \hat{\bs x}\|,
\end{equation*}
where $\bb E_{\rm e}$ is the empirical mean over several trials of the sensing context (as specified in the text). In this section the term ``VDS'' (or ``MDS'') implies the sampling strategies defined in Thm.~\ref{thm:uniform guarantee for Hadamard-Haar system} (resp.~Thm.~\ref{thm:Non-uniform guarantee for Hadamard-Haar system}). For the MDS scheme we respect the approach described in Remark~\ref{rem:mds}. We consider two algorithms for signal reconstruction: \textit{(i)} CS reconstruction, that refers to the $\ell_1$ minimization problem \eqref{eq:bpdn non-uniform} or \eqref{eq:BPDN uniform} (depending on the recovery guarantee type) for some sparsity basis to be specified later; and \textit{(ii)} Minimal Energy (ME) reconstruction \cite{candes2006robust}, which corresponds to applying the right pseudo-inverse of $\bs \Phi$ to the measurement vector. CS reconstructions \eqref{eq:bpdn non-uniform} and \eqref{eq:BPDN uniform} are performed with the Spectral Projected Gradient for $\ell_1$ minimization (SPGL1) \cite{vandenBerg:2008,spgl1:2007}. In our experiments, the parameter $\varepsilon$ in \eqref{eq:BPDN uniform} (and \eqref{eq:bpdn non-uniform}) is set to the oracle value of $\|\bs D \bs n\|$ (resp. $\|\bs n\|$).
Matrices and operators are implemented using the Spot toolbox \cite{spot}.

The MDS schemes in Thm.~\ref{thm:Non-uniform guarantee for Hadamard-Haar system} require to set the values of the local sparsity parameter $k_l$. For these simulations, when the signal of interest is not exactly sparse we perform the following procedure that is proposed by Adcock~\etal~in \cite[Eq.~2.8]{adcock2017breaking} and used in \cite{moshtaghpour2018multilevel}: \textit{(i)} given a parameter $\rho \in (0,1]$ and a signal $\bs x\in \bb C^N$ we first compute the vector of coefficients $\bs s\in \bb C^N$ in the sparsity basis $\bs \Psi$, \ie $\bs s = \bs \Psi^\top \bs x$; \textit{(ii)} the effective global sparsity value $K$ is then computed such that after applying the hard thresholding operator $\cl H_K$ to $s$, the ratio of the energy that is preserved by $K$ coefficients equals $\rho$, mathematically,
$$
K = K(\rho) = \min \{ n: \| \cl H_{n}(\bs s) \| / \| \bs s\| \ge \rho \},
$$
where we set $\rho = 0.995$ in all the experiments here ; \textit{(iii)} we finally compute the effective local sparsity values by simply localizing the number of non-zero coefficients of the hard thresholded signal $\cl H_K(\bs s)$, \ie for all $l$
$$
k_l = k_l(\rho) = |{\rm supp}(\bs P_{\cl S_l} \cl H_{K(\rho)}(\bs s))|.
$$
Note that this procedure does not sparsify the signal $\bs x$ in the basis $\bs \Psi$, as it is only used to estimate the parameters $k_l$.
\subsection{1-D signal recovery}
\begin{figure}
	\centering
	\scalebox{0.85}{
%
%
\definecolor{mycolor1}{rgb}{1.00000,0.00000,1.00000}%
\definecolor{mycolor2}{rgb}{0.00000,1.00000,1.00000}%
\begin{tikzpicture}
\begin{axis}[%
width=3in,
height=2in,
at={(0.762in,0.486in)},
scale only axis,
xmin=1,
xmax=11,
xlabel={$\text{Measurement ratio}~(M/N)$},
xtick={1,2,3,4,5,6,7,8,9,10,11},
xticklabels={0.02,0.1,0.2,0.3,0.4,0.5,0.6,0.7,0.8,0.9,1},
xlabel style={at = {(0.5,0.02)},font=\fontsize{12}{1}\selectfont},
ticklabel style={font=\fontsize{10}{1}\selectfont},
ymin=0,
ymax=30,
ytick={0,5,10,15,20,25,30,35},
yticklabels={0,5,10,15,20,25,30,35},
ylabel={SRE (dB)},
ylabel style={at = {(0.04,0.5)},font=\fontsize{12}{1}\selectfont},
axis background/.style={fill=white},
legend columns=4,
legend style={at={(0.5,1)},anchor=south,legend cell align=left,align=left,fill=none,draw=none,font=\fontsize{7}{1}\selectfont,/tikz/every even column/.style={column sep=6mm}},
legend entries={{$\sigma = 128$}, {$\sigma = 64$},{$\sigma = 32$},{$\sigma = 16$}}
]
\addlegendimage{only marks,mark=+,black},
\addlegendimage{only marks,mark=o,red},
\addlegendimage{only marks,mark=asterisk,mycolor1},
\addlegendimage{only marks,mark=*,blue,mark size =1},
\addplot [color=black,solid,line width = 0.5,mark=+,mark options={solid},forget plot]
  table[row sep=crcr]{%
1	4.33934239986875\\
2	14.4481142297582\\
3	18.4883476652375\\
4	19.9264417645857\\
5	20.6953375233484\\
6	21.1588706659058\\
7	21.4795624199772\\
8	21.7933944862933\\
9	22.1416862887092\\
10	22.4153780774386\\
11	22.6615039523468\\
};
\addplot [color=red,solid,line width = 0.5,mark=o,mark options={solid},forget plot]
  table[row sep=crcr]{%
1	3.19320652999207\\
2	11.9957088674362\\
3	15.7952686490079\\
4	18.0937686079295\\
5	19.2063730197599\\
6	19.6066947778029\\
7	20.4758964906739\\
8	21.3230690890752\\
9	21.4317881527443\\
10	21.7418401510721\\
11	21.934777576953\\
};
\addplot [color=mycolor1,solid,line width = 0.5,mark=asterisk,mark options={solid},forget plot]
  table[row sep=crcr]{%
1	2.43949404387245\\
2	7.98889866957292\\
3	12.4386231737417\\
4	15.9607174846738\\
5	18.10808567007\\
6	18.8511895893658\\
7	19.4346584434146\\
8	20.2589703367094\\
9	20.9694919919021\\
10	21.2030716140555\\
11	21.4381513412321\\
};
\addplot [color=blue,solid,line width = 0.5,mark=*,mark size =1,mark options={solid},forget plot]
  table[row sep=crcr]{%
1	1.57588832807248\\
2	6.18795701100689\\
3	10.3770911739982\\
4	13.5291153120489\\
5	15.5415935519083\\
6	17.4493794557886\\
7	18.8043680753433\\
8	19.501928804322\\
9	19.9779858758375\\
10	20.5021388399222\\
11	20.8343732643308\\
};
\addplot [color=black,dotted,line width = 0.5,mark=+,mark options={solid},forget plot]
  table[row sep=crcr]{%
1	0.0669578828556433\\
2	0.490799861091625\\
3	0.601476176339921\\
4	1.58581207955679\\
5	1.9644231284382\\
6	2.33658053256825\\
7	2.75647484010907\\
8	2.86152799104001\\
9	2.93279656328319\\
10	3.37780401130463\\
11	4.01062856276151\\
};
\addplot [color=red,dotted,line width = 0.5,mark=o,mark options={solid},forget plot]
  table[row sep=crcr]{%
1	0.128355506691435\\
2	0.332705076824772\\
3	1.1149323981129\\
4	1.319476053505\\
5	1.70490989931115\\
6	2.05990388802489\\
7	2.14123388663115\\
8	2.79677192767357\\
9	3.38300614815872\\
10	3.51452963971789\\
11	3.65117790212629\\
};
\addplot [color=mycolor1,dotted,line width = 0.5,mark=asterisk,mark options={solid},forget plot]
  table[row sep=crcr]{%
1	0.0280479823386323\\
2	0.440638695215407\\
3	0.922956087896894\\
4	1.30522291959064\\
5	1.69836098109432\\
6	2.22656679236441\\
7	2.65728189114386\\
8	2.99583239586619\\
9	3.40621868665237\\
10	4.25316578801273\\
11	4.69425002776027\\
};
\addplot [color=blue,dotted,line width = 0.5,mark=*,mark size =1,mark options={solid},forget plot]
  table[row sep=crcr]{%
1	0.075374798062238\\
2	0.50850338129398\\
3	0.851111253702095\\
4	1.31773488600812\\
5	2.05070523275027\\
6	2.49908667803546\\
7	3.04316578276759\\
8	3.63564498702864\\
9	4.20929184983858\\
10	4.78391018857234\\
11	5.71551956580075\\
};
\addplot [color=black,dashed,line width = 0.5,mark=+,mark options={solid},forget plot]
  table[row sep=crcr]{%
1 22.003465842775\\
2 25.706\\
3 25.7276615804493\\
4 25.802\\
5 25.8210483420707\\
6 25.928\\
7 25.806\\
8 25.7825012791686\\
9 25.6999873189033\\
10 25.9448543580803\\
11 25.966\\
};
\addplot [color=red,dashed,line width = 0.5,mark=o,mark options={solid},forget plot]
  table[row sep=crcr]{%
1 16.5998619015942\\
2 25.008\\
3 25.0288228868231\\
4 24.935004923356\\
5 25.0645251675844\\
6 24.9308621532797\\
7 24.8919280382659\\
8 25.0082442734962\\
9 25.0113654662628\\
10 25.3299806960014\\
11 25.1361493929278\\
};
\addplot [color=mycolor1,dashed,line width = 0.5,mark=asterisk,mark options={solid},forget plot]
  table[row sep=crcr]{%
1	11.3342432955196\\
2	21.0912746543126\\
3	24.4238237132856\\
4	24.473225657823\\
5	24.5551804559726\\
6	24.4322073765635\\
7	24.6933251420248\\
8	24.5811234095876\\
9	24.9417615004777\\
10	24.8332963607811\\
11	24.902718391911\\
};
\addplot [color=blue,dashed,line width = 0.5,mark=*,mark size =1,mark options={solid},forget plot]
  table[row sep=crcr]{%
1	6.49645659313761\\
2	14.3224585577759\\
3	21.4439381603429\\
4	23.8359282503556\\
5	24.1847080201546\\
6	24.1244589498989\\
7	24.4714633136597\\
8	24.5467756973245\\
9	24.5387285862304\\
10	24.7739664005429\\
11	24.6691270872694\\
};
\draw[black,line width=.5pt,anchor=center] (90,40) ellipse (4pt and 8pt) ;
\node[black,anchor=south]  at (90,52) {\fontsize{8}{1}\selectfont UDS};

\draw[black,line width=.5pt,anchor=north] (90,214) ellipse (4pt and 8pt) ;
\node[black,anchor=south]  at (90,173) {\fontsize{8}{1}\selectfont VDS};

\draw[black,line width=.5pt,anchor=south] (90,253) ellipse (4pt and 8pt) ;
\node[black,anchor=south]  at (90,266) {\fontsize{8}{1}\selectfont MDS};
\end{axis}
\end{tikzpicture}%
}
	\caption{The reconstruction performance comparison of the proposed MDS and VDS schemes with the traditional UDS scheme.}
	\label{fig:res_1d_gaussian}
\end{figure}
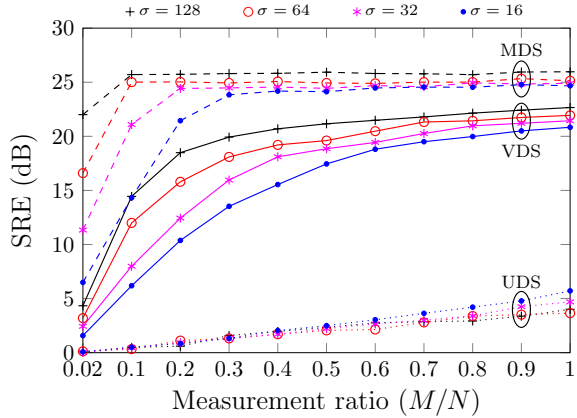
We here examine the VDS and MDS schemes defined in Thm.~\ref{thm:uniform guarantee for Hadamard-Haar system} and Thm.~\ref{thm:Non-uniform guarantee for Hadamard-Haar system} by comparing their SRE values with the one achieved by UDS scheme for different signals. In this part, the sensing and sparsity bases are set to the 1-D Hadamard and DHW bases, respectively, and the signals are recovered via only CS reconstruction. In the first simulation, a Gaussian-shape signal $\bs x \in \bb R^N$ of size $N=512$, \ie
$$
\ts x_i = \frac{1}{\sigma\sqrt{2\pi}}\exp \Big(-\frac{(i-i_0)^2}{2\sigma^2}\Big),~\forall i \in \range{N},
$$
is generated as the ground truth. The variables $i_0$ and $\sigma$ determine the center and the width of the Gaussian curve. Essentially, by increasing $\sigma$ the coefficients of the signal in Haar wavelet domain become sparser. The variable $\sigma \in \{16,32,64,128\}$ and the parameter $i_0$ is generated uniformly at random in the range $[\sigma,N-\sigma]$. We set the variance of the noise to read an SNR of 20 dB. Fig.~\ref{fig:res_1d_gaussian} displays the reconstruction quality of the generated signals as a function of the measurement ratio ($M/N$) for different values of $\sigma$ and sampling strategies (UDS, VDS, and MDS). Each point of the curves in Fig.~\ref{fig:res_1d_gaussian} is an average of 100 trials (\ie over random generation of the noise, subsampling set $\Omega$, and parameter~$i_0$). 

In the simulations here with MDS scheme, the effective local sparsities $k_l(\rho)$ are fixed for each value of $\sigma$ a priori. In particular, given $\sigma$ we first generate 100 Gaussian-shape signals (different from the ones to be recovered) whose locations $i_0$ are selected uniformly at random; and then compute their effective local sparsities as prescribed above. Finally, we consider the worst local sparsity values $k_l$ with $l \in \range{r}$ over all 100 trials for designing our MDS scheme. This approach gives a near-optimal MDS strategy, yet it is of practical interest where the true values of the local sparsity are not accessible.

From Fig.~\ref{fig:res_1d_gaussian}, we can make the following observations: \textit{(i)} by increasing the value of $\sigma$ the signal becomes sparser in the Haar domain, and thus, all reconstructions yield better SRE values; \textit{(ii)} the UDS scheme yields a poor reconstruction quality; this is aligned with the large value of the global coherence between the Hadamard and Haar bases, which drives the UDS sample-complexity in \eqref{eq:sample complexity uds}; \textit{(iii)} the VDS scheme provides a stable and robust signal recovery (with respect to the change of sparsity and noise level); \textit{(iv)} the SRE of the Hadamard-Haar system is further increased by using the MDS scheme, since it adjusts the sampling strategy to the sparsity structure of the signal;  \textit{(v)} although the MDS scheme here is not designed based on the ground truth signal, the dashed lines show significant SRE improvement compared to the VDS strategy.

In Fig.~\ref{fig:res_special_1d_signals}, we apply similar tests on four other functions, \ie  the ``Blocks'', ``Bumps'', ``HeaviSine'', and ``Doppler'' signals taken from~\cite{donoho1994ideal}. These signals display various behaviors, hence allowing us to test our scheme in a broader context. They are generated by evenly sampling the continuous functions specified in \cite{donoho1994ideal} over $N= 2048$ samples.

The reconstructed signals from 20\% subsampled Hadamard measurements using MDS, VDS, and UDS schemes are displayed in Fig.~\ref{fig:res_special_1d_signals}. As can be seen, the UDS strategy does not allow signal recovery. Note that these signals (except the Blocks signal) are not well-compressible in the Haar basis. As a consequence, most reconstructions have blocky artifacts and the VDS scheme does not provide a high quality reconstruction.  The MDS scheme, which leverages the local compressibility of the signal, achieves a much higher reconstruction quality in all examples.
\begin{figure}[t]
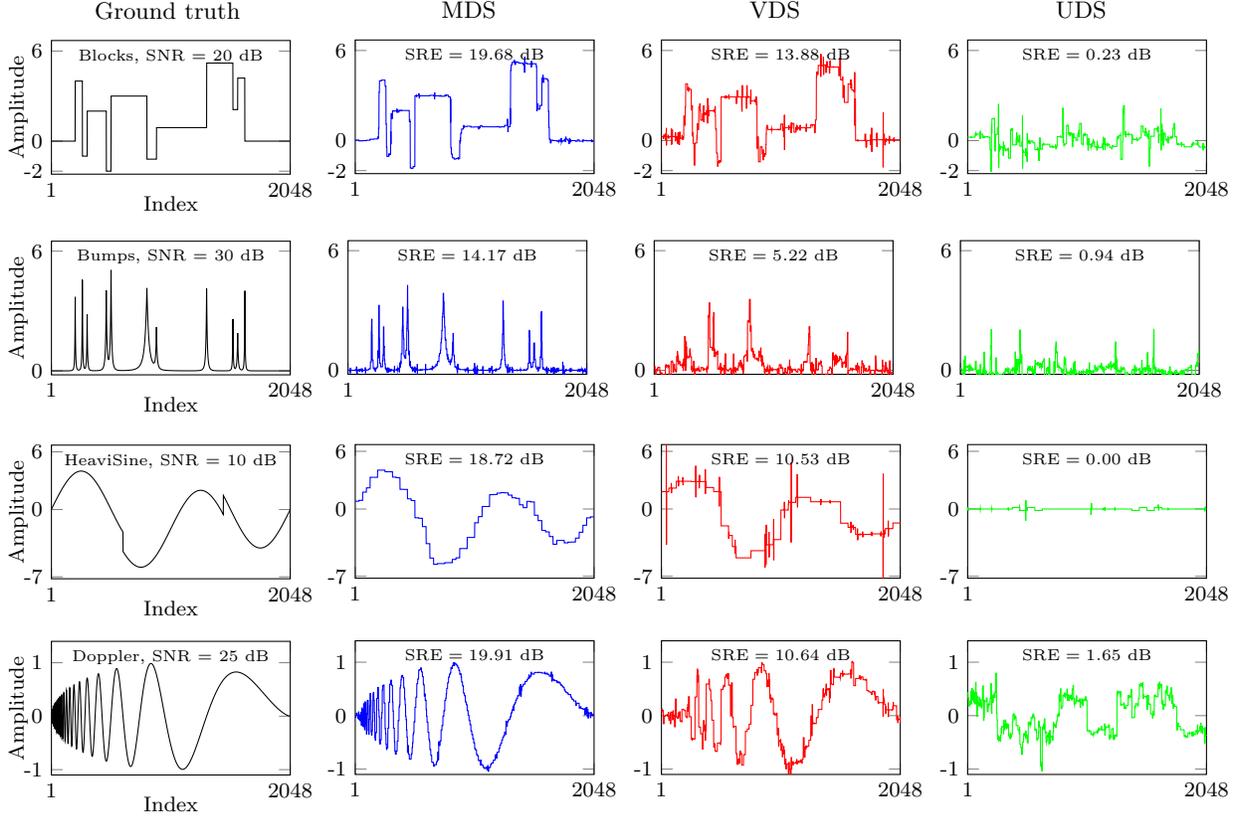

\centering \footnotesize
	\begin{minipage}{\columnwidth}
		\begin{minipage}{0.25\columnwidth}
		\centering
		\hspace{3mm}Ground truth
		\end{minipage}
		\begin{minipage}{0.24\columnwidth}
		\centering
		MDS
		\end{minipage}
		\begin{minipage}{0.24\columnwidth}
		\centering
		VDS
		\end{minipage}
		\begin{minipage}{0.24\columnwidth}
		\centering
		UDS
		\end{minipage}
	\end{minipage}
	\begin{minipage}{\columnwidth}
		\begin{minipage}{0.25\columnwidth}
		\centering\vspace{2mm}
		\scalebox{1}{\input{Figures/Res_Blocks_GroundTruth.tex}}
		\end{minipage}
		\begin{minipage}{0.24\columnwidth}
		\centering
		\scalebox{1}{\input{Figures/Res_Blocks_MDS.tex}}
		\end{minipage}
		\begin{minipage}{0.24\columnwidth}
		\centering
		\scalebox{1}{\input{Figures/Res_Blocks_VDS.tex}}
		\end{minipage}
		\begin{minipage}{0.24\columnwidth}
		\centering
		\scalebox{1}{\input{Figures/Res_Blocks_UDS.tex}}
		\end{minipage}
	\end{minipage}
	\begin{minipage}{\columnwidth}
		\begin{minipage}{0.25\columnwidth}
		\centering\vspace{2mm}
		\scalebox{1}{\input{Figures/Res_Bumps_GroundTruth.tex}}
		\end{minipage}
		\begin{minipage}{0.24\columnwidth}
		\centering
		\scalebox{1}{\input{Figures/Res_Bumps_MDS.tex}}
		\end{minipage}
		\begin{minipage}{0.24\columnwidth}
		\centering
		\scalebox{1}{\input{Figures/Res_Bumps_VDS.tex}}
		\end{minipage}
		\begin{minipage}{0.24\columnwidth}
		\centering
		\scalebox{1}{\input{Figures/Res_Bumps_UDS.tex}}
		\end{minipage}
	\end{minipage}
	\begin{minipage}{\columnwidth}
		\begin{minipage}{0.25\columnwidth}
		\centering\vspace{2mm}
		\scalebox{1}{\input{Figures/Res_Heavisin_GroundTruth.tex}}
		\end{minipage}
		\begin{minipage}{0.24\columnwidth}
		\centering
		\scalebox{1}{\input{Figures/Res_Heavisin_MDS.tex}}
		\end{minipage}
		\begin{minipage}{0.24\columnwidth}
		\centering
		\scalebox{1}{\input{Figures/Res_Heavisin_VDS.tex}}
		\end{minipage}
		\begin{minipage}{0.24\columnwidth}
		\centering
		\scalebox{1}{\input{Figures/Res_Heavisin_UDS.tex}}
		\end{minipage}
	\end{minipage}
	\begin{minipage}{\columnwidth}
		\begin{minipage}{0.25\columnwidth}
		\centering\vspace{2mm}
		\scalebox{1}{\input{Figures/Res_Doppler_GroundTruth.tex}}
		\end{minipage}
		\begin{minipage}{0.24\columnwidth}
		\centering
		\scalebox{1}{\input{Figures/Res_Doppler_MDS.tex}}
		\end{minipage}
		\begin{minipage}{0.24\columnwidth}
		\centering
		\scalebox{1}{\input{Figures/Res_Doppler_VDS.tex}}
		\end{minipage}
		\begin{minipage}{0.24\columnwidth}
		\centering
		\scalebox{1}{\input{Figures/Res_Doppler_UDS.tex}}
		\end{minipage}
	\end{minipage}
	\caption{Recovering four special 1-D signals from 20\% Hadamard measurements.}
	\label{fig:res_special_1d_signals}
\end{figure}

\subsection{2-D signal recovery}
\begin{figure}[!t]
	\centering
	\begin{minipage}{\columnwidth}
		\begin{minipage}{0.49\columnwidth}
			\scalebox{0.82}{
%
%
\definecolor{mycolor1}{rgb}{1.00000,0.00000,1.00000}%
\definecolor{mycolor2}{rgb}{0.00000,1.00000,1.00000}%
\begin{tikzpicture}
\begin{axis}[%
width=3in,
height=2in,
at={(0.762in,0.486in)},
scale only axis,
xmin=0,
xmax=1,
xlabel={$\text{Measurement ratio}~(M/N^2)$},
xtick={0.1,0.2,0.3,0.4,0.5,0.6,0.7,0.8,0.9,1},
xticklabels={0.1,0.2,0.3,0.4,0.5,0.6,0.7,0.8,0.9,1},
xlabel style={at = {(0.5,0.02)},font=\fontsize{12}{1}\selectfont},
ticklabel style={font=\fontsize{10}{1}\selectfont},
ymin=0,
ymax=35,
ytick={0,5,10,15,20,25,30,35},
yticklabels={0,5,10,15,20,25,30,35},
ylabel={SRE (dB)},
ylabel style={at = {(0.04,0.5)},font=\fontsize{12}{1}\selectfont},
axis background/.style={fill=white},
legend columns=7,
legend style={at={(0,1.05)},anchor=south west,legend cell align=left,align=left,fill=none,draw=none,font=\fontsize{7}{1}\selectfont,/tikz/every even column/.style={column sep=6mm}},
legend entries={{CS reconstruction:},{$N = 2048$}, {$N = 1024$},{$N = 512$},{$N = 256$},{$N = 128$},{$N = 64$},
{ME reconstruction:},{$N = 2048$}, {$N = 1024$},{$N = 512$},{$N = 256$},{$N = 128$},{$N = 64$}}
]
\addlegendimage{empty legend},
\addlegendimage{solid,mark=square,mycolor2},
\addlegendimage{solid,mark=x,black},
\addlegendimage{solid,mark=*,red,mark size = 1},
\addlegendimage{solid,mark=asterisk,mycolor1},
\addlegendimage{solid,mark=o,blue},
\addlegendimage{solid,mark=+,green},
\addlegendimage{empty legend},
\addlegendimage{dotted,mark=square,mycolor2,mark options={solid}},
\addlegendimage{dotted,mark=x,black,mark options={solid}},
\addlegendimage{dotted,mark=*,red,mark size = 1,mark options={solid}},
\addlegendimage{dotted,mark=asterisk,mycolor1,mark options={solid}},
\addlegendimage{dotted,mark=o,blue,mark options={solid}},
\addlegendimage{dotted,mark=+,green,mark options={solid}},
\addplot [color=green,solid,line width = 0.5,mark=+,mark options={solid},forget plot]
  table[row sep=crcr]{%
0.1	0.396664849778234\\
0.2	1.20881062996377\\
0.3	1.94670259000947\\
0.4	2.75434960634166\\
0.5	3.60709471762024\\
0.6	4.1428471377833\\
0.7	4.72224708705245\\
0.8	5.09402252377545\\
0.9	5.96130165032594\\
1	7.15833603206855\\
};
\addplot [color=green,dotted,line width = 0.5,mark=+,mark options={solid},forget plot]
  table[row sep=crcr]{%
0.1	0.493701553877155\\
0.2	0.882341064042084\\
0.3	1.04427923425583\\
0.4	1.36920031242136\\
0.5	1.87095362949117\\
0.6	2.22315764750807\\
0.7	2.59842909898713\\
0.8	2.85782525726506\\
0.9	3.47454971977059\\
1	4.25059806631552\\
};
\addplot [color=blue,solid,line width = 0.5,mark=o,mark options={solid},forget plot]
  table[row sep=crcr]{%
0.1	0.347994448762029\\
0.2	1.37815602008077\\
0.3	2.60333293079452\\
0.4	3.52170044058023\\
0.5	4.11787385174485\\
0.6	4.45438057026238\\
0.7	5.65774009982079\\
0.8	5.91564223883437\\
0.9	6.53109711550186\\
1	7.5812781546161\\
};
\addplot [color=blue,dotted,line width = 0.5,mark=o,mark options={solid},forget plot]
  table[row sep=crcr]{%
0.1	0.323000765174549\\
0.2	0.709345249461\\
0.3	1.29667974213248\\
0.4	1.77413299531251\\
0.5	2.19055550399572\\
0.6	2.3661918770742\\
0.7	3.17694562242581\\
0.8	3.39728456539693\\
0.9	3.85944196286039\\
1	4.45254300347946\\
};
\addplot [color=mycolor1,solid,line width = 0.5,mark=asterisk,mark options={solid},forget plot]
  table[row sep=crcr]{%
0.1	0.669286287622988\\
0.2	1.80049292707039\\
0.3	2.39423918026424\\
0.4	3.38960596454056\\
0.5	4.56301959979897\\
0.6	5.40488251319194\\
0.7	5.80930455134181\\
0.8	5.34958970995908\\
0.9	6.71447805282724\\
1	7.51963302330594\\
};
\addplot [color=mycolor1,dotted,line width = 0.5,mark=asterisk,mark options={solid},forget plot]
  table[row sep=crcr]{%
0.1	0.471795293200461\\
0.2	1.00060452696434\\
0.3	1.1550038519916\\
0.4	1.72295816703165\\
0.5	2.31894650885781\\
0.6	2.86516374850371\\
0.7	3.14636666799337\\
0.8	3.04067429400982\\
0.9	3.91742141800286\\
1	4.44440993114986\\
};
\addplot [color=red,dotted,line width = 0.5,mark size=1,mark=*,mark options={solid},forget plot]
  table[row sep=crcr]{%
0.1	0.356334378109805\\
0.2	0.879163677172935\\
0.3	1.33105019878415\\
0.4	1.88306078009029\\
0.5	2.11097788669906\\
0.6	2.44501825430073\\
0.7	3.13865996920694\\
0.8	3.44254764583086\\
0.9	4.29213754026499\\
1	4.11642284515205\\
};
\addplot [color=red,solid,line width = 0.5,mark=*,mark options={solid},forget plot]
  table[row sep=crcr]{%
0.1	0.576313909112883\\
0.2	1.70651460192472\\
0.3	2.70069143568864\\
0.4	3.76118780386234\\
0.5	4.12915234690145\\
0.6	4.61739701611359\\
0.7	5.72157449782934\\
0.8	6.16239205199992\\
0.9	7.5849420117156\\
1	6.75824475294086\\
};
\addplot [color=black,solid,line width = 0.5,mark=x,mark options={solid},forget plot]
  table[row sep=crcr]{%
0.1	0.599700054157332\\
0.2	1.99107231249492\\
0.3	2.44864634504677\\
0.4	3.19558497133619\\
0.5	3.97413576340644\\
0.6	4.74902285206429\\
0.7	6.09410791104067\\
0.8	8.02266531826421\\
0.9	5.97577927942468\\
1	7.19260532096939\\
};
\addplot [color=black,dotted,line width = 0.5,mark=x,mark options={solid},forget plot]
  table[row sep=crcr]{%
0.1	0.344535502588356\\
0.2	1.07879058698652\\
0.3	1.11189795202955\\
0.4	1.5897416185003\\
0.5	1.98081839986068\\
0.6	2.46326712733295\\
0.7	3.28851361877339\\
0.8	4.46966176268251\\
0.9	3.5361087194815\\
1	4.23375833849373\\
};
\addplot [color=mycolor2,dotted,line width = 0.5,mark=square,mark options={solid},forget plot]
  table[row sep=crcr]{%
0.1	0.333601746190277\\
0.2	0.929501144218086\\
0.3	1.1652900481095\\
0.4	1.53899600714095\\
0.5	1.96249214119854\\
0.6	2.06477150079868\\
0.7	3.07745437134575\\
0.8	3.72724891514495\\
0.9	4.16086963876897\\
1	4.39160704259381\\
};
\addplot [color=mycolor2,solid,line width = 0.5,mark=square,mark options={solid},forget plot]
  table[row sep=crcr]{%
0.1	0.5596544020074\\
0.2	1.82204733658349\\
0.3	2.5043414051917\\
0.4	3.23511876339843\\
0.5	3.94882048705289\\
0.6	3.99289014923088\\
0.7	5.73404305460123\\
0.8	6.61099256427484\\
0.9	7.44336589915934\\
1	7.32770657262912\\
};
\node[align=left, text=black, draw=none,anchor=north west]  at (0,340) {\fontsize{8}{1}\selectfont $\{{\rm UDS},~\bs \Psi_{\rm idhw}\}$};
\end{axis}
\end{tikzpicture}%
}
		\end{minipage}
		\begin{minipage}{0.49\columnwidth}\vspace{7mm}
			\scalebox{0.82}{
%
%
\definecolor{mycolor1}{rgb}{1.00000,0.00000,1.00000}%
\definecolor{mycolor2}{rgb}{0.00000,1.00000,1.00000}%
\begin{tikzpicture}
\begin{axis}[%
width=3in,
height=2in,
at={(0.762in,0.486in)},
scale only axis,
xmin=0,
xmax=1,
xlabel={$\text{Measurement ratio}~(M/N^2)$},
xtick={0.1,0.2,0.3,0.4,0.5,0.6,0.7,0.8,0.9,1},
xticklabels={0.1,0.2,0.3,0.4,0.5,0.6,0.7,0.8,0.9,1},
xlabel style={at = {(0.5,0.02)},font=\fontsize{12}{1}\selectfont},
ticklabel style={font=\fontsize{10}{1}\selectfont},
ymin=0,
ymax=35,
ytick={0,5,10,15,20,25,30,35},
yticklabels={0,5,10,15,20,25,30,35},
ylabel={SRE (dB)},
ylabel style={at = {(0.04,0.5)},font=\fontsize{12}{1}\selectfont},
axis background/.style={fill=white},
]
\addplot [color=mycolor1,solid,line width = 0.5,mark=square,mark options={solid},forget plot]
  table[row sep=crcr]{%
0.1	0.465077878002061\\
0.2	1.50747120133762\\
0.3	2.16798048934664\\
0.4	2.85645650561051\\
0.5	3.55017568111822\\
0.6	3.71429822862719\\
0.7	5.21531088745798\\
0.8	6.07256274199431\\
0.9	6.57519973194705\\
1	6.78954776886252\\
};
\addplot [color=red,solid,line width = 0.5,mark size=1pt,mark=*,mark options={solid},forget plot]
  table[row sep=crcr]{%
0.1	0.494135541380963\\
0.2	1.43368714586014\\
0.3	2.32524921818819\\
0.4	3.29412849343814\\
0.5	3.64362531386338\\
0.6	4.20939071482533\\
0.7	5.12135911292307\\
0.8	5.48873230641599\\
0.9	6.8726116483001\\
1	6.35317875934781\\
};
\addplot [color=mycolor2,solid,line width = 0.5,mark=asterisk,mark options={solid},forget plot]
  table[row sep=crcr]{%
0.1	0.552270057912505\\
0.2	1.53216065491338\\
0.3	2.05810949820719\\
0.4	3.00946471609076\\
0.5	3.99597768865969\\
0.6	4.81437803762274\\
0.7	5.20398073460859\\
0.8	4.94652980905824\\
0.9	6.28296671836071\\
1	7.06002198709548\\
};
\addplot [color=blue,solid,line width = 0.5,mark=o,mark options={solid},forget plot]
  table[row sep=crcr]{%
0.1	0.327737558343314\\
0.2	1.11769884338085\\
0.3	2.15306250624554\\
0.4	2.96428392895498\\
0.5	3.60505547558506\\
0.6	3.91970334757241\\
0.7	5.02686644941666\\
0.8	5.30554332669658\\
0.9	5.97748518776931\\
1	6.74383793331636\\
};
\addplot [color=green,solid,line width = 0.5,mark=+,mark options={solid},forget plot]
  table[row sep=crcr]{%
0.1	0.522198737132907\\
0.2	1.19542351659956\\
0.3	1.70391852636437\\
0.4	2.33748500525901\\
0.5	3.08086795096766\\
0.6	3.62931522799528\\
0.7	4.12038495321291\\
0.8	4.45947775490518\\
0.9	5.2607711297438\\
1	6.35291545265889\\
};
\addplot [color=green,dotted,line width = 0.5,mark=+,mark options={solid},forget plot]
  table[row sep=crcr]{%
0.1	0.493701553877155\\
0.2	0.882341064042084\\
0.3	1.04427923425583\\
0.4	1.36920031242136\\
0.5	1.87095362949117\\
0.6	2.22315764750807\\
0.7	2.59842909898713\\
0.8	2.85782525726506\\
0.9	3.47454971977059\\
1	4.25059806631552\\
};
\addplot [color=blue,dotted,line width = 0.5,mark=o,mark options={solid},forget plot]
  table[row sep=crcr]{%
0.1	0.323000765174549\\
0.2	0.709345249461\\
0.3	1.29667974213248\\
0.4	1.77413299531251\\
0.5	2.19055550399572\\
0.6	2.3661918770742\\
0.7	3.17694562242581\\
0.8	3.39728456539693\\
0.9	3.85944196286039\\
1	4.45254300347946\\
};
\addplot [color=mycolor2,dotted,line width = 0.5,mark=asterisk,mark options={solid},forget plot]
  table[row sep=crcr]{%
0.1	0.471795293200461\\
0.2	1.00060452696434\\
0.3	1.1550038519916\\
0.4	1.72295816703165\\
0.5	2.31894650885781\\
0.6	2.86516374850371\\
0.7	3.14636666799337\\
0.8	3.04067429400982\\
0.9	3.91742141800286\\
1	4.44440993114986\\
};
\addplot [color=mycolor2,dotted,line width = 0.5,mark=*,mark options={solid},forget plot]
  table[row sep=crcr]{%
0.1	0.356334378109805\\
0.2	0.879163677172935\\
0.3	1.33105019878415\\
0.4	1.88306078009029\\
0.5	2.11097788669906\\
0.6	2.44501825430073\\
0.7	3.13865996920694\\
0.8	3.44254764583086\\
0.9	4.29213754026499\\
1	4.11642284515205\\
};
\addplot [color=black,dotted,line width = 0.5,mark=x,mark options={solid},forget plot]
  table[row sep=crcr]{%
0.1	0.344535502588356\\
0.2	1.07879058698652\\
0.3	1.11189795202955\\
0.4	1.5897416185003\\
0.5	1.98081839986068\\
0.6	2.46326712733295\\
0.7	3.28851361877339\\
0.8	4.46966176268251\\
0.9	3.5361087194815\\
1	4.23375833849373\\
};
\addplot [color=mycolor1,dotted,line width = 0.5,mark=square,mark options={solid},forget plot]
  table[row sep=crcr]{%
0.1	0.333601746190277\\
0.2	0.929501144218086\\
0.3	1.1652900481095\\
0.4	1.53899600714095\\
0.5	1.96249214119854\\
0.6	2.06477150079868\\
0.7	3.07745437134575\\
0.8	3.72724891514495\\
0.9	4.16086963876897\\
1	4.39160704259381\\
};
\addplot [color=black,solid,line width = 0.5,mark=x,mark options={solid},forget plot]
  table[row sep=crcr]{%
0.1	0.484549884299283\\
0.2	1.70612951287051\\
0.3	2.05279769184676\\
0.4	2.88459280655664\\
0.5	3.55257892536182\\
0.6	4.24695940506421\\
0.7	5.51055198672079\\
0.8	7.44040594928803\\
0.9	5.62596809324189\\
1	6.60107929224926\\
};
\node[align=left, text=black, draw=none,anchor=north west]  at (0,340) {\fontsize{8}{1}\selectfont $\{{\rm UDS},~\bs \Psi_{\rm adhw}\}$};
\end{axis}
\end{tikzpicture}%
}
		\end{minipage}
	\end{minipage}
	\begin{minipage}{\columnwidth}
		\begin{minipage}{0.49\columnwidth}
			\scalebox{0.82}{
%
%
\definecolor{mycolor1}{rgb}{1.00000,0.00000,1.00000}%
\definecolor{mycolor2}{rgb}{0.00000,1.00000,1.00000}%
\begin{tikzpicture}
\begin{axis}[%
width=3in,
height=2in,
at={(0.762in,0.486in)},
scale only axis,
xmin=0,
xmax=1,
xlabel={$\text{Measurement ratio}~(M/N^2)$},
xtick={0.1,0.2,0.3,0.4,0.5,0.6,0.7,0.8,0.9,1},
xticklabels={0.1,0.2,0.3,0.4,0.5,0.6,0.7,0.8,0.9,1},
xlabel style={at = {(0.5,0.02)},font=\fontsize{12}{1}\selectfont},
ticklabel style={font=\fontsize{10}{1}\selectfont},
ymin=0,
ymax=35,
ytick={0,5,10,15,20,25,30,35},
yticklabels={0,5,10,15,20,25,30,35},
ylabel={SRE (dB)},
ylabel style={at = {(0.04,0.5)},font=\fontsize{12}{1}\selectfont},
axis background/.style={fill=white}
]
\addplot [color=green,solid,line width=0.5pt,mark=+,mark options={solid}]
  table[row sep=crcr]{%
0.1	2.5378568339999\\
0.2	3.00883099080989\\
0.3	3.4842717910668\\
0.4	4.12613688469197\\
0.5	4.64383919716151\\
0.6	5.38130730068278\\
0.7	6.33573564631075\\
0.8	6.88039183916596\\
0.9	7.47793107548249\\
1	8.66986752162103\\
};

\addplot [color=blue,solid,line width=0.5pt,mark=o,mark options={solid}]
  table[row sep=crcr]{%
0.1	3.58248799135195\\
0.2	4.96051154039629\\
0.3	6.24150041956734\\
0.4	7.52318349248444\\
0.5	8.54686876857805\\
0.6	9.3472486885404\\
0.7	9.9921790720442\\
0.8	11.0025738202852\\
0.9	11.8045311103312\\
1	12.9458602672677\\
};
\addplot [color=mycolor1,solid,line width=0.5pt,mark=asterisk,mark options={solid}]
  table[row sep=crcr]{%
0.1	6.24281056199772\\
0.2	8.15243376521893\\
0.3	9.50228995749784\\
0.4	10.6709250719432\\
0.5	11.9146989864412\\
0.6	12.9511605152616\\
0.7	14.0359378862299\\
0.8	14.6565727712959\\
0.9	15.8981804990191\\
1	16.8137744768021\\
};

\addplot [color=red,solid,line width=0.5pt,mark=*,mark size =1,mark options={solid}]
  table[row sep=crcr]{%
0.1	9.4352135295664\\
0.2	11.7994598336727\\
0.3	13.437819816732\\
0.4	14.6985788965347\\
0.5	15.8040467208319\\
0.6	16.9106853557547\\
0.7	17.6772335845229\\
0.8	18.7582309028051\\
0.9	19.5086429003962\\
1	20.1357073625491\\
};

\addplot [color=green,dotted,line width=0.5pt,mark=+,mark options={solid},forget plot]
  table[row sep=crcr]{%
0.1	2.66672083641251\\
0.2	3.22965622846998\\
0.3	3.56969163882521\\
0.4	3.8759870533545\\
0.5	4.07081450409744\\
0.6	4.359971391059\\
0.7	4.61274644791465\\
0.8	4.78504778349622\\
0.9	4.99566117694315\\
1	5.18919142397762\\
};
\addplot [color=blue,dotted,line width=0.5pt,mark=o,mark options={solid},forget plot]
  table[row sep=crcr]{%
0.1	3.59034726444695\\
0.2	4.29897698906865\\
0.3	4.78088189130077\\
0.4	5.30979139875972\\
0.5	5.59681991435474\\
0.6	5.94730079434318\\
0.7	6.13674613955693\\
0.8	6.40276269516428\\
0.9	6.58620951497195\\
1	6.82253835208584\\
};
\addplot [color=red,dotted,line width=0.5pt,mark=*,mark options={solid},mark size =1,forget plot]
  table[row sep=crcr]{%
0.1	6.56967755278011\\
0.2	7.34772520286359\\
0.3	7.77413496317267\\
0.4	8.08975414674204\\
0.5	8.30500790955947\\
0.6	8.483534878206\\
0.7	8.66579008318969\\
0.8	8.76930861398536\\
0.9	8.85961662344853\\
1	8.94280080305273\\
};
\addplot [color=mycolor1,dotted,line width=0.5pt,mark=asterisk,mark options={solid},forget plot]
  table[row sep=crcr]{%
0.1	4.81564605682968\\
0.2	5.81275258159429\\
0.3	6.47547547646619\\
0.4	6.80967417116961\\
0.5	7.11088410918886\\
0.6	7.33132019923618\\
0.7	7.54701785821626\\
0.8	7.65778381630877\\
0.9	7.81527708161045\\
1	7.95755910684212\\
};
\addplot [color=black,dotted,line width=0.5pt,mark=x,mark options={solid},forget plot]
  table[row sep=crcr]{%
0.1	7.90746939998834\\
0.2	8.50988232723476\\
0.3	8.84943564372127\\
0.4	9.02106467751074\\
0.5	9.14632286019926\\
0.6	9.24216904471914\\
0.7	9.3147032975665\\
0.8	9.36973480700991\\
0.9	9.42996876221689\\
1	9.44360753568167\\
};
\addplot [color=black,solid,line width=0.5pt,mark=x,mark options={solid},forget plot]
  table[row sep=crcr]{%
0.1	13.248867326364\\
0.2	15.5389678799654\\
0.3	17.143140531005\\
0.4	18.3593303920336\\
0.5	19.3941755095295\\
0.6	20.2802687993034\\
0.7	21.0234076238757\\
0.8	21.6723620522793\\
0.9	22.1796014252056\\
1	22.6705363541687\\
};
\addplot [color=mycolor2,solid,line width=0.5pt,mark=square,mark options={solid}]
  table[row sep=crcr]{%
0.1	16.9045673104146\\
0.2	18.9193033370569\\
0.3	20.2210028860266\\
0.4	21.253152643245\\
0.5	22.095020512702\\
0.6	22.8151083189775\\
0.7	23.4841620707399\\
0.8	24.0318812609776\\
0.9	24.5186763965474\\
1	24.9613832073515\\
};

\addplot [color=mycolor2,dotted,line width=0.5pt,mark=square,mark options={solid}]
  table[row sep=crcr]{%
0.1	9.02076905594292\\
0.2	9.31135698501751\\
0.3	9.46641781945507\\
0.4	9.54674579657137\\
0.5	9.57738018788348\\
0.6	9.65906343278919\\
0.7	9.66969056399579\\
0.8	9.67803692676681\\
0.9	9.72425774218086\\
1	9.74678602021124\\
};
\node (e1) at (80,180) {};\node (e2) at (80,220) {};\draw[<->] (e1) -- (e2);
\node[align=left, text=black, draw=none,anchor=west]  at (79,205) {\fontsize{6}{1}\selectfont $\ \approx\text{3 dB}$};

\node[align=left, text=black, draw=none,anchor=north west]  at (0,340) {\fontsize{8}{1}\selectfont $\{{\rm VDS},~\bs \Psi_{\rm idhw}\}$};

\draw[red,line width=.5pt,anchor=center] (10,169) ellipse (4pt and 4pt);
\node[red,anchor=east]  at (10,169) {\fontsize{8}{1}\selectfont $P_2$};

\draw[red,line width=.5pt,anchor=center] (10,133) ellipse (4pt and 4pt);
\node[red,anchor=east]  at (10,133) {\fontsize{8}{1}\selectfont $P_4$};
\end{axis}
\end{tikzpicture}%
}
		\end{minipage}
		\begin{minipage}{0.49\columnwidth}
			\scalebox{0.82}{
%
%
\definecolor{mycolor1}{rgb}{1.00000,0.00000,1.00000}%
\definecolor{mycolor2}{rgb}{0.00000,1.00000,1.00000}%
\begin{tikzpicture}
\begin{axis}[%
width=3in,
height=2in,
at={(0.762in,0.486in)},
scale only axis,
xmin=0,
xmax=1,
xlabel={$\text{Measurement ratio}~(M/N^2)$},
xtick={0.1,0.2,0.3,0.4,0.5,0.6,0.7,0.8,0.9,1},
xticklabels={0.1,0.2,0.3,0.4,0.5,0.6,0.7,0.8,0.9,1},
xlabel style={at = {(0.5,0.02)},font=\fontsize{12}{1}\selectfont},
ticklabel style={font=\fontsize{10}{1}\selectfont},
ymin=0,
ymax=35,
ytick={0,5,10,15,20,25,30,35},
yticklabels={0,5,10,15,20,25,30,35},
ylabel={SRE (dB)},
ylabel style={at = {(0.04,0.5)},font=\fontsize{12}{1}\selectfont},
axis background/.style={fill=white},
legend style={at={(0,1)},anchor=north west,legend cell align=left,align=left,fill=none,draw=none,font=\fontsize{6}{1}\selectfont}
]
\addplot [color=black,solid,line width=0.5pt,mark=x,mark options={solid},forget plot]
  table[row sep=crcr]{%
0.1	11.8857814181832\\
0.2	13.8828311837916\\
0.3	15.0296965512288\\
0.4	15.834925528942\\
0.5	16.4406903811139\\
0.6	16.9580273789355\\
0.7	17.401027427959\\
0.8	17.7883459037769\\
0.9	18.1371992780692\\
1	18.4633049707176\\
};
\addplot [color=red,solid,line width=0.5pt,mark=*,mark options={solid},mark size = 1,forget plot]
  table[row sep=crcr]{%
0.1	8.8198618666348\\
0.2	10.5819352948466\\
0.3	11.8011700477231\\
0.4	12.6359681499011\\
0.5	13.380277412495\\
0.6	13.9860443735599\\
0.7	14.4781578915801\\
0.8	14.923102873106\\
0.9	15.3234871091523\\
1	15.7297869642424\\
};
\addplot [color=green,solid,line width=0.5pt,mark=+,mark options={solid},forget plot]
  table[row sep=crcr]{%
0.1	2.70598445780684\\
0.2	3.32547992952275\\
0.3	3.76562997384317\\
0.4	4.1197105020547\\
0.5	4.53739648188852\\
0.6	5.15049661408412\\
0.7	5.34115029245529\\
0.8	5.82434988721889\\
0.9	6.40572053757494\\
1	6.96902676682294\\
};
\addplot [color=blue,solid,line width=0.5pt,mark=o,mark options={solid},forget plot]
  table[row sep=crcr]{%
0.1	3.92612459719082\\
0.2	5.02513086001984\\
0.3	5.9167133078253\\
0.4	6.82753902703664\\
0.5	7.36260701212513\\
0.6	8.12061216377141\\
0.7	8.64532965450462\\
0.8	9.19300558083019\\
0.9	9.6005767191165\\
1	9.96310402532156\\
};
\addplot [color=mycolor1,solid,line width=0.5pt,mark=asterisk,mark options={solid},forget plot]
  table[row sep=crcr]{%
0.1	5.73667176610102\\
0.2	7.52763569598833\\
0.3	8.7356442479528\\
0.4	9.49511198212862\\
0.5	10.1209212686296\\
0.6	10.7371323486953\\
0.7	11.2740995717307\\
0.8	11.750842882866\\
0.9	12.2745773084813\\
1	12.6440309614363\\
};
\addplot [color=mycolor2,solid,line width=0.5pt,mark=square,mark options={solid},forget plot]
  table[row sep=crcr]{%
0.1	15.2639607703154\\
0.2	16.8976488426131\\
0.3	17.8071827301172\\
0.4	18.4562591956078\\
0.5	18.9513890867452\\
0.6	19.3790064721035\\
0.7	19.7526429270418\\
0.8	20.0770890264191\\
0.9	20.3683778028977\\
1	20.6380821756603\\
};
\addplot [color=mycolor2,dotted,line width=0.5pt,mark=square,mark options={solid},forget plot]
  table[row sep=crcr]{%
0.1	8.94128957815845\\
0.2	9.24703369561115\\
0.3	9.44797125035134\\
0.4	9.5496755767807\\
0.5	9.60176090125684\\
0.6	9.77189317265765\\
0.7	9.7840267868783\\
0.8	9.89174846784951\\
0.9	9.98137773189814\\
1	10.0635751604431\\
};
\addplot [color=black,dotted,line width=0.5pt,mark=x,mark options={solid},forget plot]
  table[row sep=crcr]{%
0.1	8.18891430274246\\
0.2	8.72248647987414\\
0.3	8.91948828207549\\
0.4	9.06216475112563\\
0.5	9.16114233452301\\
0.6	9.24501938883154\\
0.7	9.256962197929\\
0.8	9.3825459515482\\
0.9	9.37062262152929\\
1	9.44378322483985\\
};
\addplot [color=red,dotted,line width=0.5pt,mark=*,mark options={solid},mark size = 1,,forget plot]
  table[row sep=crcr]{%
0.1	6.76939921000581\\
0.2	7.69476508100707\\
0.3	8.14118086956127\\
0.4	8.39873042324003\\
0.5	8.60497971297462\\
0.6	8.71099733867512\\
0.7	8.8283252611348\\
0.8	8.89056971193055\\
0.9	8.94522725565525\\
1	9.03674598170349\\
};
\addplot [color=mycolor1,dotted,line width=0.5pt,mark=asterisk,mark options={solid},forget plot]
  table[row sep=crcr]{%
0.1	4.92527605921337\\
0.2	5.95717250889417\\
0.3	6.661958620474\\
0.4	7.13264176431257\\
0.5	7.44061157052797\\
0.6	7.71285372641995\\
0.7	7.95523860811657\\
0.8	8.09906777059321\\
0.9	8.23760884558089\\
1	8.32841074760103\\
};
\addplot [color=blue,dotted,line width=0.5pt,mark=o,mark options={solid},forget plot]
  table[row sep=crcr]{%
0.1	3.72118321417104\\
0.2	4.5402281307691\\
0.3	5.00289814326093\\
0.4	5.60211063792909\\
0.5	5.89653338809384\\
0.6	6.18995912724606\\
0.7	6.50077142281613\\
0.8	6.75024901664556\\
0.9	6.9556536365386\\
1	7.16222180990763\\
};
\addplot [color=green,dotted,line width=0.5pt,mark=+,mark options={solid},forget plot]
  table[row sep=crcr]{%
0.1	2.7110773725446\\
0.2	3.34253373254998\\
0.3	3.73866401560142\\
0.4	4.00526709966368\\
0.5	4.2607398583499\\
0.6	4.62671683435642\\
0.7	4.75999358829418\\
0.8	5.00205691525894\\
0.9	5.14459466245978\\
1	5.41193830669549\\
};
\node (e1) at (80,142) {};\node (e2) at (80,183) {};\draw[<->] (e1) -- (e2);
\node[align=left, text=black, draw=none,anchor=west]  at (79,165) {\fontsize{6}{1}\selectfont $\ \approx\text{3 dB}$};

\node[align=left, text=black, draw=none,anchor=north west]  at (0,340) {\fontsize{8}{1}\selectfont $\{{\rm VDS},~\bs \Psi_{\rm adhw}\}$};
\end{axis}
\end{tikzpicture}%
}
		\end{minipage}
	\end{minipage}
	\begin{minipage}{\columnwidth}
		\begin{minipage}{0.49\columnwidth}
			\scalebox{0.82}{
%
%
\definecolor{mycolor1}{rgb}{1.00000,0.00000,1.00000}%
\definecolor{mycolor2}{rgb}{0.00000,1.00000,1.00000}%
\begin{tikzpicture}
\begin{axis}[%
width=3in,
height=2in,
at={(0.762in,0.486in)},
scale only axis,
xmin=0,
xmax=1,
xlabel={$\text{Measurement ratio}~(M/N^2)$},
xtick={0.1,0.2,0.3,0.4,0.5,0.6,0.7,0.8,0.9,1},
xticklabels={0.1,0.2,0.3,0.4,0.5,0.6,0.7,0.8,0.9,1},
xlabel style={at = {(0.5,0.02)},font=\fontsize{12}{1}\selectfont},
ticklabel style={font=\fontsize{10}{1}\selectfont},
ymin=0,
ymax=35,
ytick={0,5,10,15,20,25,30,35},
yticklabels={0,5,10,15,20,25,30,35},
ylabel={SRE (dB)},
ylabel style={at = {(0.04,0.5)},font=\fontsize{12}{1}\selectfont},
axis background/.style={fill=white},
legend style={at={(0,1)},anchor=north west,legend cell align=left,align=left,fill=none,draw=none,font=\fontsize{6}{1}\selectfont}
]
%

%

\addplot [color=mycolor1,solid,line width=0.5pt,mark=asterisk,mark options={solid},forget plot]
  table[row sep=crcr]{%
0.1	9.14076447385601\\
0.2	14.50652744684\\
0.3	18.8681628024022\\
0.4	21.3540040306875\\
0.5	22.7569912467568\\
0.6	24.218811575257\\
0.7	25.1046944121401\\
0.8	25.6006233040155\\
0.9	25.9322318653295\\
1	26.1122193862358\\
};
\addplot [color=blue,solid,line width=0.5pt,mark=o,mark options={solid},forget plot]
  table[row sep=crcr]{%
0.1	4.52516316537661\\
0.2	9.5693285843974\\
0.3	14.5211798727298\\
0.4	18.3824175736438\\
0.5	20.5289338932656\\
0.6	21.9494514633437\\
0.7	23.0336470270359\\
0.8	23.5883584876905\\
0.9	24.0396415606005\\
1	24.334570522426\\
};
\addplot [color=green,solid,line width=0.5pt,mark=+,mark options={solid},forget plot]
  table[row sep=crcr]{%
0.1	1.93574184988621\\
0.2	4.90332757912361\\
0.3	8.01172183263088\\
0.4	12.6790311211322\\
0.5	16.257660094629\\
0.6	18.7328817894311\\
0.7	20.5806572013786\\
0.8	21.4823477727917\\
0.9	22.0967306381659\\
1	22.5240036204229\\
};
\addplot [color=black,dotted,line width=0.5pt,mark=x,mark options={solid},forget plot]
  table[row sep=crcr]{%
0.1	12.7580866637528\\
0.2	14.8349403745798\\
0.3	15.9796048387629\\
0.4	17.3429134616477\\
0.5	17.6836653736922\\
0.6	18.0686373728016\\
0.7	18.4706016292616\\
0.8	18.9258419176452\\
0.9	19.4316229433621\\
1	19.9996789086735\\
};
\addplot [color=red,dotted,line width=0.5pt,mark=*,mark options={solid},mark size = 1,forget plot]
  table[row sep=crcr]{%
0.1	9.37696045925134\\
0.2	11.7765744722006\\
0.3	12.7132720364312\\
0.4	13.937322112097\\
0.5	15.3823486044104\\
0.6	16.0032747536109\\
0.7	16.6968126457655\\
0.8	17.5438909370341\\
0.9	18.5934337091364\\
1	20.0020029658777\\
};
\addplot [color=mycolor1,dotted,line width=0.5pt,mark=asterisk,mark options={solid},forget plot]
  table[row sep=crcr]{%
0.1	5.90568136207786\\
0.2	7.95313405706556\\
0.3	9.88084528625807\\
0.4	11.0259330009103\\
0.5	12.5776943172543\\
0.6	13.510276299083\\
0.7	14.4763612642159\\
0.8	15.64002059464\\
0.9	17.2882414156054\\
1	19.9832555001848\\
};
\addplot [color=blue,dotted,line width=0.5pt,mark=o,mark options={solid},forget plot]
  table[row sep=crcr]{%
0.1	3.86422389799107\\
0.2	5.34560214447579\\
0.3	6.87643406521743\\
0.4	8.01800067059439\\
0.5	9.39978699333222\\
0.6	11.423816278332\\
0.7	12.4900881393172\\
0.8	13.7744205407356\\
0.9	15.86589972106\\
1	19.9957123536569\\
};
\addplot [color=green,dotted,line width=0.5pt,mark=+,mark options={solid},forget plot]
  table[row sep=crcr]{%
0.1	2.09141772536397\\
0.2	4.07489006151781\\
0.3	4.99509066003715\\
0.4	5.82675583391466\\
0.5	6.9791537940371\\
0.6	8.15094500564494\\
0.7	9.33536184229551\\
0.8	11.0179525619894\\
0.9	13.6293847471639\\
1	20.0116866043505\\
};
\addplot [color=mycolor2,solid,line width=0.5pt,mark=square,mark options={solid},forget plot]
  table[row sep=crcr]{%
0.1	20.5905261467849\\
0.2	24.383\\
0.3	26.724\\
0.4	28.661\\
0.5	30.2829923626855\\
0.6	31.5451736614876\\
0.7	32.2422380269431\\
0.8	32.6166601081268\\
0.9	32.8148477736842\\
1	32.9037836219817\\
};
\addplot [color=black,solid,line width=0.5pt,mark=x,mark options={solid},forget plot]
  table[row sep=crcr]{%
0.1	18.0614795522577\\
0.2	21.7940814737732\\
0.3	24.6174631636657\\
0.4	26.2599963521233\\
0.5	27.8168547446791\\
0.6	29.0773456076135\\
0.7	29.7994967884568\\
0.8	30.199567484191\\
0.9	30.4127036997416\\
1	30.5253078488778\\
};
\addplot [color=red,solid,line width=0.5pt,mark=*,mark options={solid},mark size = 1,forget plot]
  table[row sep=crcr]{%
0.1	13.8839357628236\\
0.2	18.7147841244248\\
0.3	22.348780992499\\
0.4	24.1604940567336\\
0.5	25.3915096583457\\
0.6	26.650243506612\\
0.7	27.3899622624957\\
0.8	27.8448796516286\\
0.9	28.1000042204336\\
1	28.2214825003386\\
};
\addplot [color=mycolor2,dotted,line width=0.5pt,mark=square,mark options={solid},forget plot]
  table[row sep=crcr]{%
0.1	15.5391006437052\\
0.2	17.5504287723406\\
0.3	18.490739014527\\
0.4	19.5539607930758\\
0.5	19.6286108167178\\
0.6	19.6969273372423\\
0.7	19.7722645818758\\
0.8	19.8460516707525\\
0.9	19.9240670703679\\
1	19.999903818324\\
};
\node (e1) at (80,273) {};\node (e2) at (80,307) {};\draw[<->] (e1) -- (e2);
\node[align=left, text=black, draw=none,anchor=west]  at (79,290) {\fontsize{6}{1}\selectfont $\ \approx\text{2.5 dB}$};

\node[align=left, text=black, draw=none,anchor=north west]  at (0,340) {\fontsize{8}{1}\selectfont $\{{\rm MDS},~ \bs \Psi_{\rm idhw}, \cl T^{\rm iso}\}$};

\draw[red,line width=.5pt,anchor=center] (10,206) ellipse (4pt and 4pt);
\node[red,anchor=east]  at (10,206) {\fontsize{8}{1}\selectfont $P_1$};

\draw[red,line width=.5pt,anchor=center] (10,181) ellipse (4pt and 4pt);
\node[red,anchor=east]  at (10,181) {\fontsize{8}{1}\selectfont $P_3$};
\end{axis}
\end{tikzpicture}%
}
		\end{minipage}
		\begin{minipage}{0.49\columnwidth}
			\scalebox{0.82}{
%
%
\definecolor{mycolor1}{rgb}{1.00000,0.00000,1.00000}%
\definecolor{mycolor2}{rgb}{0.00000,1.00000,1.00000}%
\begin{tikzpicture}
\begin{axis}[%
width=3in,
height=2in,
at={(0.762in,0.486in)},
scale only axis,
xmin=0,
xmax=1,
xlabel={$\text{Measurement ratio}~(M/N^2)$},
xtick={0.1,0.2,0.3,0.4,0.5,0.6,0.7,0.8,0.9,1},
xticklabels={0.1,0.2,0.3,0.4,0.5,0.6,0.7,0.8,0.9,1},
xlabel style={at = {(0.5,0.02)},font=\fontsize{12}{1}\selectfont},
ticklabel style={font=\fontsize{10}{1}\selectfont},
ymin=0,
ymax=35,
ytick={0,5,10,15,20,25,30,35},
yticklabels={0,5,10,15,20,25,30,35},
ylabel={SRE (dB)},
ylabel style={at = {(0.04,0.5)},font=\fontsize{12}{1}\selectfont},
axis background/.style={fill=white},
legend style={at={(0,1)},anchor=north west,legend cell align=left,align=left,fill=none,draw=none,font=\fontsize{6}{1}\selectfont}
]
\addplot [color=green,solid,line width=0.5pt,mark=+,mark options={solid},forget plot]
  table[row sep=crcr]{%
0.1	1.25845551210852\\
0.2	4.62160165240744\\
0.3	7.10198633276568\\
0.4	9.74727820427153\\
0.5	12.7000291328555\\
0.6	15.7816398607739\\
0.7	17.3816211980636\\
0.8	19.0291668631024\\
0.9	20.2855097554761\\
1	21.0356\\
};
\addplot [color=blue,solid,line width=0.5pt,mark=o,mark options={solid},forget plot]
  table[row sep=crcr]{%
0.1	4.82188256285706\\
0.2	8.49479449115904\\
0.3	11.8342258367561\\
0.4	14.7270628360165\\
0.5	16.7371697377468\\
0.6	18.4265710975764\\
0.7	19.8384146192783\\
0.8	20.710167744664\\
0.9	21.280290238481\\
1	22.5331086761129\\
};
\addplot [color=black,dotted,line width=0.5pt,mark=x,mark options={solid},forget plot]
  table[row sep=crcr]{%
0.1	13.407154563415\\
0.2	15.5417947713215\\
0.3	16.4580328713433\\
0.4	17.3687619514874\\
0.5	18.1010384261505\\
0.6	18.5473366133015\\
0.7	18.8261443259184\\
0.8	19.013145554437\\
0.9	19.1730081123728\\
1	19.9972270023053\\
};
\addplot [color=red,dotted,line width=0.5pt,mark=*,mark options={solid},mark size =1,forget plot]
  table[row sep=crcr]{%
0.1	9.35671196082487\\
0.2	12.1742050102869\\
0.3	13.6956836154317\\
0.4	14.8571935468028\\
0.5	15.7117949298782\\
0.6	16.4649657874527\\
0.7	17.1128748093477\\
0.8	17.6045190487003\\
0.9	17.9315323843072\\
1	19.9936792446994\\
};
\addplot [color=mycolor1,dotted,line width=0.5pt,mark=asterisk,mark options={solid},forget plot]
  table[row sep=crcr]{%
0.1	5.88268592255294\\
0.2	7.91604651281894\\
0.3	9.84976471650357\\
0.4	11.3170390975581\\
0.5	12.5980262513404\\
0.6	13.5846764628163\\
0.7	14.736763314936\\
0.8	16.1349777375997\\
0.9	17.319329491716\\
1	19.9877042519565\\
};
\addplot [color=blue,dotted,line width=0.5pt,mark=o,mark options={solid},forget plot]
  table[row sep=crcr]{%
0.1	4.3278953399151\\
0.2	5.79991418079457\\
0.3	7.15391747666594\\
0.4	8.49841890469245\\
0.5	10.1018544961471\\
0.6	11.4913711810555\\
0.7	13.1105709704673\\
0.8	14.9840701650604\\
0.9	16.7200012997163\\
1	19.9708738262418\\
};
\addplot [color=green,dotted,line width=0.5pt,mark=+,mark options={solid},forget plot]
  table[row sep=crcr]{%
0.1	2.12293116223928\\
0.2	4.05022070961676\\
0.3	5.19354594652312\\
0.4	6.14240636106582\\
0.5	7.35153941956745\\
0.6	8.62380823080183\\
0.7	9.91435791612642\\
0.8	12.02805132802\\
0.9	14.3260295674346\\
1	19.9709330430668\\
};
\addplot [color=mycolor2,dotted,line width=0.5pt,mark=square,mark options={solid},forget plot]
  table[row sep=crcr]{%
0.1	17.1133572473827\\
0.2	18.353147744526\\
0.3	19.0588408867494\\
0.4	19.553681691753\\
0.5	19.7816647505456\\
0.6	19.8138565397956\\
0.7	19.8216080562673\\
0.8	19.8297446693547\\
0.9	19.83823723467\\
1	19.999903818324\\
};
\addplot [color=mycolor2,solid,line width=0.5pt,mark=square,mark options={solid},forget plot]
  table[row sep=crcr]{%
0.1	20.026753053179\\
0.2	22.1340834162349\\
0.3	23.8158321444194\\
0.4	25.4831411521771\\
0.5	26.8517138374874\\
0.6	28.0156693768445\\
0.7	28.8658778639693\\
0.8	29.4040297858399\\
0.9	29.614102504515\\
1	29.7792407877135\\
};
\addplot [color=black,solid,line width=0.5pt,mark=x,mark options={solid},forget plot]
  table[row sep=crcr]{%
0.1	16.8322231379474\\
0.2	19.7293813152763\\
0.3	21.9436845794776\\
0.4	23.6084694813411\\
0.5	24.9141384017697\\
0.6	25.9454334190241\\
0.7	26.7017528638588\\
0.8	27.1923840107799\\
0.9	27.4573069051356\\
1	27.7597681206385\\
};
\addplot [color=red,solid,line width=0.5pt,mark=*,mark options={solid},mark size = 1,forget plot]
  table[row sep=crcr]{%
0.1	12.749890953912\\
0.2	16.2007981358857\\
0.3	18.7833729701137\\
0.4	20.8658841335615\\
0.5	22.4193409406961\\
0.6	23.684213477188\\
0.7	24.6091770658324\\
0.8	25.1707353783276\\
0.9	25.7096427345489\\
1	25.7933371339029\\
};
\addplot [color=mycolor1,solid,line width=0.5pt,mark=asterisk,mark options={solid},forget plot]
  table[row sep=crcr]{%
0.1	8.18344418772522\\
0.2	12.5153420230508\\
0.3	15.5081167511169\\
0.4	17.854986835736\\
0.5	19.8885026427861\\
0.6	21.3555515939964\\
0.7	22.6298938916811\\
0.8	23.4585633721384\\
0.9	23.9699163206512\\
1	24.08297838408\\
};
\node (e1) at (80,245) {};\node (e2) at (80,277) {};\draw[<->] (e1) -- (e2);
\node[align=left, text=black, draw=none,anchor=west]  at (79,263) {\fontsize{6}{1}\selectfont $\ \approx\text{2 dB}$};

\node[align=left, text=black, draw=none,anchor=north west]  at (0,340) {\fontsize{8}{1}\selectfont $\{{\rm MDS},~ \bs \Psi_{\rm adhw}, \cl T^{\rm aniso}\}$};
\end{axis}
\end{tikzpicture}%
}
		\end{minipage}
	\end{minipage}
	\caption{The SRE of phantom image recovery from subsampled Hadamard measurements.}
	\label{fig:res_phantom_2d}
\end{figure}
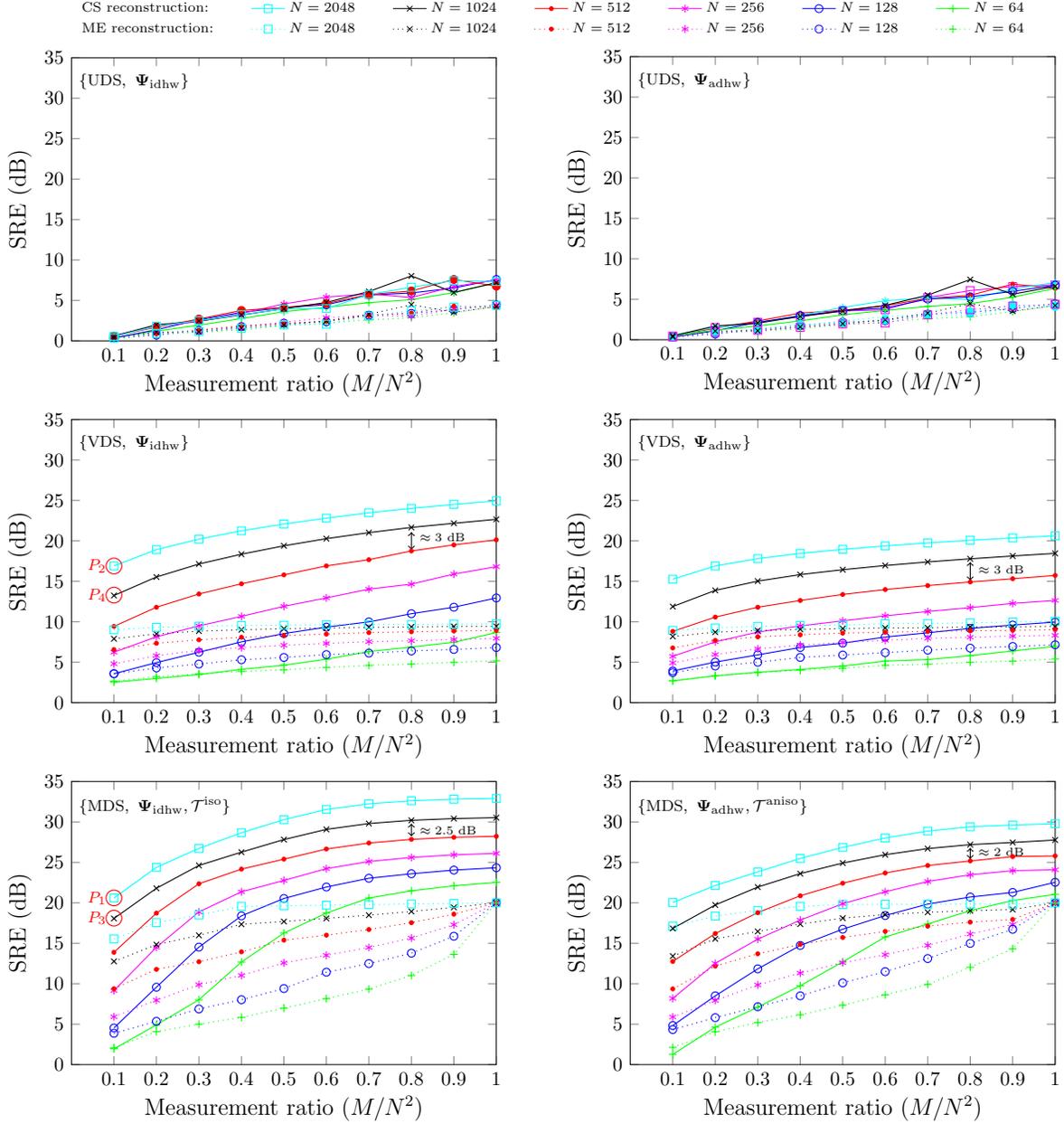
\begin{figure}
	\centering
	\begin{minipage}{0.49\columnwidth}
		\scalebox{0.82}{\input{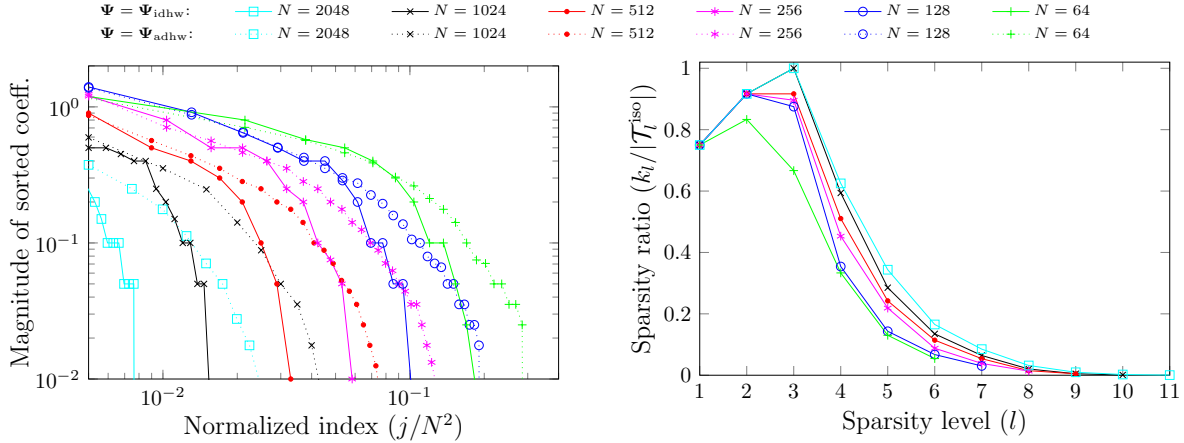}}
	\end{minipage}
	\begin{minipage}{0.49\columnwidth}\vspace{7mm}
		\scalebox{0.82}{
%
%
\definecolor{mycolor1}{rgb}{1.00000,0.00000,1.00000}%
\definecolor{mycolor2}{rgb}{0.00000,1.00000,1.00000}%
\begin{tikzpicture}
\begin{axis}[%
width=3in,
height=2in,
at={(0.762in,0.486in)},
scale only axis,
xmin=1,
xmax=11,
xlabel={Sparsity level $(l)$},
xtick={1,2,3,4,5,6,7,8,9,10,11},
xticklabels={1,2,3,4,5,6,7,8,9,10,11},
xlabel style={at = {(0.5,0.02)},font=\fontsize{12}{1}\selectfont},
ticklabel style={font=\fontsize{10}{1}\selectfont},
ymin=0,
ymax=1.02,
ytick={0,0.2,0.4,0.6,0.8,1},
yticklabels={0,0.2,0.4,0.6,0.8,1},
ylabel={Sparsity ratio $(k_l/|\cl T_l^{\rm iso}|)$},
ylabel style={at = {(0.04,0.5)},font=\fontsize{12}{1}\selectfont},
axis background/.style={fill=white},
]
\addplot [color=black,solid,mark=x,mark options={solid},forget plot]
  table[row sep=crcr]{%
1	0.75\\
2	0.916666666666667\\
3	1\\
4	0.59375\\
5	0.28515625\\
6	0.134765625\\
7	0.0643717447916667\\
8	0.0202840169270833\\
9	0.004852294921875\\
10	0.00087738037109375\\
};
\addplot [color=red,solid,mark size=1pt,mark=*,mark options={solid},forget plot]
  table[row sep=crcr]{%
1	0.75\\
2	0.916666666666667\\
3	0.916666666666667\\
4	0.510416666666667\\
5	0.2421875\\
6	0.1142578125\\
7	0.053955078125\\
8	0.014892578125\\
9	0.004669189453125\\
};
\addplot [color=mycolor1,solid,mark=asterisk,mark options={solid},forget plot]
  table[row sep=crcr]{%
1	0.75\\
2	0.916666666666667\\
3	0.895833333333333\\
4	0.453125\\
5	0.21875\\
6	0.087890625\\
7	0.0384928385416667\\
8	0.0133870442708333\\
};
\addplot [color=blue,solid,mark=o,mark options={solid},forget plot]
  table[row sep=crcr]{%
1	0.75\\
2	0.916666666666667\\
3	0.875\\
4	0.354166666666667\\
5	0.143229166666667\\
6	0.0673828125\\
7	0.0302734375\\
};
\addplot [color=green,solid,mark=+,mark options={solid},forget plot]
  table[row sep=crcr]{%
1	0.75\\
2	0.833333333333333\\
3	0.666666666666667\\
4	0.333333333333333\\
5	0.130208333333333\\
6	0.0543619791666667\\
};
\addplot [color=mycolor2,solid,mark=square,mark options={solid},forget plot]
  table[row sep=crcr]{%
1	0.75\\
2	0.916666666666667\\
3	1\\
4	0.625\\
5	0.34375\\
6	0.165364583333333\\
7	0.0850423177083333\\
8	0.0317586263020833\\
9	0.00969441731770833\\
10	0.00238545735677083\\
11	0.000444412231445313\\
};
\end{axis}
\end{tikzpicture}%
}
	\end{minipage}
	\caption{Global (left) and local (right) sparsity of the phantom image in 2-D Haar wavelet basis. On the right figure, in order to obtain meaningful curves we assumed $\cl T^{\rm iso}_0 = \emptyset$ and $\cl T^{\rm iso}_1 = \{1,2\}$.}
	\label{fig:res_phantom_sparsity}
\end{figure}
We now test the performance of the proposed VDS and MDS schemes in an imaging context. We generate synthetic Shepp-Logan phantom images \cite{Shepp1974TheFR} of size $N\times N$ with $N =2^r$ and $r\in\{7,\cdots,11\}$ as the ground truth. The variance of the noise amounts to an SNR of 20 dB. Fig.~\ref{fig:res_phantom_2d} illustrates the SRE values as a function of the measurement ratio ($M/N^2$) for different resolutions $N$, sampling strategies (UDS, VDS, and MDS), sparsity bases (IDHW and ADHW), and recovery algorithms (CS and ME). The results are averaged over 10 trials (\ie over the random generation of both the noise and random selection of the subsampling set $\Omega$ according to the sampling strategy). We note that in Fig.~\ref{fig:res_phantom_2d} and in the UDS and VDS cases, since there are repeated indices in the subsampled set $\Omega$, even for $M/N^2=1$, we cannot reach the recovery quality of fully-sampled (or Nyquist) Hadamard measurements. On the contrary, since MDS scheme does not allow repeated indices, the recovery quality of the Nyquist Hadamard-Haar system happens when $M/N^2=1$. Not surprisingly, ME reconstruction yields SRE = SNR = 20 dB when the Hadamard measurements are fully-sampled. Fig.~\ref{fig:res_phantom_sparsity} displays the global and local sparsity of the phantom images of different sizes in 2-D Haar wavelet basis. On the left, the sorted coefficients in IDHW and ADHW bases are plotted versus the normalized index axis. Fig.~\ref{fig:res_phantom_sparsity}-right shows an experiment in which we computed the local sparsity ratios for the phantom image of different sizes using IDHW sparsity basis. 

From Fig.~\ref{fig:res_phantom_2d} and Fig.~\ref{fig:res_phantom_sparsity} we can do the following observations. First, similar to the 1-D signal recovery, the UDS scheme performs poorly. Second, the CS reconstruction always outperforms the ME reconstruction, as the latter does not take into account the sparsity prior information. Third, by increasing the resolution of the signal (or the size of the problem) one can obtain a higher SRE value (up to 3 dB), regardless of the CS or ME reconstruction method. Essentially, by going higher in resolution the signal becomes (asymptotically) sparser in the wavelet domain, as represented in Fig.~\ref{fig:res_phantom_sparsity}-left. In this figure, the decay rate of the curves increases as $N$ grows. As already stressed in, \eg \cite{roman2014asymptotic}, the MDS scheme is thus expected to express its efficacy in high-dimensional applications. Fourth, the IDHW basis yields better SRE values in comparison to the ADHW basis because the phantom image is more compressible in the IDHW basis. Concretely, by comparing the solid and dotted lines in Fig.~\ref{fig:res_phantom_sparsity}-left, we conclude that the phantom image reaches higher compressibility in the IDHW basis, which further increases the quality of the signal recovery. Fifth, the MDS scheme is resolution dependent: following the sample-complexity bounds in Thm.~\ref{thm:Non-uniform guarantee for Hadamard-Haar system}, the values in Fig.~\ref{fig:res_phantom_sparsity}-right determine the required number of measurements at each level. Finally, since the MDS scheme leverages the sparsity structure of the signal, it outperforms the VDS scheme in the sense of recovery quality. 

An example of the reconstructed images in the simulation above, marked by points $P_1, P_2, P_3$, and $P_4$, is depicted in Fig.~\ref{fig:res_2d_example}. In this figure we notice the effect of the resolution on the MDS strategy and on the image recovery quality.
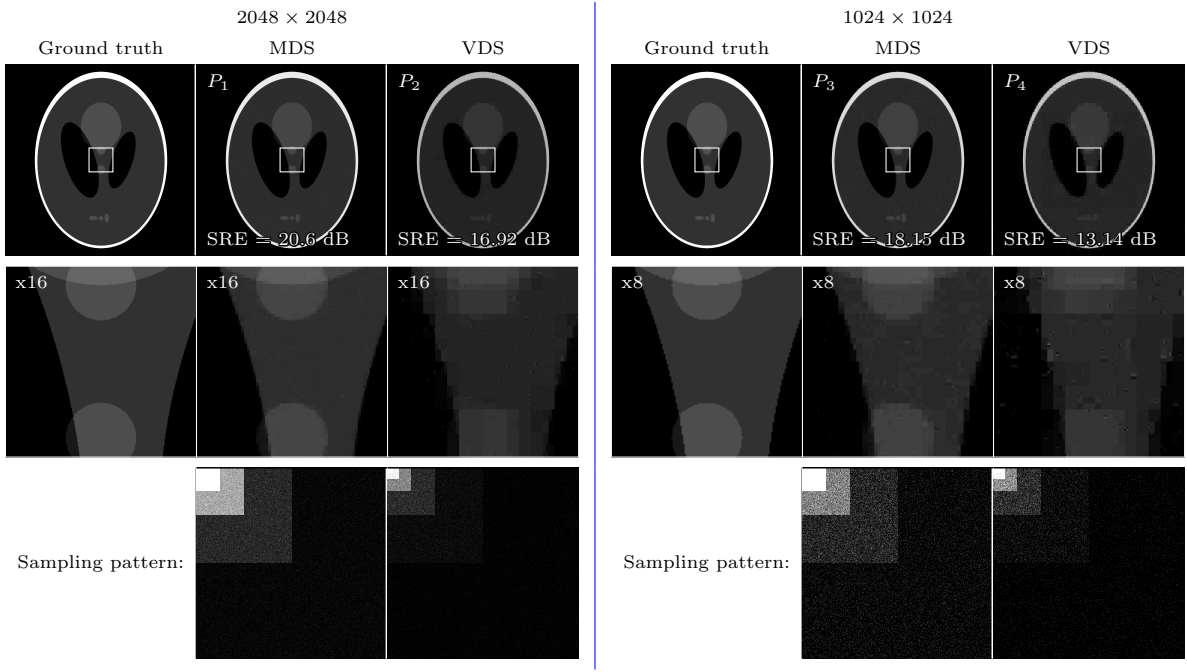
\begin{figure} 
	\centering
	\arrayrulecolor{blue}
	\begin{minipage}{1\columnwidth}
	\centering
	\begin{tabular}{c|c}
		\scalebox{1}{
%
%

\begin{tikzpicture}
\begin{axis}[%
width=1in,
height=1in,
at={(0,0)},
scale only axis,
axis on top,
xmin=0.5,
xmax=2048.5,
y dir=reverse,
ymin=0.5,
ymax=2048.5,
axis background/.style={fill=white},
hide axis
]
\addplot [forget plot] graphics [xmin=0.5,xmax=2048.5,ymin=0.5,ymax=2048.5] {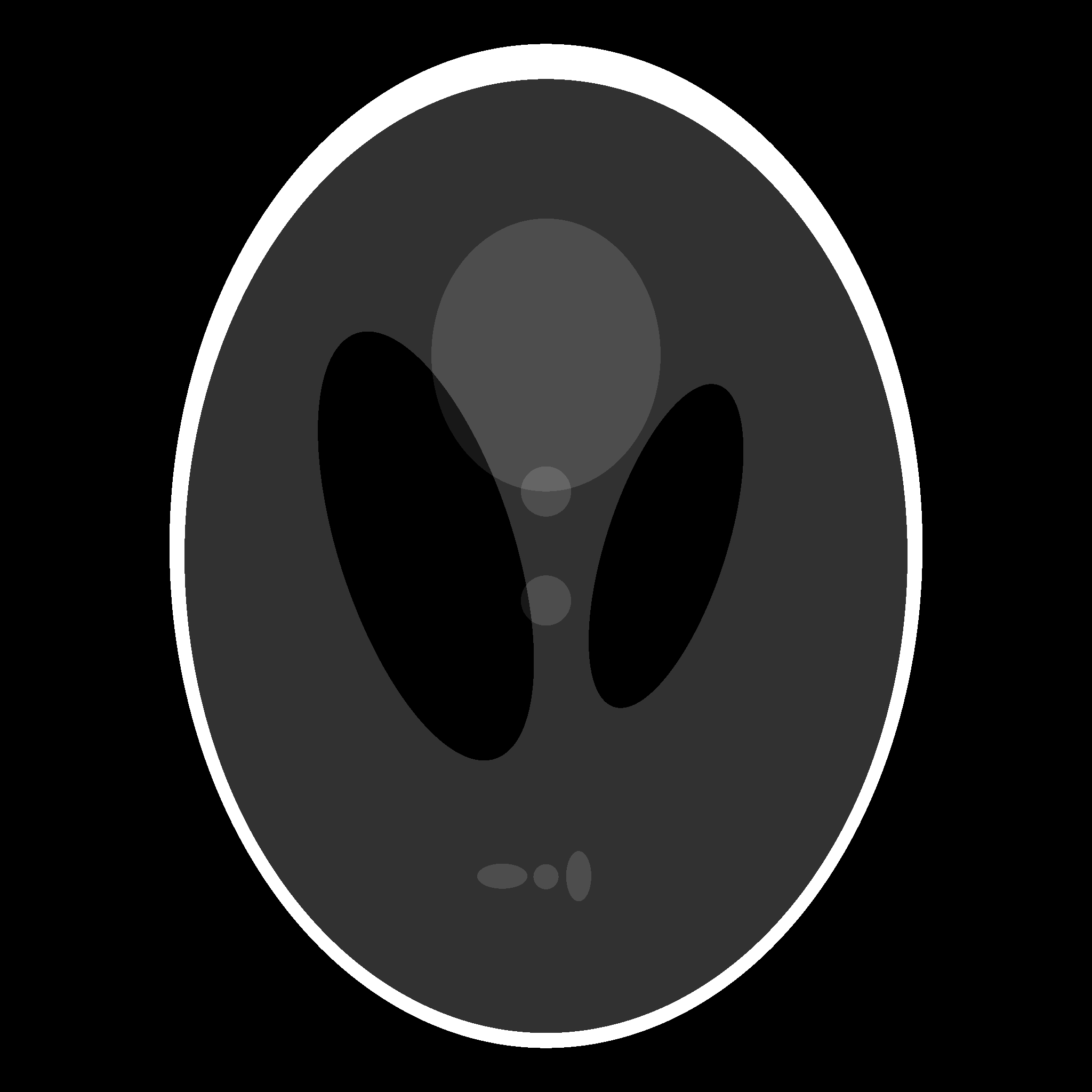};
\draw[color = white, fill = none] (axis cs:896.5,896.5)rectangle (axis cs:1152.5,1152.5);
\draw [color=white,line width=0.5mm] (axis cs:2048,0) -- (axis cs: 2048,2048);
\end{axis}
\node[draw,fill = none,text=black, draw=none,anchor = south] at (rel axis cs:0.5,-1) {\fontsize{7}{1}\selectfont Ground truth};
\end{tikzpicture}%
%
%
\begin{tikzpicture}

\begin{axis}[%
width=1in,
height=1in,
at={(0,0)},
scale only axis,
axis on top,
xmin=0.5,
xmax=2048.5,
y dir=reverse,
ymin=0.5,
ymax=2048.5,
axis background/.style={fill=white},
hide axis
]
\addplot [forget plot] graphics [xmin=0.5,xmax=2048.5,ymin=0.5,ymax=2048.5] {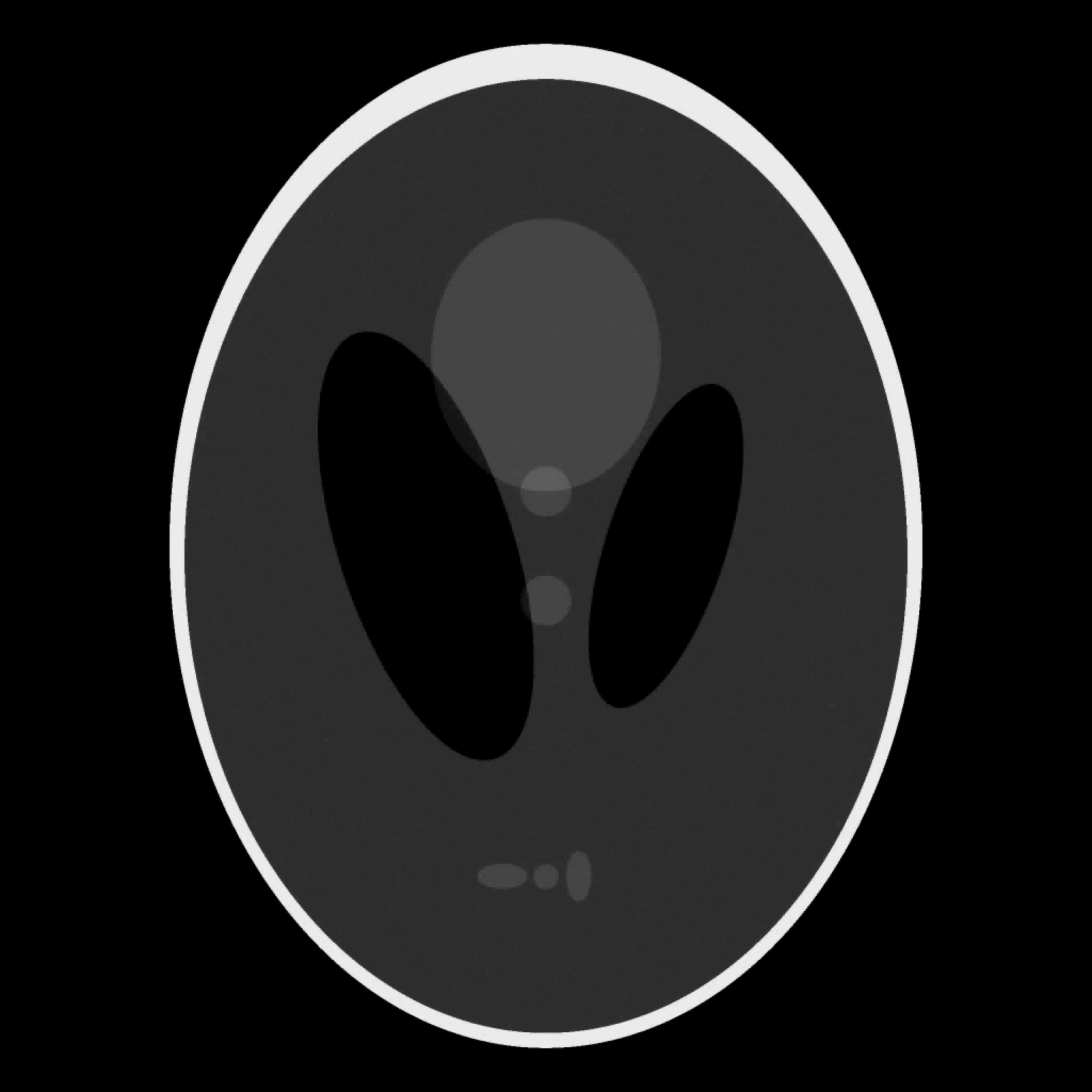};
\draw[color = white, fill = none] (axis cs:896.5,896.5)rectangle (axis cs:1152.5,1152.5);
\node[draw,fill = none,text=white, draw=none,anchor = north west] at (rel axis cs:0,1) {\fontsize{7}{1}\selectfont $P_1$};
\draw [color=white,line width=0.5mm] (axis cs:2048,0) -- (axis cs: 2048,2048);
\end{axis}
\node[draw,fill = none,text=black, draw=none,anchor = south west] at (rel axis cs:0,0) {\fontsize{7}{1}\selectfont \contourlength{.5pt}\textcolor{white}{\contour{black}{SRE = 20.6 dB}}}; 
\node[draw,fill = none,text=black, draw=none,anchor = south] at (rel axis cs:0.5,-1) {\fontsize{7}{1}\selectfont MDS};
\node[draw,fill = none,text=black, draw=none,anchor = south] at (rel axis cs:0.5,-1.15) {\fontsize{7}{1}\selectfont $2048 \times 2048$};
\end{tikzpicture}%
%
%
\begin{tikzpicture}

\begin{axis}[%
width=1in,
height=1in,
at={(0,0)},
scale only axis,
axis on top,
xmin=0.5,
xmax=2048.5,
y dir=reverse,
ymin=0.5,
ymax=2048.5,
axis background/.style={fill=white},
hide axis
]
\addplot [forget plot] graphics [xmin=0.5,xmax=2048.5,ymin=0.5,ymax=2048.5] {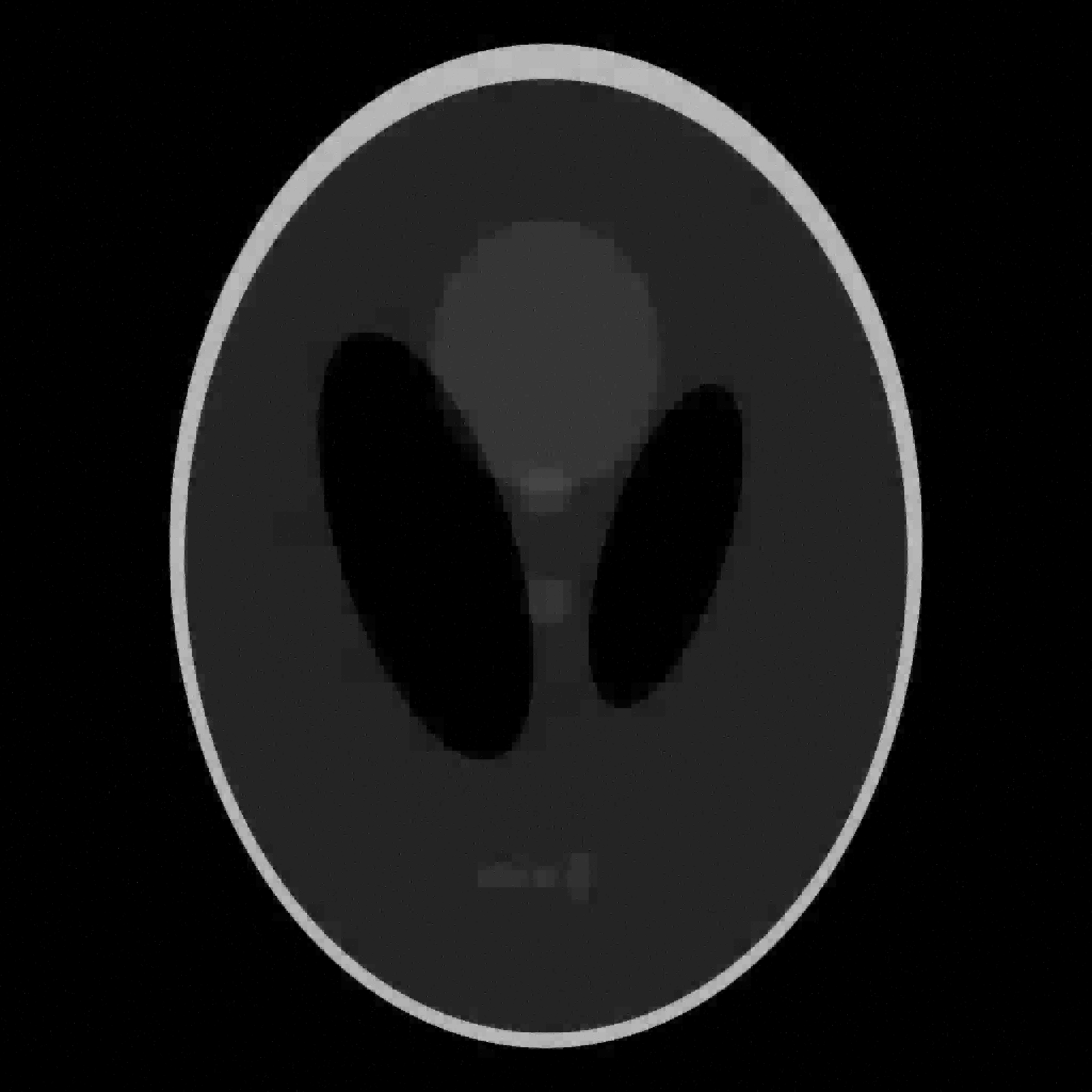};
\draw[color = white, fill = none] (axis cs:896.5,896.5)rectangle (axis cs:1152.5,1152.5);
\node[draw,fill = none,text=white, draw=none,anchor = north west] at (rel axis cs:0,1) {\fontsize{7}{1}\selectfont $P_2$};
\end{axis}
\node[draw,fill = none,text=black, draw=none,anchor = south west] at (rel axis cs:0,0) {\fontsize{7}{1}\selectfont \contourlength{.5pt}\textcolor{white}{\contour{black}{SRE = 16.92 dB}}}; 
\node[draw,fill = none,text=black, draw=none,anchor = south] at (rel axis cs:0.5,-1) {\fontsize{7}{1}\selectfont VDS};
\end{tikzpicture}%
}
		&
		\scalebox{1}{
%
%

\begin{tikzpicture}
\begin{axis}[%
width=1in,
height=1in,
at={(0,0)},
scale only axis,
axis on top,
xmin=0.5,
xmax=1024.5,
y dir=reverse,
ymin=0.5,
ymax=1024.5,
axis background/.style={fill=white},
hide axis
]
\addplot [forget plot] graphics [xmin=0.5,xmax=1024.5,ymin=0.5,ymax=1024.5] {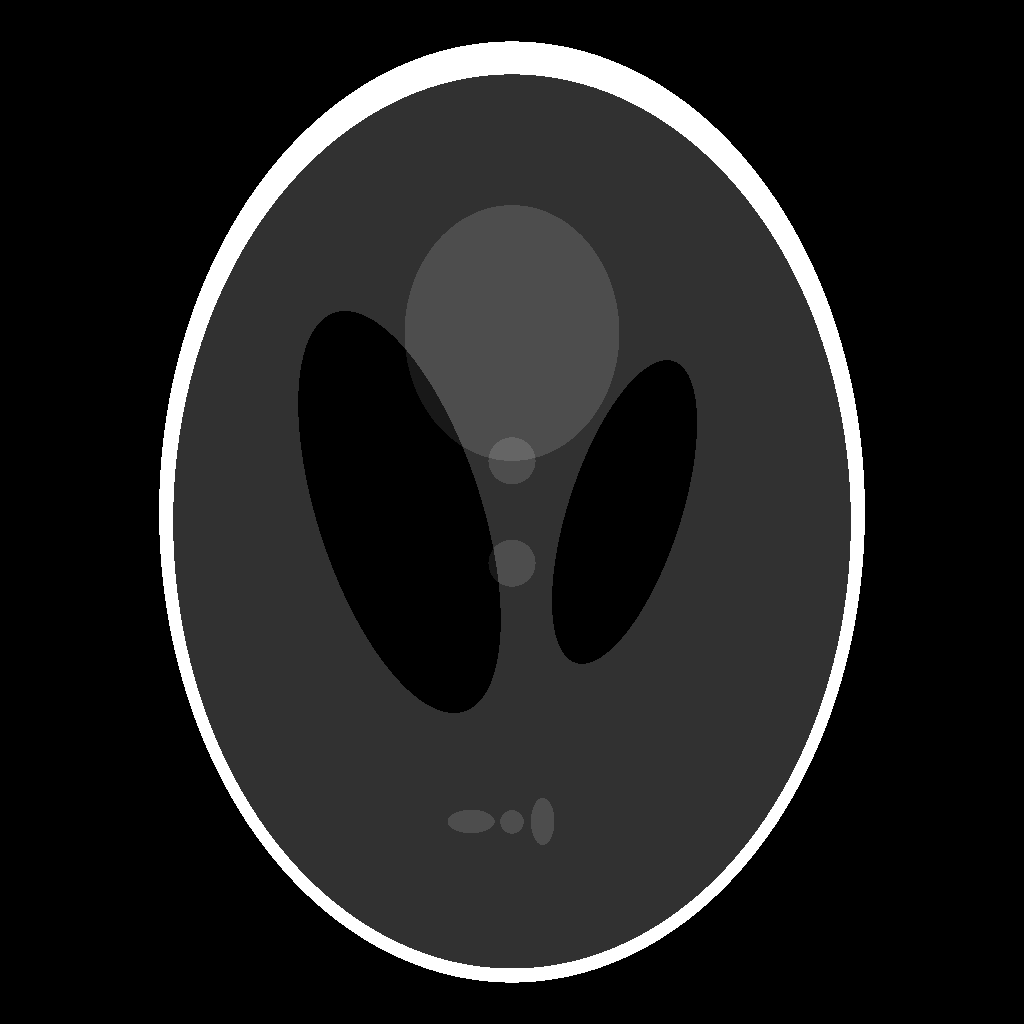};
\draw[color = white, fill = none] (axis cs:448.5,448.5)rectangle (axis cs:576.5,576.5);
\draw [color=white,line width=0.5mm] (axis cs:1024,0) -- (axis cs: 1024,1024);
\end{axis}
\node[draw,fill = none,text=black, draw=none,anchor = south] at (rel axis cs:0.5,-1) {\fontsize{7}{1}\selectfont Ground truth};
\end{tikzpicture}%
%
%
\begin{tikzpicture}

\begin{axis}[%
width=1in,
height=1in,
at={(0,0)},
scale only axis,
axis on top,
xmin=0.5,
xmax=1024.5,
y dir=reverse,
ymin=0.5,
ymax=1024.5,
axis background/.style={fill=white},
hide axis
]
\addplot [forget plot] graphics [xmin=0.5,xmax=1024.5,ymin=0.5,ymax=1024.5] {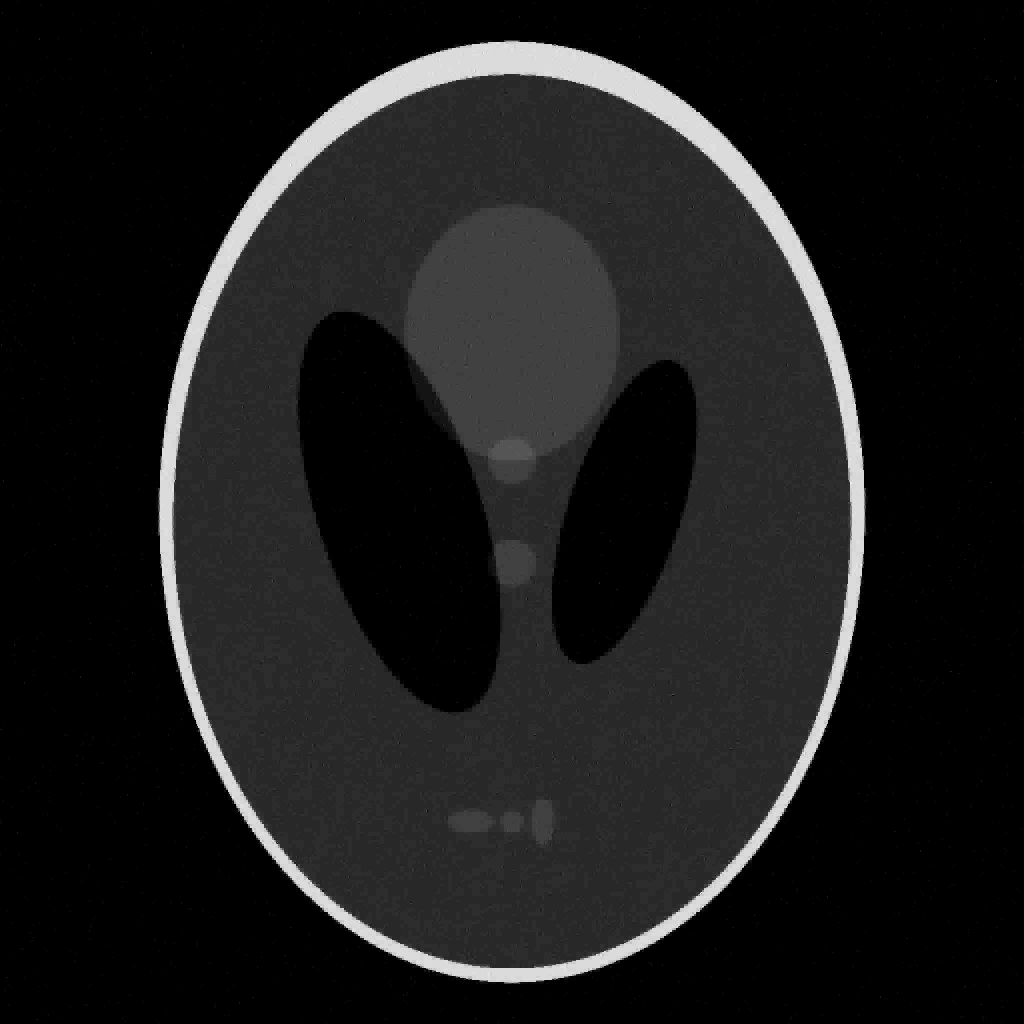};
\draw[color = white, fill = none] (axis cs:448.5,448.5)rectangle (axis cs:576.5,576.5);
\node[draw,fill = none,text=white, draw=none,anchor = north west] at (rel axis cs:0,1) {\fontsize{7}{1}\selectfont $P_3$};
\draw [color=white,line width=0.5mm] (axis cs:1024,0) -- (axis cs: 1024,1024);
\end{axis}
\node[draw,fill = none,text=black, draw=none,anchor = south west] at (rel axis cs:0,0) {\fontsize{7}{1}\selectfont \contourlength{.5pt}\textcolor{white}{\contour{black}{SRE = 18.15 dB}}}; 
\node[draw,fill = none,text=black, draw=none,anchor = south] at (rel axis cs:0.5,-1) {\fontsize{7}{1}\selectfont MDS};
\node[draw,fill = none,text=black, draw=none,anchor = south] at (rel axis cs:0.5,-1.15) {\fontsize{7}{1}\selectfont $1024 \times 1024$};
\end{tikzpicture}%
%
%
\begin{tikzpicture}

\begin{axis}[%
width=1in,
height=1in,
at={(0,0)},
scale only axis,
axis on top,
xmin=0.5,
xmax=1024.5,
y dir=reverse,
ymin=0.5,
ymax=1024.5,
axis background/.style={fill=white},
hide axis
]
\addplot [forget plot] graphics [xmin=0.5,xmax=1024.5,ymin=0.5,ymax=1024.5] {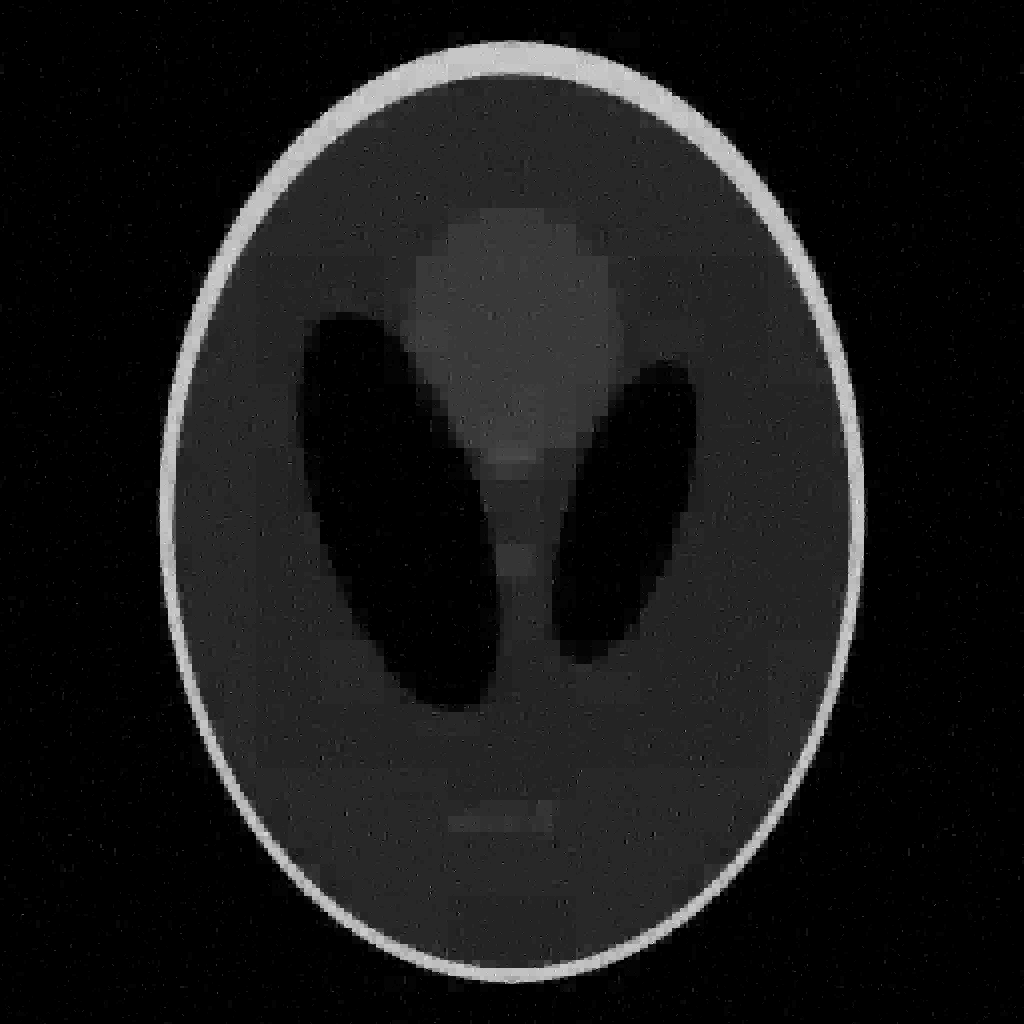};
\draw[color = white, fill = none] (axis cs:448.5,448.5)rectangle (axis cs:576.5,576.5);
\node[draw,fill = none,text=white, draw=none,anchor = north west] at (rel axis cs:0,1) {\fontsize{7}{1}\selectfont $P_4$};
\end{axis}
\node[draw,fill = none,text=black, draw=none,anchor = south west] at (rel axis cs:0,0) {\fontsize{7}{1}\selectfont \contourlength{.5pt}\textcolor{white}{\contour{black}{SRE = 13.14 dB}}}; 
\node[draw,fill = none,text=black, draw=none,anchor = south] at (rel axis cs:0.5,-1) {\fontsize{7}{1}\selectfont VDS};
\end{tikzpicture}%
}
	\\
		\scalebox{1}{
%
%
\begin{tikzpicture}

\begin{axis}[%
width=1in,
height=1in,
at={(0,0)},
scale only axis,
axis on top,
xmin=896.5,
xmax=1152.5,
y dir=reverse,
ymin=896.5,
ymax=1152.5,
axis background/.style={fill=white},
hide axis
]
\addplot [forget plot] graphics [xmin=0.5,xmax=2048.5,ymin=0.5,ymax=2048.5] {Res_Phantom_GT_2048-1.png};
\node[draw,fill = none,text=white, draw=none,anchor = north west] at (rel axis cs:0,1) {\fontsize{7}{1}\selectfont x16};
\draw[color = white, fill = none] (axis cs:896.5,896.5)rectangle (axis cs:1152.5,1152.5);
\end{axis}
\end{tikzpicture}%

%
%
\begin{tikzpicture}

\begin{axis}[%
width=1in,
height=1in,
at={(0,0)},
scale only axis,
axis on top,
xmin=896.5,
xmax=1152.5,
y dir=reverse,
ymin=896.5,
ymax=1152.5,
axis background/.style={fill=white},
hide axis
]
\addplot [forget plot] graphics [xmin=0.5,xmax=2048.5,ymin=0.5,ymax=2048.5] {Res_Phantom_CS_MDS_10_2048-1.png};
\node[draw,fill = none,text=white, draw=none,anchor = north west] at (rel axis cs:0,1) {\fontsize{7}{1}\selectfont x16};
\draw[color = white, fill = none] (axis cs:896.5,896.5)rectangle (axis cs:1152.5,1152.5);
\end{axis}
\end{tikzpicture}%

%
%
\begin{tikzpicture}

\begin{axis}[%
width=1in,
height=1in,
at={(0,0)},
scale only axis,
axis on top,
xmin=896.5,
xmax=1152.5,
y dir=reverse,
ymin=896.5,
ymax=1152.5,
axis background/.style={fill=white},
hide axis
]
\addplot [forget plot] graphics [xmin=0.5,xmax=2048.5,ymin=0.5,ymax=2048.5] {Res_Phantom_CS_VDS_10_2048-1.png};
\node[draw,fill = none,text=white, draw=none,anchor = north west] at (rel axis cs:0,1) {\fontsize{7}{1}\selectfont x16};
\draw[color = white, fill = none] (axis cs:896.5,896.5)rectangle (axis cs:1152.5,1152.5);
\end{axis}
\end{tikzpicture}%
}
		&
		\scalebox{1}{
%
%
\begin{tikzpicture}

\begin{axis}[%
width=1in,
height=1in,
at={(0,0)},
scale only axis,
axis on top,
xmin=448.5,
xmax=576.5,
y dir=reverse,
ymin=448.5,
ymax=576.5,
axis background/.style={fill=white},
hide axis
]
\addplot [forget plot] graphics [xmin=0.5,xmax=1024.5,ymin=0.5,ymax=1024.5] {Res_Phantom_GT_1024-1.png};
\node[draw,fill = none,text=white, draw=none,anchor = north west] at (rel axis cs:0,1) {\fontsize{7}{1}\selectfont x8};
\draw[color = white, fill = none] (axis cs:448.5,448.5)rectangle (axis cs:576.5,576.5);
\end{axis}
\end{tikzpicture}%

%
%
\begin{tikzpicture}

\begin{axis}[%
width=1in,
height=1in,
at={(0,0)},
scale only axis,
axis on top,
xmin=448.5,
xmax=576.5,
y dir=reverse,
ymin=448.5,
ymax=576.5,
axis background/.style={fill=white},
hide axis
]
\addplot [forget plot] graphics [xmin=0.5,xmax=1024.5,ymin=0.5,ymax=1024.5] {Res_Phantom_CS_MDS_10_1024-1.png};
\node[draw,fill = none,text=white, draw=none,anchor = north west] at (rel axis cs:0,1) {\fontsize{7}{1}\selectfont x8};
\draw[color = white, fill = none] (axis cs:448.5,448.5)rectangle (axis cs:576.5,576.5);
\end{axis}
\end{tikzpicture}%

%
%
\begin{tikzpicture}

\begin{axis}[%
width=1in,
height=1in,
at={(0,0)},
scale only axis,
axis on top,
xmin=448.5,
xmax=576.5,
y dir=reverse,
ymin=448.5,
ymax=576.5,
axis background/.style={fill=white},
hide axis
]
\addplot [forget plot] graphics [xmin=0.5,xmax=1024.5,ymin=0.5,ymax=1024.5] {Res_Phantom_CS_VDS_10_1024-1.png};
\node[draw,fill = none,text=white, draw=none,anchor = north west] at (rel axis cs:0,1) {\fontsize{7}{1}\selectfont x8};
\draw[color = white, fill = none] (axis cs:448.5,448.5)rectangle (axis cs:576.5,576.5);
\end{axis}
\end{tikzpicture}%
}
	\\
		\scalebox{1}{
%
%

\begin{tikzpicture}
\begin{axis}[%
width=1in,
height=1in,
at={(0,0)},
scale only axis,
axis on top,
xmin=0.5,
xmax=2048.5,
y dir=reverse,
ymin=0.5,
ymax=2048.5,
axis background/.style={fill=white},
hide axis
]
\draw [color=black,line width=0.2mm] (axis cs:2048,0) -- (axis cs: 2048,2048);
\end{axis}
\node[draw,fill = none,text=black, draw=none,anchor = center] at (rel axis cs:0.5,-0.5) {\fontsize{7}{1}\selectfont Sampling pattern:};
\end{tikzpicture}%
%
%
\begin{tikzpicture}

\begin{axis}[%
width=1in,
height=1in,
at={(0,0)},
scale only axis,
axis on top,
xmin=0.5,
xmax=2048.5,
y dir=reverse,
ymin=0.5,
ymax=2048.5,
axis background/.style={fill=white},
hide axis
]
\addplot [forget plot] graphics [xmin=0.5,xmax=2048.5,ymin=0.5,ymax=2048.5] {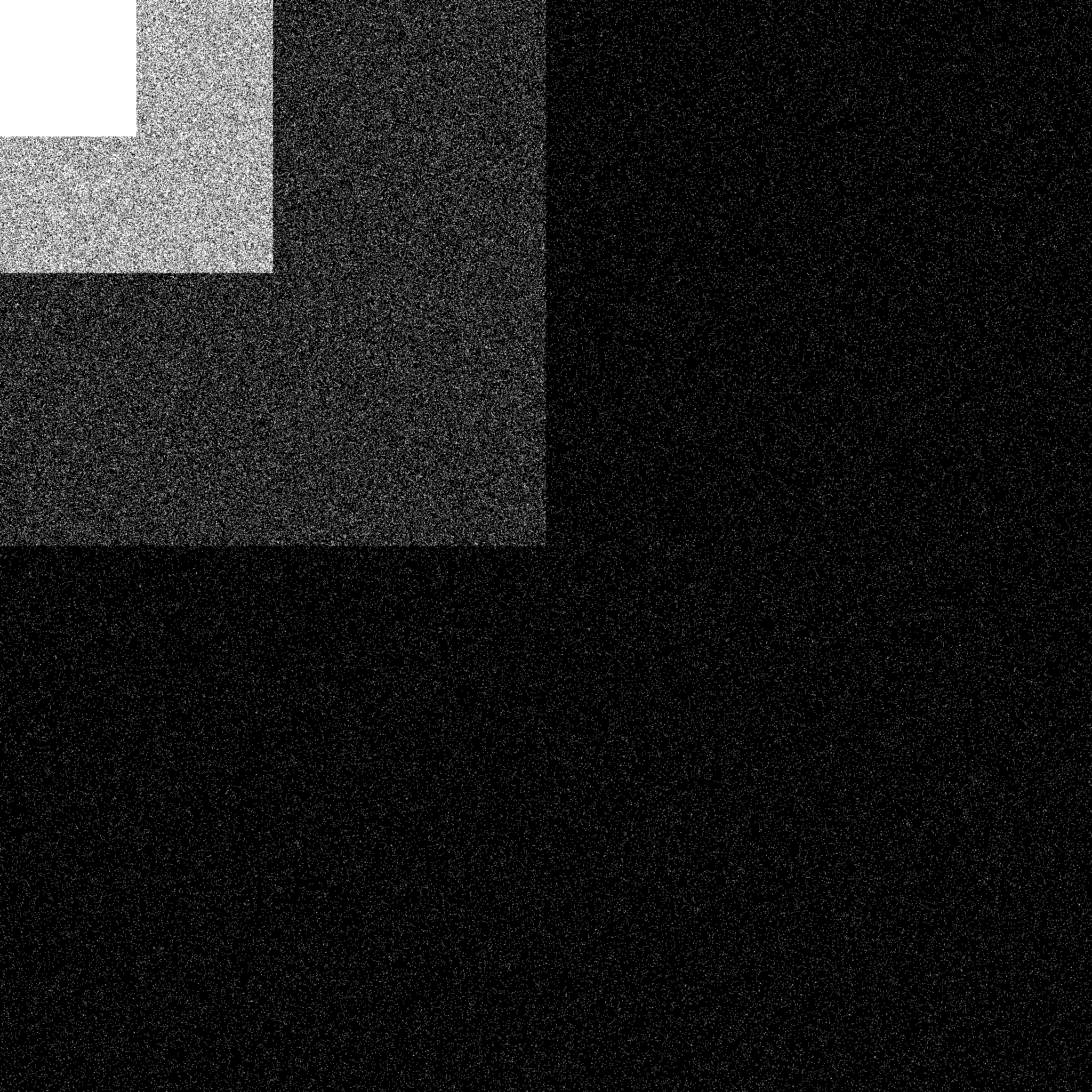};
\draw [color=white,line width=0.5mm] (axis cs:2048,0) -- (axis cs: 2048,2048);
\draw [color=black,line width=0.2mm] (axis cs:0,0) -- (axis cs: 2048,0);
\end{axis}
\end{tikzpicture}%
%
%
\begin{tikzpicture}

\begin{axis}[%
width=1in,
height=1in,
at={(0,0)},
scale only axis,
axis on top,
xmin=0.5,
xmax=2048.5,
y dir=reverse,
ymin=0.5,
ymax=2048.5,
axis background/.style={fill=white},
hide axis
]
\addplot [forget plot] graphics [xmin=0.5,xmax=2048.5,ymin=0.5,ymax=2048.5] {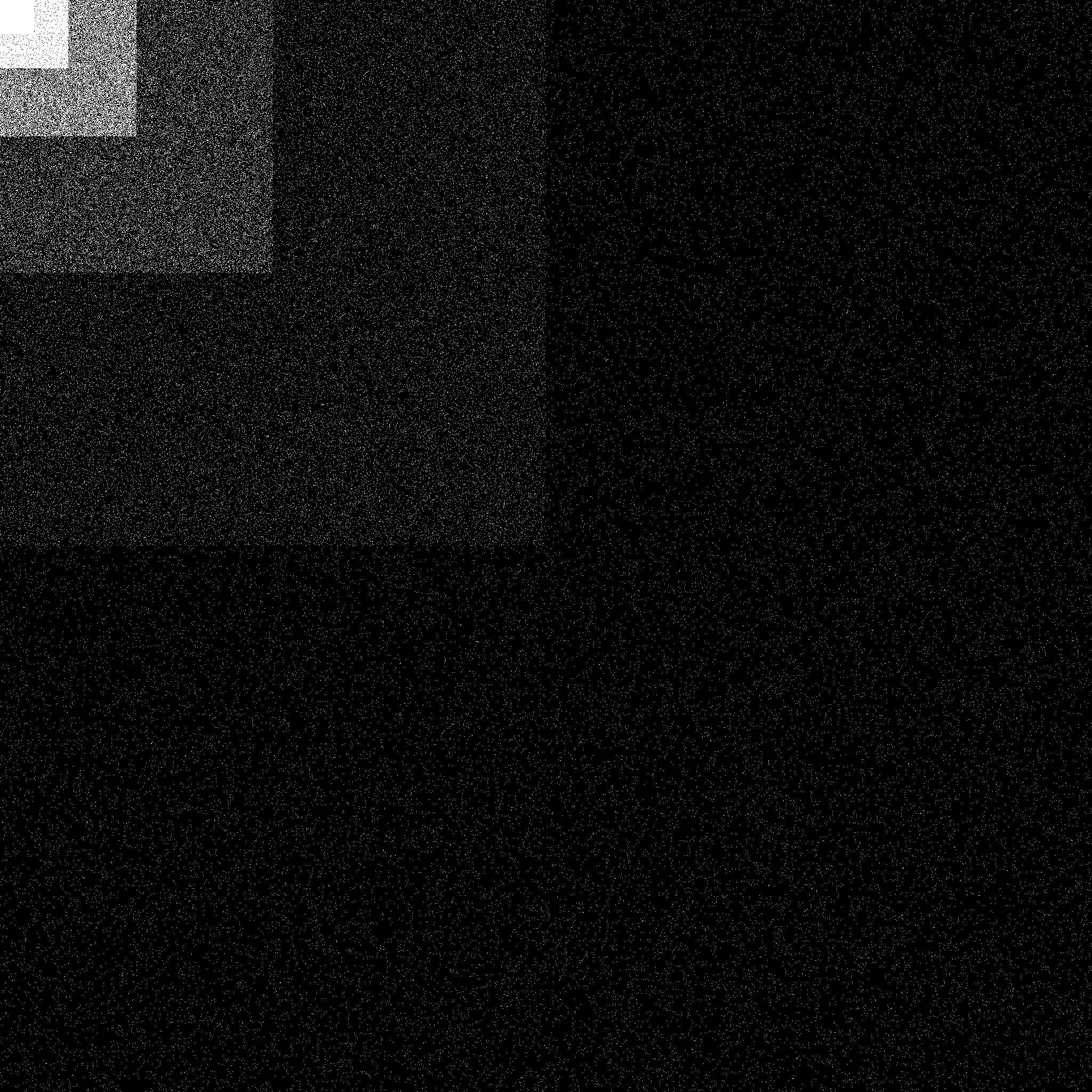};
\draw [color=black,line width=0.2mm] (axis cs:0,0) -- (axis cs: 2048,0);
\end{axis}
\end{tikzpicture}%
}
		&
		\scalebox{1}{
%
%

\begin{tikzpicture}
\begin{axis}[%
width=1in,
height=1in,
at={(0,0)},
scale only axis,
axis on top,
xmin=0.5,
xmax=1024.5,
y dir=reverse,
ymin=0.5,
ymax=1024.5,
axis background/.style={fill=white},
hide axis
]
\draw [color=black,line width=0.2mm] (axis cs:1024,0) -- (axis cs: 1024,1024);
\end{axis}
\node[draw,fill = none,text=black, draw=none,anchor = center] at (rel axis cs:0.5,-0.5) {\fontsize{7}{1}\selectfont Sampling pattern:};
\end{tikzpicture}%
%
%
\begin{tikzpicture}

\begin{axis}[%
width=1in,
height=1in,
at={(0,0)},
scale only axis,
axis on top,
xmin=0.5,
xmax=1024.5,
y dir=reverse,
ymin=0.5,
ymax=1024.5,
axis background/.style={fill=white},
hide axis
]
\addplot [forget plot] graphics [xmin=0.5,xmax=1024.5,ymin=0.5,ymax=1024.5] {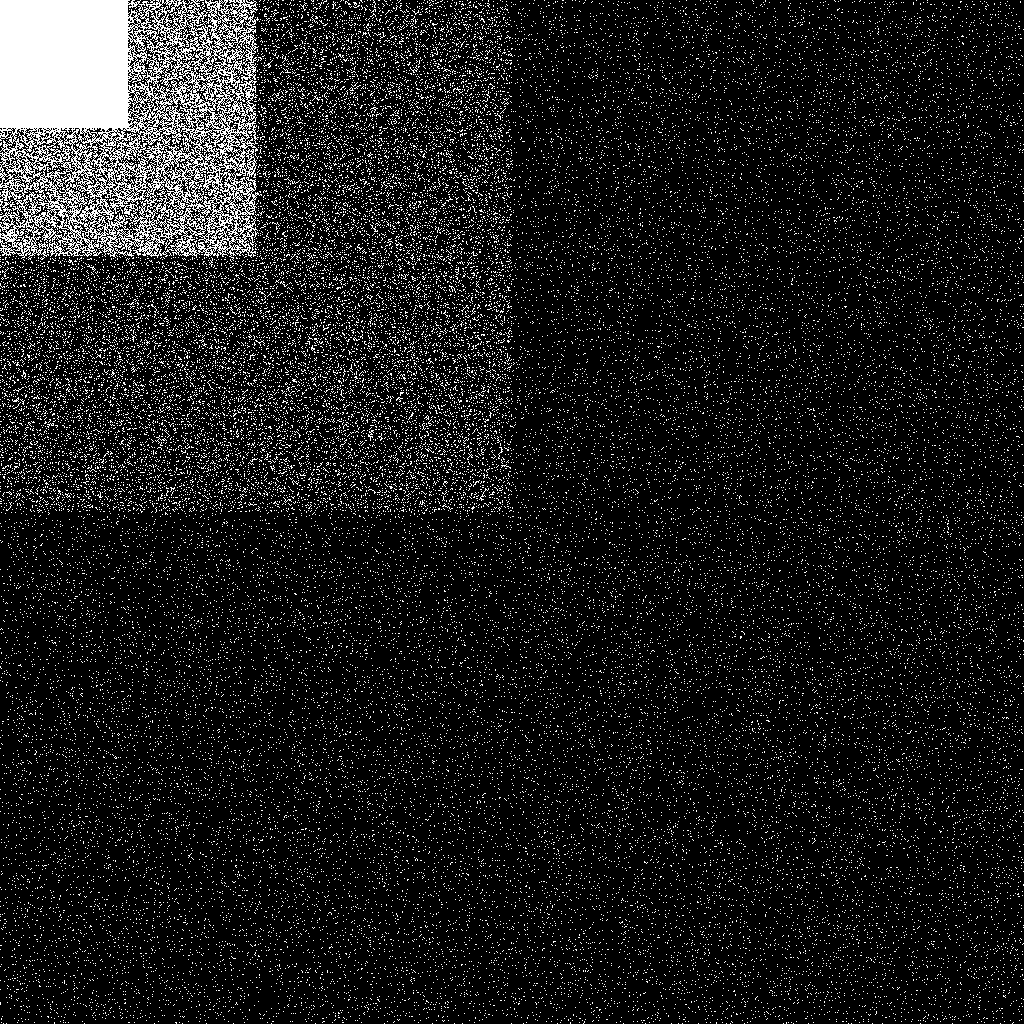};
\draw [color=white,line width=0.5mm] (axis cs:1024,0) -- (axis cs: 1024,1024);
\draw [color=black,line width=0.2mm] (axis cs:0,0) -- (axis cs: 1024,0);
\end{axis}
\end{tikzpicture}%
%
%
\begin{tikzpicture}

\begin{axis}[%
width=1in,
height=1in,
at={(0,0)},
scale only axis,
axis on top,
xmin=0.5,
xmax=1024.5,
y dir=reverse,
ymin=0.5,
ymax=1024.5,
axis background/.style={fill=white},
hide axis
]
\addplot [forget plot] graphics [xmin=0.5,xmax=1024.5,ymin=0.5,ymax=1024.5] {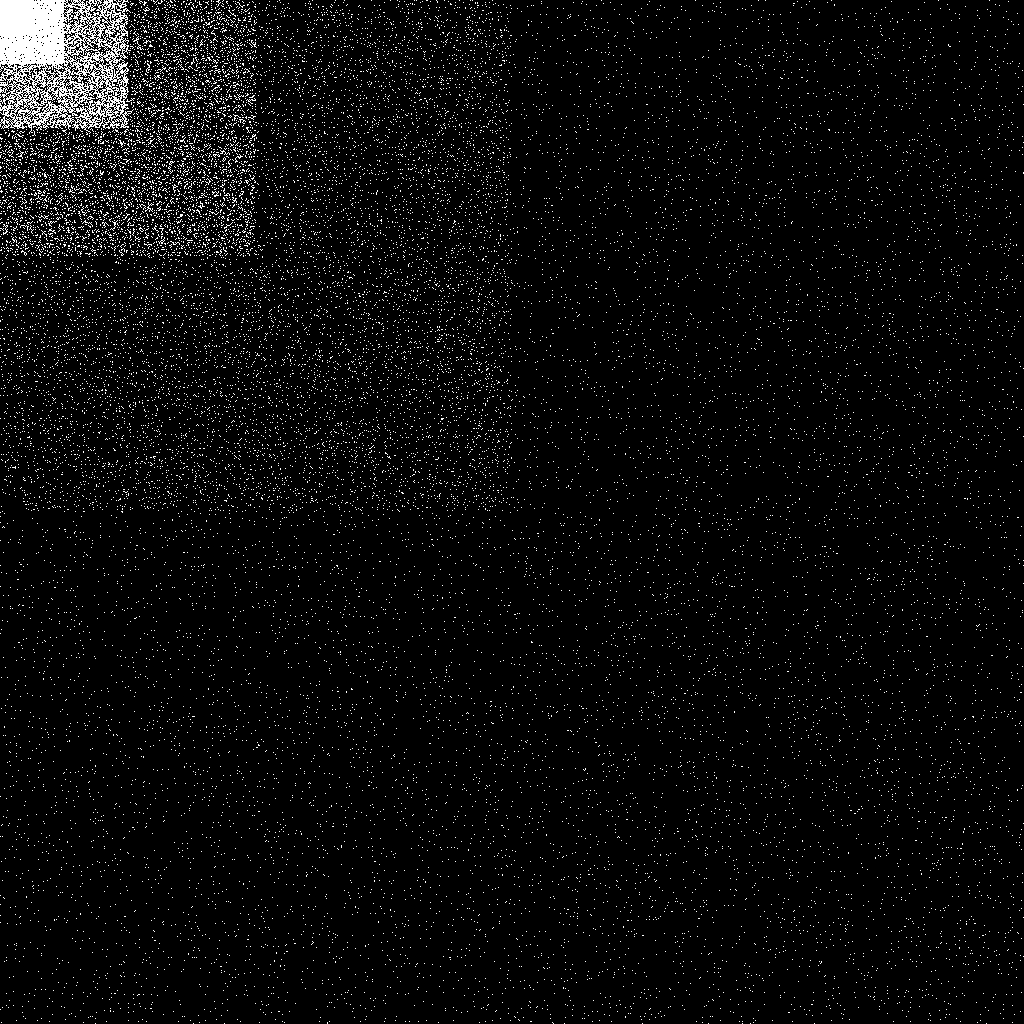};
\draw [color=black,line width=0.2mm] (axis cs:0,0) -- (axis cs: 1024,0);
\end{axis}
\end{tikzpicture}%
}
	\\
	\end{tabular}
	\end{minipage}
	\caption{An example of the reconstructed images from $10\%$ of the Hadamard measurements. These images correspond to the points $P_1, P_2, P_3$, and $P_4$ in Fig.~\ref{fig:res_phantom_2d}. Superior quality of the CS reconstruction is obvious both visually and quantitatively. We also recall the repetition in the selected indices in VDS scheme which results in less white points in the sampling pattern.}
	\label{fig:res_2d_example}
\end{figure}
\section{Proofs}\label{sec:proofs}
We now turn our attention to the proofs of the main results. We present first a few auxiliary lemmas used later in this section.
\begin{lem}\label{lem:kronecker_cartesian_relation}
Let $\bs u \in \bb C^{N}$, $\bs u' \in \bb C^{N'}$, and $\bs v = \bs u' \otimes \bs u \in \bb C^{\bar{N}}$ with $\bar{N} = NN'$. For two sets $\cl S\subset \range{N}$ and $\cl S'\subset\range{N'}$, and $\cl S = \cl S \times \cl S'$, we have
\begin{equation}
\bs P_{\overline{\cl S}}\,\bs v = (\bs P_{\cl S'}\bs u') \otimes (\bs P_{\cl S}\bs u).
\end{equation}
\end{lem}
\begin{proof}
Defining $\bs e_i \coloneqq (\bs I_N)_i$, $\bs e'_j \coloneqq (\bs I_{N'})_j$, and $\bar{\bs e}_l\coloneqq (\bs I_{\bar{N}})_l$, we first note that
$$
\ts \bs u' \otimes \bs u = \sum_{i=1}^{N_1} \sum_{j=1}^{N_2} u_i u'_j \left(\bs e'_j \otimes \bs e_i\right).
$$
Therefore,
\small
\begin{align*}
\ts \bs P_{\overline{\cl S}}\,\bs v =&  \sum_{l\in \overline{\cl S}}\bs v_l \bar{\bs e}_l 
= \sum_{(i,j)\in \cl S}  u_i  u'_j \left(\bs e'_j \otimes \bs e_i\right)
= \sum_{i\in \cl S}  u_i \Big(\sum_{j\in \cl S'}  u'_j \bs e'_j\Big) \otimes \bs e_i \\
=& \sum_{i\in \cl S} u_i \big(\left(\bs P_{\cl S'}\bs u_2\right) \otimes \bs e_i\big)
= \left(\bs P_{\cl S'}\bs u'\right) \otimes \Big(\sum_{i\in \cl S}u_i \bs e_i\Big) = \left(\bs P_{\cl S'}\bs u'\right) \otimes\left(\bs P_{\cl S}\bs u\right),
\end{align*}
\normalsize
where in the first line we used the fact that $ u_l= u_i  u_j$ and $\bar{\bs e}_l = \bs e'_j \otimes \bs e_i$ for $l \xrightleftharpoons{N_1,N_2}(i,j)$.
\end{proof}
\begin{lem}
For $\bs A \in \bb C^{M \times N}$ and $\bs B \in \bb C^{P \times Q}$, we have
\begin{subequations}
\begin{align}
\begin{split}\label{eq:useful_lem_1}
\mu (\bs A \otimes \bs B) &= \mu(\bs A) \cdot \mu (\bs B),
\end{split}\\
\begin{split}\label{eq:useful_lem_3}
\mu^{\rm loc}_{l} (\bs A \otimes \bs B)  &= \mu^{\rm loc}_{l_1} (\bs B) \cdot \mu^{\rm loc}_{l_2} (\bs A),~{\rm with~}\subtoind{l}{l_1}{l_2}{P,M}.
\end{split}
\end{align}
\end{subequations}
\end{lem}
\begin{proof}
From the definition of coherence in \eqref{eq:mutual coherenec}, 
$$
\mu (\bs A \otimes \bs B) = \max_{i,j} |(\bs A \otimes \bs B)_{i,j}| = \max_{i_1,j_1} |a_{i_1,j_1}|\cdot \max_{i_2,j_2} |b_{i_2,j_2}| = \mu(\bs A) \cdot \mu (\bs B).
$$
For the second relation, following the definition of the local coherence in \eqref{eq:local coherence}, we find
\begin{align*}
\ts \mu^{\rm loc}_{l} (\bs A \otimes \bs B) 
= \max_{j}|\left(\bs a_{l_2} \otimes \bs b_{l_1}\right)_j| 
= \max_{j_1,j_2}|a_{l_2,j_2} \cdot b_{l_1,j_1}|
= \max_{j_2}|a_{l_2,j_2}| \cdot \max_{j_1}|b_{l_1,j_1}|,
\end{align*}
where we used the relation $\subtoind{j}{j_1}{j_2}{Q,N}$.
\end{proof}
\subsection{Proof of Lemma \ref{lem:dhw matrix}}\label{proof:lem:dhw matrix}
Below, to get simpler notation, we write $\cl T_l$ instead of $\cl T^{\rm 1d}_l$. We first note from \eqref{eq:def:dhw matrix} and \eqref{eq:def:dhw0 matrix} that $\bs W^{(a)} \bs P^\top_{\cl T_0} = \bs 1_{2^r}$, since $\bs W^{(a)}_0 = [1]$, for $a \in \{0,1\}$. Since $\cl T_l \subset \cl T_{<l+1}$ and $\bar{\bs P}_{\cl T_{<l+1}} = \bs P^\top_{\cl T_{<l+1}}\bs P_{\cl T_{<l+1}}$, we have $\bs P_{\cl T_l} \bar{\bs P}_{\cl T_{<l+1}} = \bs P_{\cl T_l}$ and using Lemma~\ref{lem:proof:lem:dhw matrix} proved below we have
\begin{align}\label{eq:proof:lem:dhw matrix_1}
\ts \bs W^{(a)}_r \bs P^\top_{\cl T_l} = \ts \bs W^{(a)}_r \bar{\bs P}^\top_{\cl T_{<l+1}} \bs P^\top_{\cl T_l} = 2^{\frac{l-r}{2}} \left[\bs W^{(a)}_{l} \otimes \bs 1_{2^{r-l}},\bs 0\right]\bs P^\top_{\cl T_l}.
\end{align}
Inserting the recursive formulation of $\bs W^{(1)}_r$ and $\bs W^{(0)}_r$ in \eqref{eq:def:dhw matrix} and \eqref{eq:def:dhw0 matrix}, respectively, in \eqref{eq:proof:lem:dhw matrix_1}, using $\left(\bs A \otimes \bs B\right) \otimes \bs C = \bs A \otimes \left(\bs B \otimes \bs C\right)$ and $[\bs A,\bs B]\otimes \bs C = [\bs A \otimes \bs C, \bs B \otimes \bs C]$, and noting that both matrices $\bs W^{(a)}_{l-1}$ and $\bs I_{2^{l-1}}$ have $2^{l-1}$ columns and the operator $\bs P^\top_{\cl T_{l}}$ selects only the columns indexed in $\cl T_{l} = \{2^{l-1}+1,\cdots,2^{l}\}$, we get
\small
\begin{align}\label{eq:proof:lem:dhw matrix_2}
\ts \bs W^{(a)}_r \bs P^\top_{\cl T_l} 
= 2^{\frac{l-r-1}{2}} \left[\bs W^{(a)}_{l-1} \otimes \bs 1_{2^{r-l+1}},\bs I_{2^{l-1}} \otimes \begin{bmatrix} \bs 1_{2^{r-l}}  \\ (-1)^{a}\bs 1_{2^{r-l}} \end{bmatrix},\bs 0\right]\bs P^\top_{\cl T_l} 
= 2^{\frac{l-r-1}{2}}\bs I_{2^{l-1}} \otimes \begin{bmatrix}\bs 1_{2^{r-l}} \\ (-1)^{a}\bs 1_{2^{r-l}} \end{bmatrix}.
\end{align}
\normalsize
By expanding the right-hand side of \eqref{eq:proof:lem:dhw matrix_2}, the $(i,j)^{\rm th}$ component of the matrix $\ts \bs W^{(a)}_r \bs P^\top_{\cl T_l} $ reads
\begin{align}\label{eq:proof:lem:dhw matrix_3}
& \ts \left(\bs W^{(a)}_r \bs P^\top_{\cl T_l}\right)_{i,j} 
= \begin{cases}
2^{\frac{l-r-1}{2}}, & {\rm if~} (j-1)2^{r-l+1}+1 \le i <(j+\frac{1}{2})2^{r-l+1}+1,\\
(-1)^{a}2^{\frac{l-r-1}{2}}, & {\rm if~} (j+\frac{1}{2})2^{r-l+1}+1 \le i <(j+1)2^{r-l+1}+1,\\
0, & {\rm otherwise}.
\end{cases}
\end{align}
By comparing \eqref{eq:proof:lem:dhw matrix_3} with \eqref{eq:def:dhw} and \eqref{eq:def:dhw0}, we conclude that $(\bs \Psi^{(a)}_l)_{i,j} = \left(\bs W_r \bs P^\top_{\cl T_l}\right)_{i,j} = h^{(a)}_{l-1,j-1}(i-1)$, which completes the proof.
\begin{lem}\label{lem:proof:lem:dhw matrix}
For $a \in \{0,1\}$ and $q_l = 2^{r-1}\cdot \sum_{k=0}^{r-l}2^{-k}$,
$$
\ts \bs W^{(a)}_r \bar{\bs P}^\top_{\cl T_{<l}} = 2^{\frac{l-r-1}{2}} \left[\bs W^{(a)}_{l-1} \otimes \bs 1_{2^{r-l+1}},\bs 0_{2^r\times q_l}\right].
$$
\end{lem}
\begin{proof}
We prove this lemma by induction over the value of $l$. From \eqref{eq:def:dhw matrix} or \eqref{eq:def:dhw0 matrix} and the definition of $\bar{\bs P}^\top_{\cl T_{<r}}$ one can observe that the base case $\ts \bs W^{(a)}_r \bar{\bs P}^\top_{\cl T_{<r}} = 2^{\frac{-1}{2}} \left[\bs W^{(a)}_{r-1} \otimes \bs 1_{2},\bs 0_{2^r\times 2^{r-1}}\right]$ is true. We now show that if the statement of the lemma holds for $l = j+1$ (induction hypothesis), then it holds for $l =j$. Since $\cl T_{<j} \subset \cl T_{<j+1}$, we have $\bar{\bs P}_{\cl T_{<j}} \bar{\bs P}_{\cl T_{<j+1}} = \bar{\bs P}_{\cl T_{<j}}$, and using the induction hypothesis we can write
\small
\begin{align}\label{eq:lem:proof:lem:dhw matrix}
\ts \bs W^{(a)}_r \bar{\bs P}^\top_{\cl T_{<j}} &  
\ts = \bs W^{(a)}_r \bar{\bs P}^\top_{\cl T_{<j+1}} \bar{\bs P}^\top_{\cl T_{<j}} 
= 2^{\frac{j-r}{2}} \left[\bs W^{(a)}_{j} \otimes \bs 1_{2^{r-j}},\bs 0_{2^r\times q_{j+1}}\right] \bar{\bs P}^\top_{\cl T_{<j}}.
\end{align}
\normalsize
By injecting the recursion formula of $\bs W^{(0)}_r$ and $\bs W^{(1)}_r$ from \eqref{eq:def:dhw matrix} and \eqref{eq:def:dhw0 matrix} (with $r=j$) in \eqref{eq:lem:proof:lem:dhw matrix}, $\left(\bs A \otimes \bs B\right) \otimes \bs C = \bs A \otimes \left(\bs B \otimes \bs C\right)$ and $[\bs A,\bs B]\otimes \bs C = [\bs A \otimes \bs C, \bs B \otimes \bs C]$, we get
\small
\begin{align}
\ts \bs W^{(a)}_r \bar{\bs P}^\top_{\cl T_{<j}} &  
= 2^{\frac{j-r-1}{2}} \left[\bs W^{(a)}_{j-1} \otimes \bs 1_{2^{r-j+1}},\bs 0_{2^r\times 2^{j-1}},\bs 0_{2^r\times q_{j+1}}\right],
\end{align}
\normalsize
since $\cl T_{<j} = \range{2^{j-1}}$ and $\bar{\bs P}^\top_{\cl T_{<j}}$ preserves the first $2^{j-1}$ columns of $\bs W^{(a)}_r$. Noting that $q_{j+1} + 2^{j-1} = q_j$ confirms the statement of the lemma for $n=j$ and thus, completes the proof.
\end{proof}
\subsection{Proof of Prop.~\ref{prop: structure of had-haar}}\label{proof:prop:structure of had-haar}
From the definitions of the Hadamard and DHW bases in Sec.~\ref{sec:definitions}, we quickly obtain $\bs U^{(1)}_0 = \bs H^\top_0 \bs W^{(1)}_0 = [1]$ and $\bs U^{(0)}_0 = \bs H^\top_0 \bs W^{(0)}_0 = [1]$. Since $(\bs A \otimes \bs B)(\bs C \otimes \bs D) = (\bs A \bs C) \otimes (\bs B \bs D)$, we get, for $r\ge 1$,
\begin{align*}
	\ts \bs H^\top_r \bs W^{(1)}_{\! r} &= 
	\ts \frac{1}{2}\begin{bmatrix}\ts \big(\bs H^\top_{r-1} \otimes \begin{bsmallmatrix} 1  & \textcolor{white}{-}1 \end{bsmallmatrix}\big)\ts \big( \bs W^{(1)}_{r-1} \otimes \begin{bsmallmatrix} 1  \\ 1 \end{bsmallmatrix}\big)
	& \ts \left(\bs H^\top_{r-1} \otimes \begin{bsmallmatrix} 1  & \textcolor{white}{-}1 \end{bsmallmatrix}\right)\big(\bs I_{2^{r-1}} \otimes \begin{bsmallmatrix} \textcolor{white}{-}1  \\ -1 \end{bsmallmatrix}\big)\\
	\ts \left(\bs H^\top_{r-1} \otimes \begin{bsmallmatrix} 1  & -1 \end{bsmallmatrix}\right)\big(\bs W^{(1)}_{r-1} \otimes \begin{bsmallmatrix} 1  \\ 1 \end{bsmallmatrix}\big)
	& \ts \big(\bs H^\top_{r-1} \otimes \begin{bsmallmatrix} 1  & -1 \end{bsmallmatrix}\big)\big(\bs I_{2^{r-1}} \otimes \begin{bsmallmatrix} \textcolor{white}{-}1  \\ -1 \end{bsmallmatrix}\big)
	\end{bmatrix}\\
	&\ts = \frac{1}{2}\begin{bmatrix}\bs H^\top_{r-1}\bs W^{(1)}_{r-1} \otimes [2] & \bs H^\top_{r-1}\bs I_{2^{r-1}} \otimes [0]\\
	\bs H^\top_{r-1}\bs W^{(1)}_{r-1} \otimes [0]
	& \ts\bs H^\top_{r-1}\bs I_{2^{r-1}} \otimes [2]
	\end{bmatrix}= \begin{bmatrix}\bs H_{r-1}\bs W^{(1)}_{r-1} & \bs 0\\
		\bs 0
		& \bs H_{r-1}
		\end{bmatrix}.
\end{align*}
Similarly, we can write
\begin{align*}
	\ts \bs H^\top_r \bs W^{(0)}_r &= 
	 \ts \frac{1}{2}\begin{bmatrix}\big(\bs H^\top_{r-1} \otimes \begin{bsmallmatrix} 1  & \textcolor{white}{-}1 \end{bsmallmatrix}\big)\big(\bs W^{(0)}_{r-1} \otimes \begin{bsmallmatrix} 1  \\ 1 \end{bsmallmatrix}\big)
	& \big(\bs H^\top_{r-1} \otimes \begin{bsmallmatrix} 1  & \textcolor{white}{-}1 \end{bsmallmatrix}\big)\big(\bs I_{2^{r-1}} \otimes \begin{bsmallmatrix} 1  \\ 1 \end{bsmallmatrix}\big)\\
	\big(\bs H^\top_{r-1} \otimes \begin{bsmallmatrix} 1  & -1 \end{bsmallmatrix}\big)\big(\bs W^{(0)}_{r-1} \otimes \begin{bsmallmatrix} 1  \\ 1 \end{bsmallmatrix}\big)
	& \big(\bs H^\top_{r-1} \otimes \begin{bsmallmatrix} 1  & -1 \end{bsmallmatrix}\big)\big(\bs I_{2^{r-1}} \otimes \begin{bsmallmatrix} 1  \\ 1 \end{bsmallmatrix}\big)
	\end{bmatrix}\\
	&\ts = \frac{1}{2}\begin{bmatrix}\bs H^\top_{r-1}\bs W^{(0)}_{r-1} \otimes [2] & \bs H^\top_{r-1}\bs I_{2^{r-1}} \otimes [2]\\
	\bs H^\top_{r-1}\bs W^{(0)}_{r-1} \otimes [0]
	& \bs H^\top_{r-1}\bs I_{2^{r-1}} \otimes [0]
	\end{bmatrix}.
\end{align*}
By recursion, and from the definition of the 1-D dyadic levels $\cl T^{\rm 1d}$ we then get the structure described in Fig.~\ref{fig:structure in hadamard haar}. Moreover, from  Fig.~\ref{fig:structure in hadamard haar}-left and using the fact that $\bs H_r$ is symmetric, we conclude that $\bs U^{(1)}_r$ is symmetric as well. 
\subsection{Proof of Prop.~\ref{prop: structure of 2d-had-haar}}\label{proof:prop:structure of 2d-had-haar}
From Remark~\ref{rem:aniso_level_adhw} and Remark~\ref{rem:iso_aniso_level_2had} we can write, for $\subtoind{t}{t_1+1}{t_2+1}{r+1}$ and $\subtoind{l}{l_1+1}{l_2+1}{r+1}$,
\begin{align*}
\ts \bs P_{\cl T^{\rm aniso}_t}\bs \Phi^\top_{\rm 2had} \bs \Psi_{\rm adhw}\bs P^\top_{\cl T^{\rm aniso}_l}
& = \big(\bs P_{\cl T^{\rm 1d}_{t_2}} \bs H_{t_2}\bs W^{(1)}_{l_2}\bs P^\top_{\cl T^{\rm 1d}_{l_2}}\big) \otimes \big(\bs P_{\cl T^{\rm 1d}_{t_1}} \bs H_{t_1}\bs W^{(1)}_{l_1}\bs P^\top_{\cl T^{\rm 1d}_{l_1}}\big) \in \bb R^{2^{t_1+t_2-2}\times 2^{l_1+l_2-2}},
\end{align*}
and this matrix, using \eqref{eq:rem:structure of had-haar_1}, is $\bs H_{(t_2-1)_+} \otimes \bs H_{(t_1-1)_+}$, if $t_1=l_1$ and $t_2=l_2$ (and $\bs 0$ otherwise).

We now prove the second part of the proposition, and we simply write $\cl T_l$ for $\cl T^{\rm 1d}_l$. Recall from the definition of the IDHW basis and the 2-D isotropic wavelet levels in Sec.~\ref{sec:definitions} that
\begin{equation}\small\label{eq:proof_multilevelcoherence_idhw}
\ts \bs \Psi_{\rm idhw} \bs P^\top_{\cl T_l^{\rm iso}} =  \left[
\left(\bs W^{(0)}_r\bs P^\top_{\cl T_l}\right) \otimes \left(\bs W^{(1)}_r\bs P^\top_{\cl T_l}\right),
\left(\bs W^{(1)}_r\bs P^\top_{\cl T_l}\right) \otimes \left(\bs W^{(1)}_r\bs P^\top_{\cl T_l}\right),
\left(\bs W^{(1)}_r\bs P^\top_{\cl T_l}\right) \otimes \left(\bs W^{(0)}_r\bs P^\top_{\cl T_l}\right)
\right],
\end{equation}
for $l \in \range{r}$. Define $\bs U^{(1)} \coloneqq \bs H_r\bs W^{(1)}_r$ and $\bs U^{(0)} \coloneqq \bs H_r\bs W^{(0)}_r$, and $\bs V^{(t,l)} \coloneqq\bs P_{\cl T^{\rm iso}_t} \bs \Phi_{\rm 2had}^\top\bs \Psi_{\rm idhw}\bs P^\top_{\cl T^{\rm iso}_l}$. For the proof we need to compute $\bs V^{(t,l)}$ for $t,l\in\range{r}_0$. First, we assume that $t,l\in \range{r}$. From Remark \ref{rem:iso_aniso_level_2had} and \eqref{eq:proof_multilevelcoherence_idhw} we have
\small
\begin{align*}
\bs V^{(t,l)} & \!= \!\!
\begin{bsmallmatrix}
(\bs P_{\cl T_{<t}} \bs U^{(0)} \bs P^\top_{\cl T_l}) \,\otimes\, (\bs P_{\cl T_{t}} \bs U^{(1)} \bs P^\top_{\cl T_l})
&
(\bs P_{\cl T_{<t}} \bs U^{(1)} \bs P^\top_{\cl T_l})\, \otimes\, (\bs P_{\cl T_{t}} \bs U^{(1)} \bs P^\top_{\cl T_l})
&
(\bs P_{\cl T_{<t}} \bs U^{(1)} \bs P^\top_{\cl T_l})\, \otimes\, (\bs P_{\cl T_{t}} \bs U^{(0)} \bs P^\top_{\cl T_l})
\\
~\,(\bs P_{\cl T_t} \bs U^{(0)} \bs P^\top_{\cl T_l})\,\otimes\, (\bs P_{\cl T_{t}} \bs U^{(1)} \bs P^\top_{\cl T_l})
&
~\,(\bs P_{\cl T_t} \bs U^{(1)} \bs P^\top_{\cl T_l})\, \otimes\, (\bs P_{\cl T_{t}} \bs U^{(1)} \bs P^\top_{\cl T_l})
&
~\,(\bs P_{\cl T_t} \bs U^{(1)} \bs P^\top_{\cl T_l})\, \otimes\, (\bs P_{\cl T_{t}} \bs U^{(0)} \bs P^\top_{\cl T_l})
\\
~~~(\bs P_{\cl T_t} \bs U^{(0)} \bs P^\top_{\cl T_l})\,\otimes\, (\bs P_{\cl T_{<t}} \bs U^{(1)} \bs P^\top_{\cl T_l})
&
~~~(\bs P_{\cl T_t} \bs U^{(1)} \bs P^\top_{\cl T_l})\, \otimes\, (\bs P_{\cl T_{<t}} \bs U^{(1)} \bs P^\top_{\cl T_l})
&
~~~(\bs P_{\cl T_t} \bs U^{(1)} \bs P^\top_{\cl T_l})\, \otimes\, (\bs P_{\cl T_{<t}} \bs U^{(0)} \bs P^\top_{\cl T_l})
\end{bsmallmatrix}.
\end{align*}
\normalsize
From Remark \ref{rem:structure of had-haar} (with an attention to the conditions on the right-hand side of the relations) we observe that the diagonal blocks in $\bs V^{(t,l)}$ are equal to $\bs H_{t-1}\otimes \bs H_{t-1}$ if $t=l$ and $\bs 0$ otherwise. Therefore, if $t=l$,
\small
\begin{align*}
\bs V^{(t,l)} =
\begin{bmatrix}
\bs H_{{(t-1)}} \otimes \bs H_{{(t-1)}}
&
\bs 0
&
\bs 0
\\
\bs 0
&
\bs H_{{(t-1)}} \otimes \bs H_{{(t-1)}}
&
\bs 0
\\
\bs 0
&
\bs 0
&
\bs H_{{(t-1)}} \otimes \bs H_{{(t-1)}}
\end{bmatrix} = \bs I_3 \otimes \big(\bs H_{{(t-1)}} \otimes \bs H_{{(t-1)}}\big),
\end{align*}
\normalsize
while $\bs V^{(t,l)}= \bs 0$ if $t\ne l$. Second, we compute $\bs V^{(t,l)}$ for $t \in \range{r}_0$ and $l=0$. Since $\bs \Psi_{\rm idhw} \bs P^\top_{\cl T_0^{\rm iso}} = \left(\bs W^{(0)}_r\bs P^\top_{\cl T_0}\right) \otimes \left(\bs W^{(0)}_r\bs P^\top_{\cl T_0}\right)$, $\bs U^{(0)}\bs P^\top_{\cl T_0} = \bs I_{2^{r}}\bs P^\top_{\{1\}}$, and from Remark \ref{rem:iso_aniso_level_2had}, we have
\begin{align*}
\bs V^{(t,0)} & = 	
	\begin{bmatrix} 	(\bs P_{\cl T_{<t}} \bs U^{(0)}\bs P^\top_{\cl T_0}) \otimes (\bs P_{\cl T_{t}} \bs U^{(0)}\bs P^\top_{\cl T_0})
		\\
		~\,(\bs P_{\cl T_t} \bs U^{(0)}\bs P^\top_{\cl T_0}) \otimes (\bs P_{\cl T_{t}} \bs U^{(0)}\bs P^\top_{\cl T_0})
		\\
		~~~(\bs P_{\cl T_t} \bs U^{(0)}\bs P^\top_{\cl T_0}) \otimes (\bs P_{\cl T_{<t}} \bs U^{(0)}\bs P^\top_{\cl T_0})
		\end{bmatrix} 
	 =	\begin{bmatrix} 	\bs P_{\overline{\cl T_{t} \times \cl T_{<t}}} \bs I_{2^{2r}}\bs P^\top_{\{1\}}
		\\
		\bs P_{\overline{\cl T_{t} \times \cl T_{t}}} \bs I_{2^{2r}}\bs P^\top_{\{1\}} 
		\\
		\bs P_{\overline{\cl T_{<t} \times \cl T_{t}}} \bs I_{2^{2r}}\bs P^\top_{\{1\}}
		\end{bmatrix} = \bs P_{\cl T^{\rm iso}_t}[1,\bs 0]^\top.
\end{align*}
Finally, we need to compute $\bs V^{(t,l)}$ for $t = 0$ and $l\in \range{r}$, \ie
\small
\begin{align*}
\bs V^{(0,l)} & = 
\begin{bmatrix}
(\bs P_{\cl T_{0}} \bs U^{(0)} \bs P^\top_{\cl T_l}) \otimes (\bs P_{\cl T_{0}} \bs U^{(1)} \bs P^\top_{\cl T_l})
&
(\bs P_{\cl T_0} \bs U^{(1)} \bs P^\top_{\cl T_l}) \otimes (\bs P_{\cl T_{0}} \bs U^{(1)} \bs P^\top_{\cl T_l})
&
(\bs P_{\cl T_0} \bs U^{(1)} \bs P^\top_{\cl T_l}) \otimes (\bs P_{\cl T_{0}} \bs U^{(0)} \bs P^\top_{\cl T_l})
\end{bmatrix}.
\end{align*}
\normalsize
Using \eqref{eq:rem:structure of had-haar_1} and \eqref{eq:rem:structure of had-haar_3} with $t = 0$ and $l\in \range{r}$ yields $\bs V^{(0,l)} = \bs 0$. This completes the proof.
\subsection{Proof of Prop. \ref{prop:Local coherence of Had-haar}}\label{proof:prop:Local coherence of Had-haar}
In this proof we write $\cl T_l$ for $\cl T_l^{\rm 1d}$. Recall that $\mu(\bs H_{r}) = 2^{-r/2}$, and for any $k > 1$,  $k \in \cl T_{\bar{l}(k)}$ with $\bar{l}(k) \coloneqq \lfloor \log_2(k-1)\rfloor+1$, since $\cl T_l = \range{2^l}\backslash\range{2^{l-1}}$, for $l\ge1$. We first observe that $\mu^{\rm loc}_1(\bs U^{(1)}_r) = 1$, since $(\bs H_{r})_{1,i} = (\bs W^{(1)}_{r})_{1,i} = 2^{-r/2}$ for all $i \in \range{2^r}$.

To prove \eqref{eq:local_coherence_hadhaar_1d}, note that, for $k>1$, since $\bar{\bs P}_{\Omega} = \bs P^\top_{\Omega}\bs P_{\Omega}$, for any subset $\Omega$, and $\bs P_{\{k\}} \bar{\bs P}_{\cl T_{\bar{l}(k)}} = \bs P_{\{k\}}$, $|(\bs H_r)_{i,j}| = 2^{-r/2}$ for all $i,j \in \range{2^r}$ and using \eqref{eq:rem:structure of had-haar_1},
\begin{align*}
	\mu_{k}^{\rm loc}(\bs U^{(1)}_{2^r}) 
	= \mu (\bs P_{\{k\}}\bar{\bs P}_{\cl T_{\bar{l}(k)}}\bs U^{(1)}_{2^r})
	= \mu (\bs P_{\{k\}} \bs H_{{\bar{l}(k)-1}})
	= 2^{-\frac{\bar{l}(k)-1}{2}}
	= 2^{-\frac{\lfloor \log_2(k-1) \rfloor}{2}}. 
\end{align*}
In addition, $\ts \|\bs \mu^{\rm loc}(\bs U^{(1)}_r)\|_2^2
	= 1 + \sum_{k=2}^{N}2^{-\left \lfloor \log_2 (k-1) \right \rfloor} 
	= 1+\sum_{l=0}^{r-1}2^l \cdot 2^{-l}=\log_2(N)+1$.
	
Next, to prove \eqref{eq:local_coherence_hadhaar_2d_iso}, we first note that $\mu^{\rm loc}_1(\bs \Phi_{\rm 2had}^\top \bs \Psi_{\rm idhw}) = 1$, since $(\bs \Phi_{\rm 2had})_{1,i} = (\bs \Psi_{\rm idhw})_{1,i} = 2^{-r}$ for all $i \in \range{2^{2r}}$. Consider the rule $\subtoind{k}{k_1}{k_2}{N}$. Using \eqref{eq:prop: structure of 2d-had-haar_iso} and \eqref{eq:useful_lem_3}, for $1<k \in \cl T^{\rm iso}_t$,
\begin{equation}\label{eq:proof:prop:Local coherence of Had-haar_1}
\mu^{\rm loc}_k(\bs \Phi_{\rm 2had}^\top \bs \Psi_{\rm idhw}) = \mu^{\rm loc}_{k_1}(\bs H_{(t-1)}) \cdot \mu^{\rm loc}_{k_2}(\bs H_{(t-1)}) = 2^{-(t-1)}.
\end{equation}
Moreover, we find
\begin{equation}\label{eq:proof:prop:Local coherence of Had-haar_2}
k \in \cl T^{\rm iso}_t ~~~~\Leftrightarrow ~~~~\max(k_1,k_2) \in \cl T_{t}~~~~ \Leftrightarrow ~~~~ t-1 = \lfloor \log_2(\max(k_1,k_2)-1)\rfloor.
\end{equation}
Combining \eqref{eq:proof:prop:Local coherence of Had-haar_1} and \eqref{eq:proof:prop:Local coherence of Had-haar_2} implies the local coherence relation in \eqref{eq:local_coherence_hadhaar_2d_iso}.

Moreover, since $|\cl T^{\rm iso}_t| = 3\cdot 2^{2(t-1)}$ for $t\in \range{r}$,\eqref{eq:proof:prop:Local coherence of Had-haar_1} provides
\begin{align*}
	\ts \|\bs \mu^{\rm loc}(\bs \Phi_{\rm 2had}^\top \bs \Psi_{\rm idhw})\|_2^2 
	& = 1 + \sum_{t=1}^{r} \sum_{k \in \cl T^{\rm iso}_{t}} \mu_k^{\rm loc}(\bs \Phi_{\rm 2had}^\top \bs \Psi_{\rm idhw})^2
	  = 1 + \sum_{t=1}^{r} |\cl T^{\rm iso}_t| \cdot 2^{-2(t-1)}
	  = 1 + 3 \cdot r.
\end{align*}	

Finally, to prove \eqref{eq:local_coherence_hadhaar_2d_aniso}, we first observe that $\mu^{\rm loc}_1(\bs \Phi_{\rm 2had}^\top \bs \Psi_{\rm adhw}) = 1$, since $(\bs \Phi_{\rm 2had})_{1,i} = (\bs \Psi_{\rm adhw})_{1,i} = 2^{-r}$ for all $i \in \range{2^{2r}}$. Consider the rules $\subtoind{t}{t_1+1}{t_2+1}{r+1}$ and $\subtoind{k}{k_1}{k_2}{N}$. From the construction of the 2-D anisotropic levels we have, for $k>1$,
\begin{equation}\label{eq:proof:prop:Local coherence of Had-haar_3}
\ts k \in \cl T^{\rm aniso}_t ~~~~\Leftrightarrow~~~~k_1 \in \cl T_{t_1},~k_2 \in \cl T_{t_2}~~~~\Leftrightarrow ~~~~ t_1-1 = \lfloor \log_2(k_1-1)\rfloor,~t_2-1 = \lfloor \log_2(k_2-1)\rfloor.
\end{equation}
Using \eqref{eq:prop: structure of 2d-had-haar_aniso} and \eqref{eq:useful_lem_3}, for $1<k \in \cl T^{\rm aniso}_t$, we get
\begin{align}\label{eq:proof:prop:Local coherence of Had-haar_4}
	\mu^{\rm loc}_k(\bs \Phi_{\rm 2had}^\top \bs \Psi_{\rm adhw}) 
	= \mu^{\rm loc}_k(\bs H_{t_2-1} \otimes \bs H_{t_1-1}) 
	= \mu^{\rm loc}_{k_1}(\bs H_{t_1-1}) \cdot \mu^{\rm loc}_{k_2}(\bs H_{t_2-1}).
\end{align}
Combining \eqref{eq:proof:prop:Local coherence of Had-haar_3} and \eqref{eq:proof:prop:Local coherence of Had-haar_4} with the relation in \ref{eq:local_coherence_hadhaar_1d} implies the local coherence value in \eqref{eq:local_coherence_hadhaar_2d_aniso}.

In addition, using \eqref{eq:local_coherence_hadhaar_1d},
\begin{align*}
	\ts \|\bs \mu^{\rm loc}(\bs \Phi_{\rm 2had}^\top \bs \Psi_{\rm adhw})\|_2^2 
	  = \left(1+\sum_{k_1=2}^{N} 2^{-\left \lfloor \log_2 (k_1-1) \right \rfloor} \right) \cdot \left(1+\sum_{k_2=2}^{N} 2^{-\left \lfloor \log_2 (k_2-1) \right \rfloor} \right)
	  = (\log_2(N) +1)^2.
\end{align*}
\subsection{Proof of Prop. \ref{prop:Multilevel coherence of Had-haar}}\label{proof:prop:Multilevel coherence of Had-haar}
Given $\bs U^{(1)}_r = \bs H_r \bs W^{(1)}_r$, and $\cl T_l = \cl T_l^{\rm 1d}$ for $l\in\range{r}_0$, we note that $\mu(\bs H_{r}) = 2^{-r/2}$, $\mu(\bs P_{\cl W_t}\bs A) = \max_{l}\mu(\bs P_{\cl W_t}\bs A\bs P^\top_{\cl S_l})$, and for any orthonormal matrix $\bs \Phi$, $\| \bs \Phi\|_{2,2} = \max_{\|\bs v\|_2=1} \|\bs \Phi\bs v\|_2 = \|\bs v\|_2 =1$.

We first prove \eqref{eq:multilevel_coherence_hadhaar_1d}. From Remark~\ref{rem:structure of had-haar}, note that $\mu(\bs P_{\cl T_t}\bs U^{(1)}_r\bs P^\top_{\cl T_0}) = \delta_{t,0}$ and for $\l \in \range{r}$,
\begin{equation}\label{eq:prop:Multilevel coherence of Had-haar_1}
\mu(\bs P_{\cl T_t}\bs U^{(1)}_r\bs P^\top_{\cl T_l}) 
 = \mu(\bs H_{{t-1}} )\cdot \delta_{t,l} = 2^{-(t-1)/2} \cdot \delta_{t,l}.
\end{equation}
Therefore, for $t,l \in \range{r}_0$ we obtain $\mu(\bs P_{\cl T_t}\bs U^{(1)}_r )= 2^{\ts -\frac{(t-1)_+}{2}}$, $\mu^{\cl T,\cl T}_{t,l}(\bs U^{(1)}) = 2^{-(t-1)_+} \cdot \delta_{t,l}$. To compute the relative sparsity, from Lemma~\ref{lem:relative sparsity} and Remark~\ref{rem:structure of had-haar}, and since $\|\bs H_{r}\|_{2,2}=1$, we find 
\begin{equation}\label{eq:prop:Multilevel coherence of Had-haar_2}
K_t^{\cl T, \cl T}(\bs U^{(1)}, \bs k)^{1/2} \le \sum_{l=0}^{r} \|\bs P_{\cl T_t} \bs U^{(1)}\bs P^\top_{\cl T_l}\|_{2,2} \sqrt{k_l}= \|\bs H_{(t-1)_+}\|_{2,2} \sqrt{k_t} = \sqrt{k_t}.
\end{equation}

To prove \eqref{eq:multilevel_coherence_hadhaar_2d_iso}, note from \eqref{eq:prop: structure of 2d-had-haar_iso} that $\mu(\bs P_{\cl T^{\rm iso}_t} \bs \Phi_{\rm 2had}^\top\bs \Psi_{\rm idhw}\bs P^\top_{\cl T^{\rm iso}_0}) = \delta_{t,0}$, and, for $l \in \range{r}$,
\begin{equation}\label{eq:prop:Multilevel coherence of Had-haar_3}
\mu(\bs P_{\cl T^{\rm iso}_t} \bs \Phi_{\rm 2had}^\top\bs \Psi_{\rm idhw}\bs P^\top_{\cl T^{\rm iso}_l}) 
 = \mu(\bs I_3 ) \cdot \mu(\bs H_{(t-1)} ) \cdot \mu(\bs H_{(t-1)} )\cdot \delta_{t,l} = 2^{-(t-1)} \cdot \delta_{t,l},
\end{equation}
where we used the rule in \eqref{eq:useful_lem_1}. Therefore, for $t,l \in \range{r}_0$ we obtain $\mu(\bs P_{\cl T^{\rm iso}_t} \bs \Phi_{\rm 2had}^\top)= 2^{-(t-1)_+}$ and $\mu^{\cl T^{\rm iso},\cl T^{\rm iso}}_{t,l}( \bs \Phi_{\rm 2had}^\top\bs \Psi_{\rm idhw}) = 2^{-2(t-1)_+} \cdot \delta_{t,l}$. To compute the relative sparsity, from Lemma~\ref{lem:relative sparsity} and Prop.~\ref{prop: structure of 2d-had-haar}, we have
\begin{equation}\label{eq:prop:Multilevel coherence of Had-haar_4}
K^{\cl T^{\rm iso},\cl T^{\rm iso}}_{t}( \bs \Phi_{\rm 2had}^\top\bs \Psi_{\rm idhw},\bs k)^{1/2}
 \le \sqrt{k_0}\cdot \delta_{t,0} + \sum_{l=1}^{r} \|\bs I_3 \otimes \big(\bs H_{(l-1)}\otimes \bs H_{(l-1)}\big)\|_{2,2}\,\sqrt{k_l}\cdot \delta_{t,l}=   \sqrt{k_t}.
\end{equation}

We now prove \eqref{eq:multilevel_coherence_hadhaar_2d_aniso}. Consider $t,l \in \range{(r+1)^2}$ such that $\subtoind{t}{t_1+1}{t_2+1}{r+1},~\subtoind{l}{l_1+1}{l_2+1}{r+1}$ and $t_1,t_2,l_1,l_2 \in \range{r}_0$. From \eqref{eq:prop: structure of 2d-had-haar_aniso} and using \eqref{eq:useful_lem_1} we have
\begin{equation}\label{eq:prop:Multilevel coherence of Had-haar_5}
\ts \mu(\bs P_{\cl T^{\rm aniso}_t} \bs \Phi_{\rm 2had}^\top\bs \Psi_{\rm adhw}\bs P^\top_{\cl T^{\rm aniso}_l}) 
 = \mu(\bs H_{(t_1-1)_+} )\cdot \mu(\bs H_{(t_2-1)_+} )\cdot \delta_{t_1,l_1}\cdot \delta_{t_2,l_2} = 2^{\frac{-(t_1-1)_+}{2}}\cdot 2^{\frac{-(t_2-1)_+}{2}}\cdot \delta_{t_1,l_1}\cdot \delta_{t_2,l_2}.
\end{equation}
Therefore, $\mu(\bs P_{\cl T^{\rm aniso}_t} \bs \Phi_{\rm 2had}^\top)= 2^{\frac{-(t_1-1)_+}{2}}\cdot 2^{\frac{-(t_2-1)_+}{2}}$, $\mu^{\cl T^{\rm aniso},\cl T^{\rm aniso}}_{t,l}( \bs \Phi_{\rm 2had}^\top\bs \Psi_{\rm adhw}) = 2^{-(t_1-1)_+} \cdot 2^{-(t_2-1)_+}\cdot \delta_{t_1,l_1}\cdot \delta_{t_2,l_2}$. To compute the relative sparsity, from Lemma~\ref{lem:relative sparsity} and Prop.~\ref{prop: structure of 2d-had-haar}, we have
\begin{equation}\label{eq:prop:Multilevel coherence of Had-haar_6}
\ts K^{\cl T^{\rm aniso},\cl T^{\rm aniso}}_{t}( \bs \Phi_{\rm 2had}^\top\bs \Psi_{\rm adhw},\bs k)^{1/2}
 \le  \sum_{l=1}^{(r+1)^2} \|\bs H_{(l_2-1)_+} \otimes \bs H_{(l_1-1)_+}\|_{2,2}\,\sqrt{k_l}\cdot \delta_{t,l}=  \sqrt{k_t}.
\end{equation}
\subsection{Proof of Thm.~\ref{thm:Non-uniform guarantee for Hadamard-Haar system}}\label{proof:thm:Non-uniform guarantee for Hadamard-Haar system}
Following Thm.~\ref{thm:mds}, since in all cases covered by Thm.~\ref{thm:Non-uniform guarantee for Hadamard-Haar system} (\ie 1-D Hadamard-Haar, 2-D isotropic Hadamard-Haar, and 2-D anisotropic Hadamard-Haar) we have $\cl W = \cl S$,  we need to show that the sample-complexity bound for each case satisfies
\begin{align*}
	& m_t \gtrsim |\cl S_t| \cdot \Big(\sum_{l=1}^{|\cl S|}\mu_{t,l}^{\cl S,\cl S}(\bs \Phi^\top\bs \Psi) \cdot k_l\Big)\cdot \log(K\epsilon^{-1})\cdot\log(N),\\
	& m_t \gtrsim \hat{m}_t\cdot\log(K\epsilon^{-1})\cdot\log(N),
\end{align*}
where $\hat{m}_t$ must satisfy
\begin{align*}
	\sum_{t=1}^{|\cl S|} \frac{|\cl S_t|\cdot\mu_{t,l}^{\cl S,\cl S}(\bs \Phi^\top\bs \Psi)\cdot K_t^{\cl S,\cl S}(\bs \Phi^\top\bs \Psi,\bs k)}{\hat{m}_t}\  \lesssim 1,~~{\rm for}~l \in \range{|\cl S|}.
\end{align*}
Moreover, since in the three covered cases the multilevel coherence $\mu_{t,l}^{\cl S,\cl S}(\bs \Phi^\top\bs \Psi)$ vanishes for $t \ne l$, and $\mu_{t,l}^{\cl S,\cl S}(\bs \Phi^\top\bs \Psi) = |\cl S_l|^{-1}$ for $t = l$, the proof is further simplified, as the condition on $\hat{m}_t$ holds if $\hat{m}_l \gtrsim K_l^{\cl S,\cl S}(\bs \Phi^\top\bs \Psi,\bs k)$. Thus, it suffices to show that in each case
\begin{align*}
	 m_t \gtrsim \max \left(K_t^{\cl S,\cl S}(\bs \Phi^\top\bs \Psi,\bs k),k_t\right)\cdot \log(K\epsilon^{-1})\cdot\log(N).
\end{align*}
However, for the three cases, $\max\left(K_t^{\cl S,\cl S}(\bs \Phi^\top\bs \Psi,\bs k),k_t \right)= k_t$. Therefore, $m_t \gtrsim k_t \cdot \log(K\epsilon^{-1})\cdot\log(N)$ for all the three cases, which completes the proof.
\section{Discussion}
This work has studied the Hadamard-Haar systems in the context of CS theory, \ie the problem of recovering signals from subsampled Hadamard measurements using Haar wavelet sparsity basis. 

Traditional UDS scheme is inapplicable in Hadamard-Haar systems, since the Hadamard and Haar bases are maximally coherent. The new CS principles, \ie local and multilevel coherences, introduced by Krahmer and Ward \cite{krahmer2014stable} and by Adcock~\etal~\cite{adcock2016note}, respectively, inspired us to design sampling strategies that require minimum number of Hadamard measurements and in the same time allow stable and robust signal recovery. By computing the exact values of local and multilevel coherences we achieved the tight sample-complexity bounds for both uniform and non-uniform recovery guarantees. In two-dimensions, we considered two constructions of the 2-D Haar wavelet basis, \ie using either tensor product of two 1-D Haar bases or the isotropic construction of a multi-resolution analysis; and observed that an efficient design of sampling strategy for each system is unique.

Our results have been illustrated by several numerical tests for different types of signals with varying resolution, sparsity, and number of measurements. In particular, we have numerically demonstrated the impact of the resolution in signal recovery. 

Our uniform recovery guarantee in Thm.~\ref{thm:uniform guarantee for Hadamard-Haar system} is linked to the $\ell_1$ minimization problem \eqref{eq:BPDN uniform}. A variant of this problem would be to replace the $\ell_1$-norm term with the total variation norm. Following the proof of Thm.~3.1 in \cite{krahmer2014stable} we believe that the same sample-complexity bounds and sampling strategies as in Thm.~\ref{thm:uniform guarantee for Hadamard-Haar system} provides stable and robust signal recovery (from subsampled Hadamard measurements) via the total variation norm minimization problem. However, we postpone this potential extension to a future study. 

As mentioned in the introduction, Li and Adcock \cite{li2017compressed} have recently developed a uniform version of the recovery guarantee for MDS scheme in Prop.~\ref{thm:mds}. The computed multilevel coherences in Prop.~\ref{prop:Multilevel coherence of Had-haar} can be directly applied to the sample-complexity bound in \cite[Thm.~3.1]{li2017compressed}. Due to the uniform recovery nature of Thm.~3.1~in \cite{li2017compressed}, the final sample-complexity bounds for Hadamard-Haar systems, in the context of MDS scheme, would be the same as the ones in Thm.~\ref{thm:Non-uniform guarantee for Hadamard-Haar system} up to some extra log factors.

Following the uncovered cells in Table.~\ref{tab:related works}, a line of study would be to characterize the effect of the other sparsity bases on our local and multilevel coherence analysis, \eg the 2-D Daubechies wavelets.

Finally, in this respect, it is worth mentioning that the recurrence relations provided by the Kronecker factorization in \eqref{eq:def:dhw matrix} and  \eqref{eq:def:hadamard_matrix} goes beyond the Hadamard and Haar matrices. In fact, the Kronecker product has been used to describe a range of other unitary matrices, \eg the discrete Fourier transform and the related Sine, Cosine, and Hartley transforms \cite{granata1992,loan2000,regalia1989kronecker}; see also \cite{fijany1998quantum} for the factorization of the Daubechies wavelets. An interesting research would be to investigate the combinations of different sensing and sparsity bases and to find other scaling structures.
\section*{Acknowledgment}                        
\label{sec:acknowledgment}
We would like to thank Ben Adcock for the his valuable remarks during the iTWIST'18 workshop (Marseille, France).
\bibliographystyle{IEEEtran}        
\bibliography{Main}

\end{document}